\documentclass[12pt]{article}
\usepackage[utf8]{inputenc}
\usepackage[T1]{fontenc}

\usepackage[nocompress]{cite}
\usepackage{IEEEtrantools,stackengine}
\stackMath

\usepackage{authblk}
\usepackage{fullpage}
\usepackage{amssymb,amsmath}

\usepackage{siunitx}
\usepackage{csquotes}

\usepackage{xspace}
\newcommand\ie{i.\,e.\xspace}
\newcommand\eg{e.\,g.\xspace}

\usepackage{booktabs}
\usepackage{longtable}
\usepackage{multirow}
\usepackage{tabularx}

\usepackage[dvipsnames]{xcolor}

\usepackage[left]{lineno}

\usepackage[
    colorinlistoftodos,
    textsize=footnotesize,
        ]{todonotes}

\definecolor{darkgreen}{rgb}{0.0, 0.5, 0.0}

\usepackage{setspace}
\usepackage[unicode=true]{hyperref}
\hypersetup{breaklinks=true,
            bookmarks=true,
            colorlinks=true,
            pdfborder={0 0 0},
            allcolors=blue}
\usepackage[%
  sort&compress
]{cleveref}
  \Crefname{appendix}{Supplement}{Supplements}

\usepackage{times}
\usepackage{enumerate}
\usepackage{float}

\usepackage{subfig}
\usepackage{graphicx}

\usepackage{zref-xr,zref-user}
\zexternaldocument*{supplements}

\newcommand{\figletter}[1]{{{\fontfamily{\sfdefault}\selectfont \textbf{#1}}}}

\usepackage{rotating}
\usepackage{tabularx}
\usepackage{dcolumn}
\usepackage{pdflscape}
\usepackage{rotating}

\usepackage{array}

\usepackage{sectsty}
  \subsubsectionfont{\normalfont\itshape}

\makeatletter
\renewcommand{\fps@figure}{H}         
\renewcommand{\fps@table}{H}         
\makeatother

\allowdisplaybreaks

\usepackage{eurosym}

\usepackage{ntheorem}
\theoremseparator{:}

\usepackage{dirtytalk}
\usepackage{acro}

\DeclareAcronym{UN}{
  short=UN,
  long=United Nations,
}
\DeclareAcronym{IRA}{
  short=IRA,
  long=Internet Research Agency,
}
\DeclareAcronym{usa}{
  short=U.S.,
  long=United States of America,
}
\DeclareAcronym{UK}{
  short=U.K.,
  long=United Kingdom,
}

\begin{document}


\title{\centering\LARGE\singlespacing Russian propaganda on social media\\ during the 2022 invasion of Ukraine}

\renewcommand\Affilfont{\fontsize{9}{10.8}\selectfont}

\author[1,2]{Dominique Geissler}
\author[1,2]{Dominik Bär}
\author[3]{Nicolas Pröllochs}
\author[1,2]{Stefan Feuerriegel}

\affil[1]{LMU Munich, Munich, Germany}
\affil[2]{Munich Center for Machine Learning (MCML), Germany}
\affil[3]{University of Giessen, Germany}
\affil[ ]{\{d.geissler, baer\}@lmu.de, nicolas.proellochs@wi.jlug.de, feuerriegel@lmu.de}

\date{}

\maketitle

\newpage
\nolinenumbers
\begin{abstract}\normalfont
\noindent
The Russian invasion of Ukraine in February 2022 was accompanied by practices of information warfare, yet existing evidence is largely anecdotal while large-scale empirical evidence is lacking. Here, we analyze the spread of pro-Russian support on social media. For this, we collected $N = \num{349455}$ messages from Twitter with pro-Russian support. Our findings suggest that pro-Russian messages received $\sim$\num{251000} retweets and thereby reached around \num{14.4}~million users. We further provide evidence that bots played a disproportionate role in the dissemination of pro-Russian messages and amplified its proliferation in early-stage diffusion. Countries that abstained from voting on the United Nations Resolution \mbox{ES-11/1} such as India, South Africa, and Pakistan showed pronounced activity of bots. Overall, \num{20.28}\% of the spreaders are classified as bots, most of which were created at the beginning of the invasion. Together, our findings suggest the presence of a large-scale Russian propaganda campaign on social media and highlight the new threats to society that originate from it. Our results also suggest that curbing bots may be an effective strategy to mitigate such campaigns.
\end{abstract}

\flushbottom
\maketitle
\thispagestyle{empty}

\begin{center}
\begin{tabular}{p{14.5cm}}
\small
\noindent\textbf{Keywords}: social media, online spreading, propaganda, bots, Russo-Ukraine war
\end{tabular}
\end{center}

\sloppy
\raggedbottom


\newpage
\section*{Main}
\label{sec:introduction}

On February 24, 2022, Russia invaded Ukraine \cite{UN.SC.War.2022, CNN.2022}, thereby escalating the Russo-Ukrainian war that began with the annexation of Crimea in 2014 \cite{UN.SC.2016}. As of now, the war has led to a major energy crisis \cite{BBC.Energy.2022}, global food shortages \cite{TheEconomist.Food.2022}, and one of the largest refugee crises with more than \num{7} million Ukrainian refugees \cite{UNHCR.2022}. The invasion was later deplored by the United Nations (UN) General Assembly, with 141 countries approving Resolution \mbox{ES-11/1}, 5 countries voting against (e.g., Belarus, North Korea), and 35 countries abstaining (e.g., India, South Africa, and Pakistan) \cite{UN.GA.2022}.     

A widespread concern is that practices of modern warfare in form of large-scale Russian propaganda campaigns are used to shape the narrative around the war, yet corresponding research is still nascent. On the one hand, the Russian government enforced new legislation exerting power over traditional media outlets to persuade citizens to support the war. As a result, domestic media outlets are forced to adopt the official narrative \cite{Sloane.2022, Alyukov.2022, Troianovski.03042022}. On the other hand, Russian propaganda has been suspected to influence other countries outside Russia, in particular, by using social media to promote hostility against the West. Here, one goal could be to diminish the support for sanctions against Russia and to weaken the support for Ukraine, especially in countries that have abstained from approving the United Nations Resolution \mbox{ES-11/1} deploring the invasion. However, evidence of Russian propaganda campaigns from the 2022 invasion of Ukraine is so far purely anecdotal, whereas rigorous empirical evidence is missing. 

Russian propaganda has been documented in several Western countries during previous conflicts \cite{Alieva.2022, Golovchenko.2020}. Oftentimes, the underlying narratives are recycled from past propaganda campaigns \cite{Yablokov.2022, Sanovich.2017} and aim to destabilize democratic countries by sowing doubt and polarizing citizens \cite{Yablokov.2022}. With the rise of the Internet, propaganda campaigns increasingly make use of social media. This gives rise to growing concerns that social media may be strategically used to increase political division and influence public opinion as a tool of modern warfare  \cite{Ratkiewicz.2018, Bail.2020, Golovchenko.2018, DelVicario.2016}. For example, a coordinated social media campaign was launched by a Russian organization known as the {Internet Research Agency} (IRA) during the 2014 Russo-Ukrainian conflict \cite{Bail.2020, Doroshenko.2021}. The IRA has also been suspected of meddling in several elections. Among others, the IRA aimed to influence the outcomes of the 2016 U.S. presidential election \cite{Shao.2018, Badawy.2018, Guess.2020, Luceri.2020, Bessi.2016, Dutta.2021, Arif.2018}, even though the influence on voting behavior has been questioned \cite{Eady.2023}. Other examples of foreign influence operations through the IRA are, e.g., the U.K. Brexit Referendum \cite{Grcar.2017}, and the 2017 French presidential election \cite{Ferrara.2017}. Yet, the aforementioned works focus on historical tactics of the IRA, while it is likely that the tactics of Russian foreign influence operations have become more refined over time. For example, in 2016, the IRA primarily employed trolls (rather than automated accounts such as bots) to influence foreign events \cite{Golovchenko.2020b}, and Twitter has taken actions to find and remove accounts associated with the IRA \cite{Twitter.IRA}. Hence, it is likely that social media campaigns such as from Russian propaganda have become more advanced over time and employ new tactics, which thus pose the need for new, large-scale empirical evidence.

A particular threat of social media is that propaganda campaigns can reach online exposure at an unprecedented scale. While previous campaigns from the IRA relied largely upon trolls to spread propaganda \cite{Golovchenko.2020b, Twitter.IRA}, it is likely that current influence operations make increasing use of bots. Generally, bots allow producing high volumes of software-controlled social media profiles at low cost \cite{Ferrara.2016}. Previously, bots have been deployed to spread disinformation, fake news, and hate speech on social media \cite{Chen.2021, Shao.2018, Stella.2018, Caldarelli.2020}. In particular, they aid in the spread of low-credibility content (\eg, misinformation, false news) by amplifying early-stage diffusion \cite{Shao.2018}. Despite that bots post and receive less retweets than humans in social media networks, bots still attract more attention than human accounts \cite{GonzalezBailon.2021} and thus can proliferate content that would otherwise not go viral \cite{Caldarelli.2020}. An example of this was seen by the role of bots in the 2016 U.S. presidential election \cite{Bessi.2016, Badawy.2018, Badawy.2019}. The result is that bots have the potential to shape the online discourse, radicalize users, and amplify social division \cite{Stella.2018, Bail.2020}. In the context of Russian propaganda, anecdotal evidence suggests that Russia invested in automated disinformation tools and \enquote{bot farms} for many years \cite{Bail.2020, Doroshenko.2021, Stukal.2017}. This raises the concern that pro-Russian bots may fuel and amplify Russian propaganda efforts also during the 2022 Russian invasion of Ukraine.

In this paper, we analyze the spread of pro-Russian support on social media. For this, we collected $N = \num{349455}$ messages from February through July 2022 with pro-Russian content from Twitter. Our analysis is three-fold. First, we analyze the overall reach of the pro-Russian messages. 
We find that pro-Russian messages received more than $\sim$\num{251000} retweets and thereby reached $\sim$\num{14.4}~million users. Second, we analyze the strategy with which pro-Russian messages were disseminated. In particular, we document a disproportionate role of bots, which suggests the presence of a coordinated campaign: $\sim$\num{20.28}\% of the spreaders are classified as bots, and most of them were created at the beginning of the invasion. Third, we study between-country heterogeneity in the impact of bots and find pronounced bot activity in countries abstaining from voting on United Nations Resolution \mbox{ES-11/1} such as India, South Africa, and Pakistan.
Together, our findings provide evidence for a Russian propaganda campaign, which was disseminated widely on social media and was amplified by bots in the early diffusion.
Finally, our findings have important implications for designing effective counter-strategies to mitigate societal threats from propaganda in modern warfare.

\newpage
\section*{Methods}
\label{sec:methods}

\subsection*{Data collection} 
The data for this study were collected from the social media platform Twitter (\url{http://twitter.com}). Twitter was chosen because it is widely used for news consumption (in addition to entertainment) \cite{Mitchell.2021} and because of its high popularity in various parts of the world including Western, African, and Asian countries \cite{Statista.Twitter.User.2022}. This is different from other social media platforms that sometimes have only a narrow user base in a specific geographic region, whereas our choice should allow us to study cross-country heterogeneity in pro-Russian support.

We queried the Twitter API v2 (Academic Research track) \cite{Twitter.API.2022} for messages (source tweets, retweets, and replies) from February~1, 2022 through July~31, 2022. For this, we first defined a ``seed'' search query which we then expanded iteratively. Specifically, we started with the hashtag \texttt{\#istandwithrussia}, which was a widespread hallmark of pro-Russian support on Twitter and among the most trending hashtags on both March~2 and March~3, 2022. We then analyzed a random subsample of \num{1000} messages to search for other pertinent hashtags that may have been used to signal pro-Russian support. As a result, we identified three additional common hashtags with a clearly pro-Russian connotation (\ie, \texttt{\#standwithrussia}, \texttt{\#istandwithputin}, and \texttt{\#standwithputin}), and we then queried Twitter also for these hashtags.
Note that the above hashtags likely capture the bulk of messages with pro-Russian hashtags on Twitter. The reason is that other (less common) hashtags that may also be indicative of pro-Russian support are typically used in conjunction with at least one of these hashtags (see the example messages in Supplementary~Table~\ref{tab:russian_tweets}). 

We decided to use hashtags, instead of keywords, as search terms for multiple reasons. (1) The chosen hashtags went surprisingly viral in March 2022 and were suspected to be part of a larger propaganda campaign \cite{TheEconomist.2022, BBC.2022, Atlantic.2022}. Hence, to provide large-scale, empirical evidence of such a campaign, an analysis based on messages containing these hashtags is necessary. (2) The use of hashtags is more strict than the use of keywords as search terms. This way, we ensured to only record messages that were likely part of the coordinated propaganda campaign. (3) The query hashtags contain distinct pro-Russian stances that more general keywords do not cover. This ensures that we capture pro-Russian support on Twitter rather than a more general discussion of the invasion. 

Overall, our dataset consists of $N = \num{368762}$ messages (\ie, source tweets, retweets, and replies) with pro-Russian hashtags that were posted by \num{139591} different users. The majority of messages (\num{80.93}\%) was written in English.

\subsection*{Preprocessing} 
While our data collection allows for comprehensive coverage, the use of pro-Russian hashtags does not always equate to a pro-Russian stance. For example, users expressing an anti-Russian view sometimes employ pro-Russian hashtags to connect to the existing discourse. Similarly, Western news media report on the information warfare using the pro-Russian hashtags. After manual inspection, we found several false positives in our dataset, that is, Twitter messages that express an anti-Russian view or journalistic content, even though the message still uses a pro-Russian hashtag (\eg, \texttt{\#istandwithrussia}). To remove false positives, we proceeded as follows. (1)~We manually identified a list of 19 different anti-Russian and anti-Putin hashtags (\eg, \texttt{\#stopputinnow}, \texttt{\#stoprussia}). Note that we selected only hashtags that clearly shift the stance of a Twitter message, and, thus, one would not expect to find these hashtags in pro-Russian messages. The list is in Supplementary~Table~\ref{tab:anti-russian_hashtags}. (2)~We discarded all messages containing one or more of the aforementioned hashtags. (3)~We manually checked all verified accounts in our dataset and identified 44 Western news media outlets (\eg, NBC News, The Times). We used our common knowledge, as well as the biographies and queried messages of verified accounts to identify these news media outlets. The list is in Supplementary~Table~\ref{tab:western_news_media}. We then discarded all messages from the aforementioned Western news outlets (as well as retweets of those messages) as they were merely reporting on Russian propaganda on Twitter using the query hashtags.

Overall, the filtering removed \num{19307} messages (i.e., \num{5.24}\%). The resulting dataset contains $N=\num{349455}$ pro-Russian messages from \num{132131} users, out of which \num{250853} messages (\num{71.78}\%) were retweets.

\subsection*{Dataset with pro-Ukrainian support} 
To compare pro-Russian and pro-Ukrainian support on Twitter, we collected a second dataset via the Twitter API v2 \cite{Twitter.API.2022}. We performed the search analogous to the above; that is, we limited the search to the same time frame (February~1, 2022 - July~31, 2022) and used a comparable set of hashtags in our search query: \texttt{\#istandwithukraine}, \texttt{\#standwithukraine}, \texttt{\#istandwithzelensky}, and \texttt{\#standwithzelensky}. 

We applied the same preprocessing procedure to the messages with pro-Ukrainian support. To remove false positives, we identified five anti-Ukrainian hashtags that clearly shift the stance of the messages (see Supplementary~Table~\ref{tab:anti-ukrainian_hashtags}). Overall, the filtering removed \num{461} messages. This left us with $N = \num{9818566}$ messages (i.e., source tweets, retweets, and replies) posted by \num{2079198} users, which we consider as pro-Ukraine.  Unless stated otherwise, all analyses in the main paper refer to the dataset with pro-Russian support (and not to the dataset with pro-Ukrainian support).

\subsection*{Human validation}
We validated our preprocessing approach against human annotations following best practices \cite{Song.2020}. Specifically, we recruited workers from Prolific (\url{https://www.prolific.co/}) and asked them whether a tweet was pro-Russia or pro-Ukraine. The annotators could select ``pro-Russia'', ``pro-Ukraine'', or ``neutral/unclear/unrelated'' as possible answers. For both datasets, we sampled 50 messages that were removed and 50 messages that remained after preprocessing. Messages that were removed from the pro-Russian dataset were considered pro-Ukrainian and vice versa. In accordance with best practices \cite{Song.2020}, we split the validation into two batches of 100 messages each to avoid fatigue. Each dataset was annotated by three workers. The workers were subject to a strict screening procedure: residency in UK/US/AUZ, English as a first language; enrollment in an undergraduate, graduate, or doctoral degree; a minimum approval rate of 95\%; and a minimum of 500 completed submissions on Prolific. We used the majority label for the final validation.

For the Russian dataset, we obtained a moderate agreement between the human annotators (Krippendorff's $\alpha = 0.49$ and Fleiss' $\kappa = 0.49$). The majority label from the annotators and the label from our preprocessing were in fair agreement (Cohen's $\kappa = 0.36$) when we considered the neutral/unclear label. When removing messages that were labeled as neutral/unclear, we obtained substantial agreement (Cohen's $\kappa = 0.7$) between the annotators and our preprocessing labels. Similarly, we obtained moderate agreement of annotators for the pro-Ukrainian dataset (Krippendorff's $\alpha = 0.52$ and Fleiss' $\kappa = 0.51$). The annotated majority label and our preprocessing label had moderate agreement (Cohen's $\kappa = 0.56$) when considering the neutral/unclear label and substantial agreement (Cohen's $\kappa = 0.71$) without the neutral/unclear label. Overall, this validates the reliability of our preprocessing approach.

\subsection*{Bot detection} 
We followed earlier research \cite{Broniatowski.2018, Shao.2018, Wojcik.2018} and identified bots using Botometer \cite{Varol.2017}. Botometer is a supervised machine learning classifier that assesses the likelihood of an account being a bot using different features derived from the account, the friendship network, and different linguistic features. Previous research has empirically shown that bot detection via Botometer is highly accurate (area under the receiver operating curve [AUROC] of 0.96) \cite{Sayyadiharikandeh.2020}. Moreover, Botometer is well maintained, updated regularly to incorporate state-of-the-art data and methods, and has been widely adopted in research \cite{Yang.2022}. We directly accessed Botometer API \cite{Botometer.Python} maintained by the Indiana University Observatory on Social Media. Botometer then returns the probability of an account being a bot. In line with previous research \cite{Shao.2018}, we classified accounts with Botometer scores $>0.5$ as bots. Overall, the Botometer API returned bot scores for \num{82604} users (\num{62.5}\%). Accounts that could not be matched onto human vs. bot due to Twitter's content moderation efforts were excluded from analyses that specifically differentiate between bot vs. human.

We validated the share of bots detected by Botometer \cite{Varol.2017} using Bot Sentinel \cite{BotSentinel.2022}. Bot Sentinel is a machine learning classifier for inappropriate accounts on Twitter, which includes bots, trolls, and coordinated accounts (with a high accuracy of 95\% \cite{BotSentinel.2022}). It requires at least ten sample messages per account to make classifications, which highly limits the number of accounts in our dataset that can be validated. We let Bot Sentinel classify the subset that fulfilled the requirements, which amounted to \num{2661} accounts. The agreement between the classifications of Botometer and Bot Sentinel on this subset is \num{61.93}\%. This can be explained by the slightly different definitions by which accounts are flagged. Botometer is designed to detect bots, whereas Bot Sentinel is designed to detect inappropriate accounts, i.e., a much broader concept. Yet, the algorithms show a high agreement regarding the share of bots and humans: Botometer classifies \num{25.29}\% of the accounts as bots while Bot Sentinel classifies \num{26.53}\%. This suggests that the Botometer is able to accurately classify the share of bots in our data. Moreover, the main conclusions of our analysis did not change when considering only the validation set classified by Bot Sentinel in our analysis: India, South Africa, and the U.S. remain the main targets for bots.

\subsection*{Location analysis} 
To infer the geographic location where users are active, we applied the following procedure. (1)~Users sometimes directly tagged their geolocation in messages in the form of a country code. In our dataset, this allowed us to identify the country location for \num{0.6}\% of the users in our dataset. (2)~Otherwise, we analyzed the self-reported location in a user's Twitter profile. In our dataset, this information was available for around \num{59}\% of the users. We then entered the self-reported locations into Python Geocoder \cite{Geocoder.2013}, which extracts real-world locations based on the OpenStreetMap API \cite{OpenStreetMapWiki.2021}. The API returns the spatial coordinates of the real-world location and an accuracy score, i.e., an estimate of how well the model was able to match the input to a real-world location. To account for incorrect or invalid locations, we filtered the results based on the accuracy that the Python Geocoder returns. We analyzed the distribution of the accuracy scores and found a bimodal distribution with a valley at \num{0.45}. We manually inspected the self-reported locations with an accuracy below \num{0.45} and subsequently set the threshold accordingly, discarding all geocode annotations with an accuracy below \num{0.45}. (3)~For users with neither geotagged messages nor a valid, self-reported location, the location was determined using the following heuristic. Specifically, we assumed that users live in the same country as their followers and, we thus approximated a user's country location through the country location of their follower base. Hence, for the remaining users with no location, we extracted the top \num{1000} followers each using the Twitter API v2 Users Endpoint \cite{Twitter.User}. We then geocoded the self-reported locations of the followers (where possible) and computed their geometric median. Subsequently, we mapped the spatial coordinates onto country codes using the ``naturalearth geometry'' in GeoPandas v0.11.1 \cite{Geopandas.2022}, which we then used as the estimated country of residence. Overall, the steps (1)--(3) yielded location information for \num{70.19}\% of all users in our dataset.  The relative frequency of bots in each country was computed as the mean number of bots among the classified users of that country in our dataset. We later also perform robustness checks: we plot only the accounts from steps 1 and 2 without the followers proxy (see Supplementary~Figure~\ref{fig:worldmap_nofollowers}) and we plot humans, bots and accounts without bot score information separately (see Supplementary~Figure~\ref{fig:worldmap_supp}).

We further validated our approach in two steps. First, we validated the accuracy of the Python Geocoder \cite{Geocoder.2013}. For this, we sampled \num{750} users for which Twitter was able to obtain a geotag of the message and that also provided a self-reported location. Analogous to above, we entered the self-reported locations into the Python Geocoder and obtained spatial coordinates for \num{599} users after applying a threshold of \num{0.45}. We obtained an almost perfect agreement (Cohen's $\kappa = 0.92$) between the country code provided by Twitter and the country code provided by the Python Geocoder \cite{Geocoder.2013}. This proves that the country codes obtained through the Python Geocoder are highly accurate.
Second, we validated our assumption that users live in the same country as their followers. For this, we extracted the top \num{1000} followers for the same sample of \num{750} users as above. Analogously, we geocoded the self-reported locations of the followers and computed the geometric median to obtain the country of residence. This yielded an estimated location for \num{452} users, which were in almost perfect agreement (Cohen's $\kappa = 0.81$) with the true country of the validation set.

\subsection*{Retweet network} 
To visualize the retweet network, we represented individual users as nodes and retweets as edges. We colored the nodes and edges based on the country of origin of the corresponding accounts (India = purple, U.S. = blue, and South Africa = green). In our implementation, we built the network using networkx \cite{Networkx.2008} and used the software Gephi v0.9.7 \cite{Gephi.2009} for visualization. For better readability, we applied a weighted degree filter of \num{10} and used the filter ``giant components'', so that only nodes with a large number of retweets remained. Since the retweet network is undirected, the weighted degree refers to the number of in- and outgoing retweets a node has. We later also perform a robustness check where we plot the retweeting network for users where no bot score information was available (see Supplementary~Figure~\ref{fig:network_undefined}).

\clearpage
\section*{Results}
\label{sec:results}

\subsection*{Pro-Russian support on social media}

Our analysis is based on Twitter messages posted between February through July 2022 that used the hashtags \texttt{\#istandwithrussia}, \texttt{\#standwithrussia}, \texttt{\#istandwithputin}, and \texttt{\#standwithputin}. We applied further filtering rules to select only messages where the content was pro-Russian (see \nameref{sec:methods}). Overall, this yielded $N =$~\num{349455} messages. The messages further generated nearly \num{1}~million likes. To measure the global exposure to pro-Russian messages, we estimated the overall readership based on the number of unique users that followed authors of pro-Russian messages in our dataset \cite{Cha.2010}, amounting to $\sim$\num{14.4}~million users. 

The messages in our dataset are fairly diverse (see Supplementary~Table~\ref{tab:russian_tweets}). For example, some messages contain only a series of hashtags (\eg, \emph{\say{\texttt{\#IStandWithPutin}  \texttt{\#isupportrussia}  \texttt{\#Putin} \texttt{\#standforrussia}  \texttt{\#StandWithPutin}  \texttt{\#IndiaWithRussia}}}), while others state verbal affirmations of support for Putin or hate against Ukraine or NATO countries. Examples of the latter are: \emph{\say{@RWApodcast I literally love Putin. The most honest leader in the world. \texttt{\#istandwithrussia}}} and \emph{\say{US is responsible for more than 81\% conflicts in the world. The real war criminal is US. US should be completely isolated on the global stage \texttt{\#IStandWithPutin} \texttt{\#RussiaArmy} \texttt{\#IStandWithPutin}}}. By analyzing popular hashtags, we also see that several of them are unique to expressing a pro-Russian sentiment (see Supplementary~Table~\ref{tab:russian_hashtags}). Examples are, \eg, \texttt{\#hypocrisy} (posted \num{5682} times),  \texttt{\#doublestandards} (posted \num{2552} times), and \texttt{\#stopnato} (posted \num{2156} times).

\begin{figure}[H]
\centering
    \begin{minipage}{\textwidth}
    \centering
        \begin{figure}
            \centering
            \includegraphics[width=.75\textwidth]{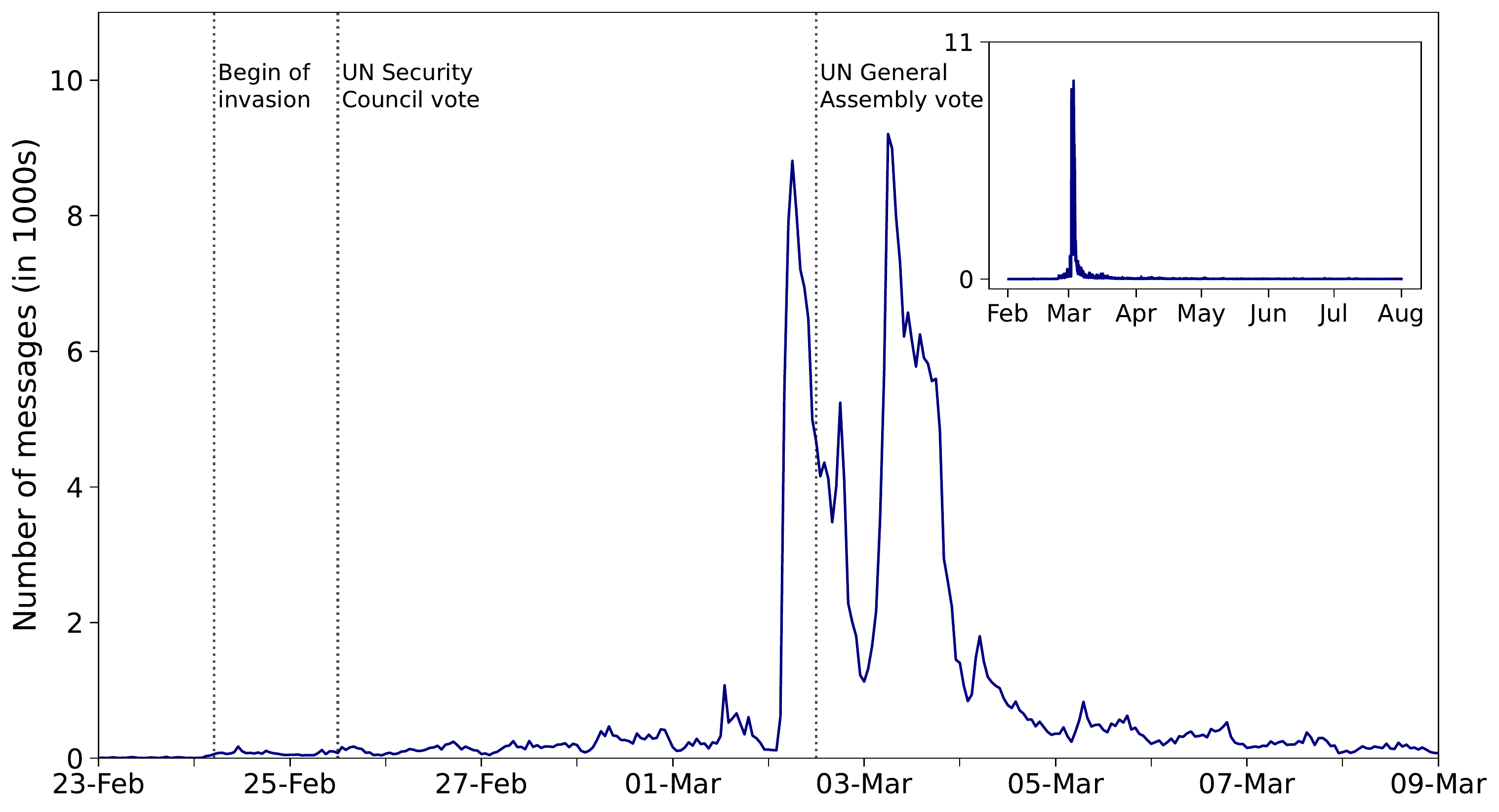}
        \end{figure}
    \end{minipage}
\caption{\textbf{Temporal dynamics of pro-Russian support.} The plot shows the number of pro-Russian messages during the first two weeks of the invasion. The peak on March~2, 2022, coincides with the day the United Nations General Assembly adopted Resolution \mbox{ES-11/1} deploring the invasion. Inset: volume of pro-Russian messages for the entire time period of the dataset.}
\label{fig:message_volume}
\end{figure}

Pro-Russian messages showed distinctive temporal patterns (see Figure~\ref{fig:message_volume}) that coincided with the day that the United Nations General Assembly adopted Resolution \mbox{ES-11/1} deploring the invasion (March~2, 2022). For example, peaks in the message volume occurred on March~2, 2022 (\num{64738} pro-Russian messages), March~3, 2022 (\num{103772} pro-Russian messages), and March~4, 2022 (\num{66794} pro-Russian messages), respectively. A fine-grained analysis showing temporal dynamics of the number of bot and human messages can be found in Supplementary~Figure~\ref{fig:message_volume_bothuman}.

Further, on the day of the UN vote (March~2, 2022), $\sim$\num{41.7}\% of the posted messages can be traced back to India, followed by Pakistan ($\sim$\num{5.9}\%) and Nigeria ($\sim$\num{2}\%). In contrast, on the day after the UN vote (March~3, 2022), the majority of the messages were posted from the U.S. ($\sim$\num{14.1}\%), Nigeria ($\sim$\num{10.5}\%), and India ($\sim$\num{10}\%). Apparently, messages from the U.S. were surprisingly rare on the day of the UN vote, despite that the majority of the Twitter user base is from the U.S. \cite{Statista.Twitter.User.2022}. This suggests that pro-Russian support was potentially disseminated through a campaign targeting specific countries, for which we provide evidence in the following. 

\subsection*{Spreading dynamics of pro-Russian support}

Pro-Russian messages have been spread by \num{132131} accounts (see Supplementary~Table~\ref{tab:influential_indegree_users} for a list of influential accounts). To analyze the role of bots in the spread of pro-Russian messages, we used Botometer \cite{Varol.2017} to classify accounts according to humans and bots. For each account, we computed a bot score ($\rho \in [0, 1]$), which can be interpreted as the level of automation of that account \cite{Shao.2018}. A threshold of 0.5 is typically used to classify an account as likely human or likely bot (see \nameref{sec:methods} for details). Using this method, \num{20.28}\% of the accounts were categorized as bots. Hence, bots played a critical role in spreading pro-Russian messages.

\begin{figure}[H]
\centering
    \begin{minipage}{0.45\textwidth}
        \begin{figure}
            \centering
            \raggedright\figletter{a}\\
            \includegraphics[width=\textwidth]{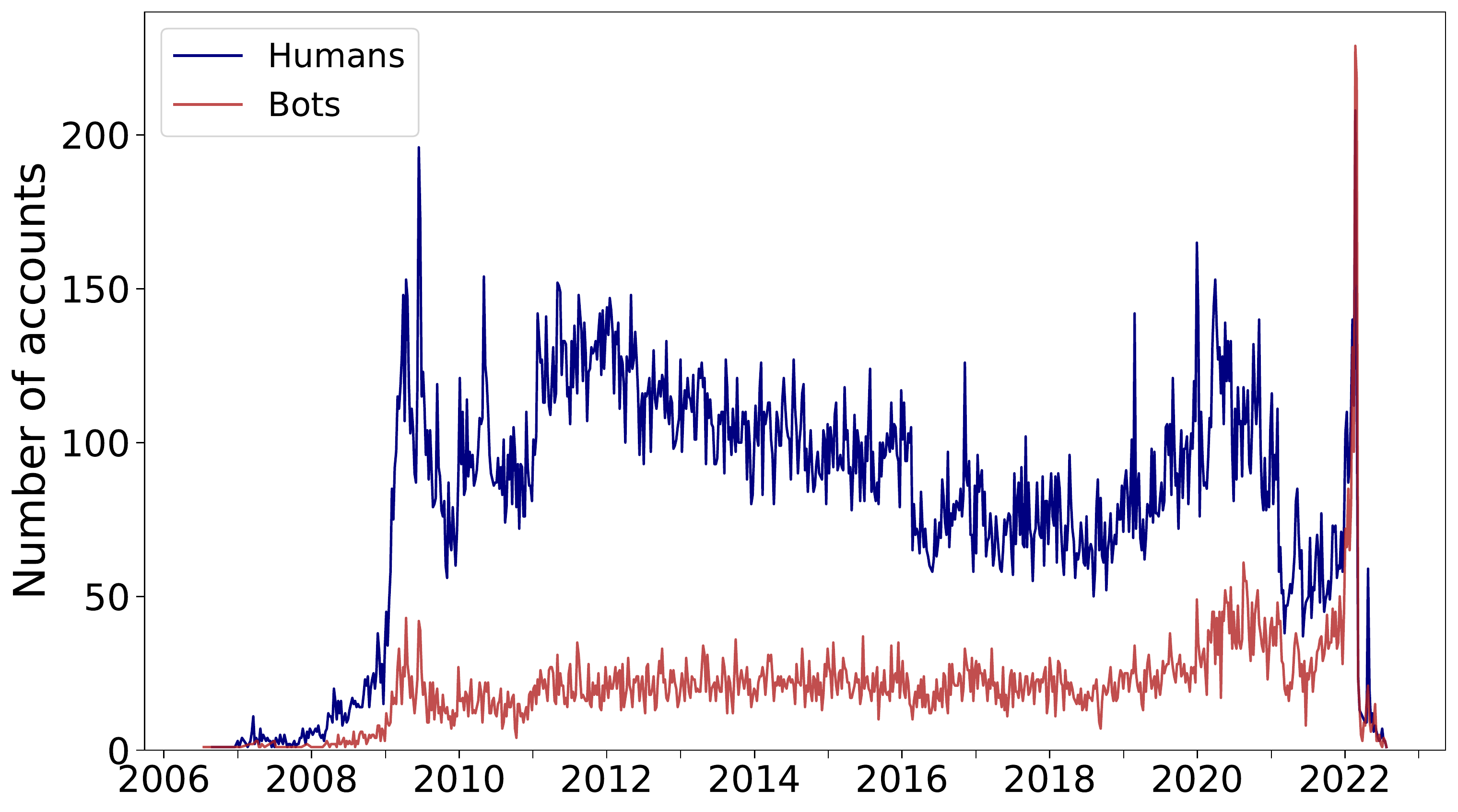}
        \end{figure}
    \end{minipage}
    \quad\quad    
        \begin{minipage}{0.45\textwidth}
        \begin{figure}
            \centering
            \raggedright\figletter{b}\\
            \includegraphics[width=\textwidth]{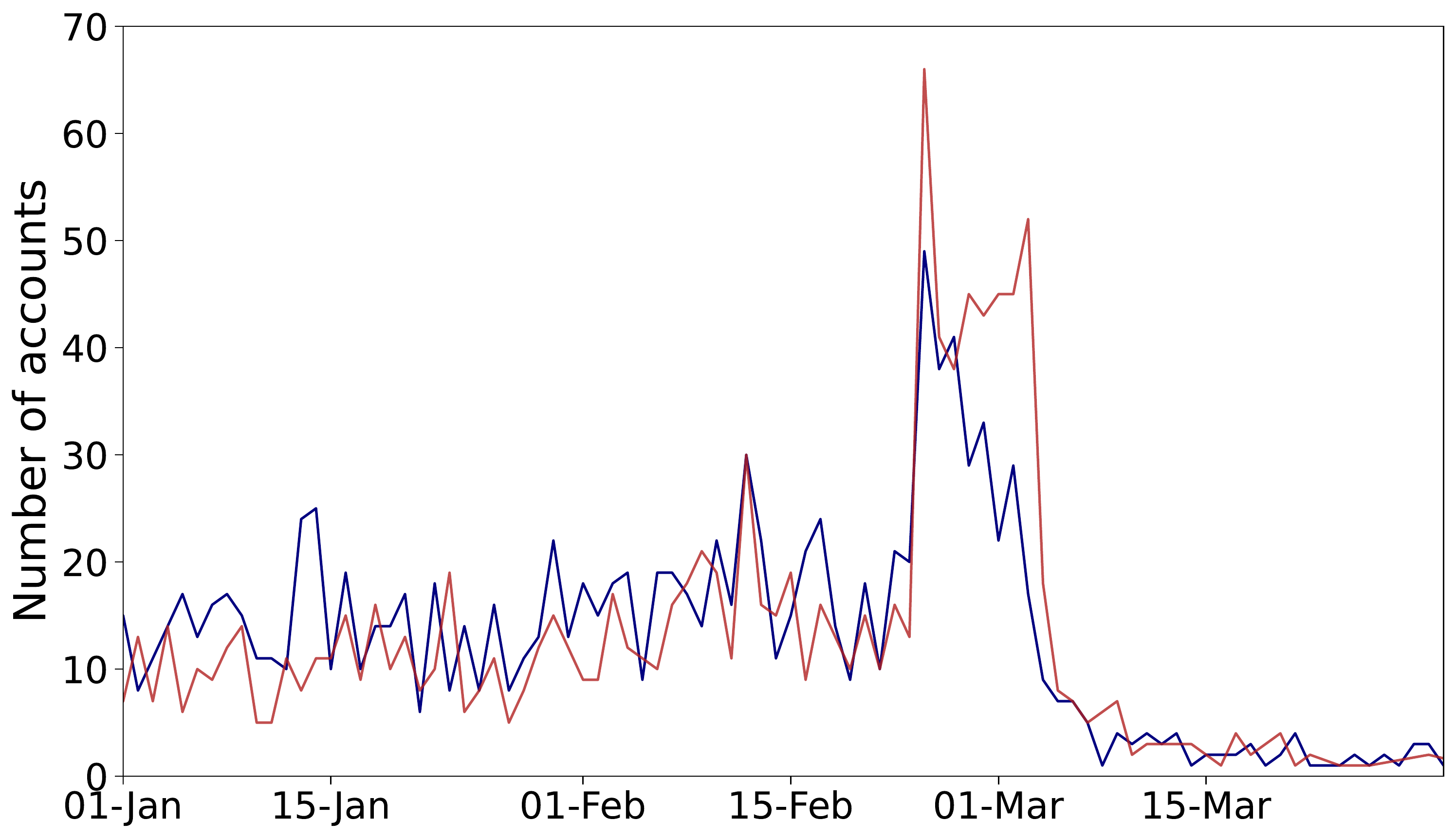}
        \end{figure}
    \end{minipage}
\caption{\textbf{Spreaders of pro-Russian messages.} \figletter{a},~Dates on which accounts were created. Here, the time axis starts with the inception of Twitter in 2006. \figletter{b},~Dates on which accounts were created. Here, the time axis starts shortly before the beginning of the 2022 Russian invasion.}
\label{fig:spreaders_pro_russia}
\end{figure}

Accounts from humans and bots showed a clear difference in when the accounts were created (Figure~\ref{fig:spreaders_pro_russia}a). Accounts classified as bots tended to have been created more recently than accounts classified as humans. Notably, there also was a clear peak in the number of newly created bots, which coincided with the beginning of the invasion on February~24, 2022 (Figure~\ref{fig:spreaders_pro_russia}b). A robustness check showing the creation dates of accounts for which a bot score could not be assigned is provided in Supplementary~Figure~\ref{fig:spreaders_pro_russia_undefined}.

\begin{figure}[H]
\centering
    \begin{minipage}{0.35
    \textwidth}
    \centering
        \begin{figure}
            \centering
            \raggedright\figletter{a}\\
            \includegraphics[width=\textwidth]{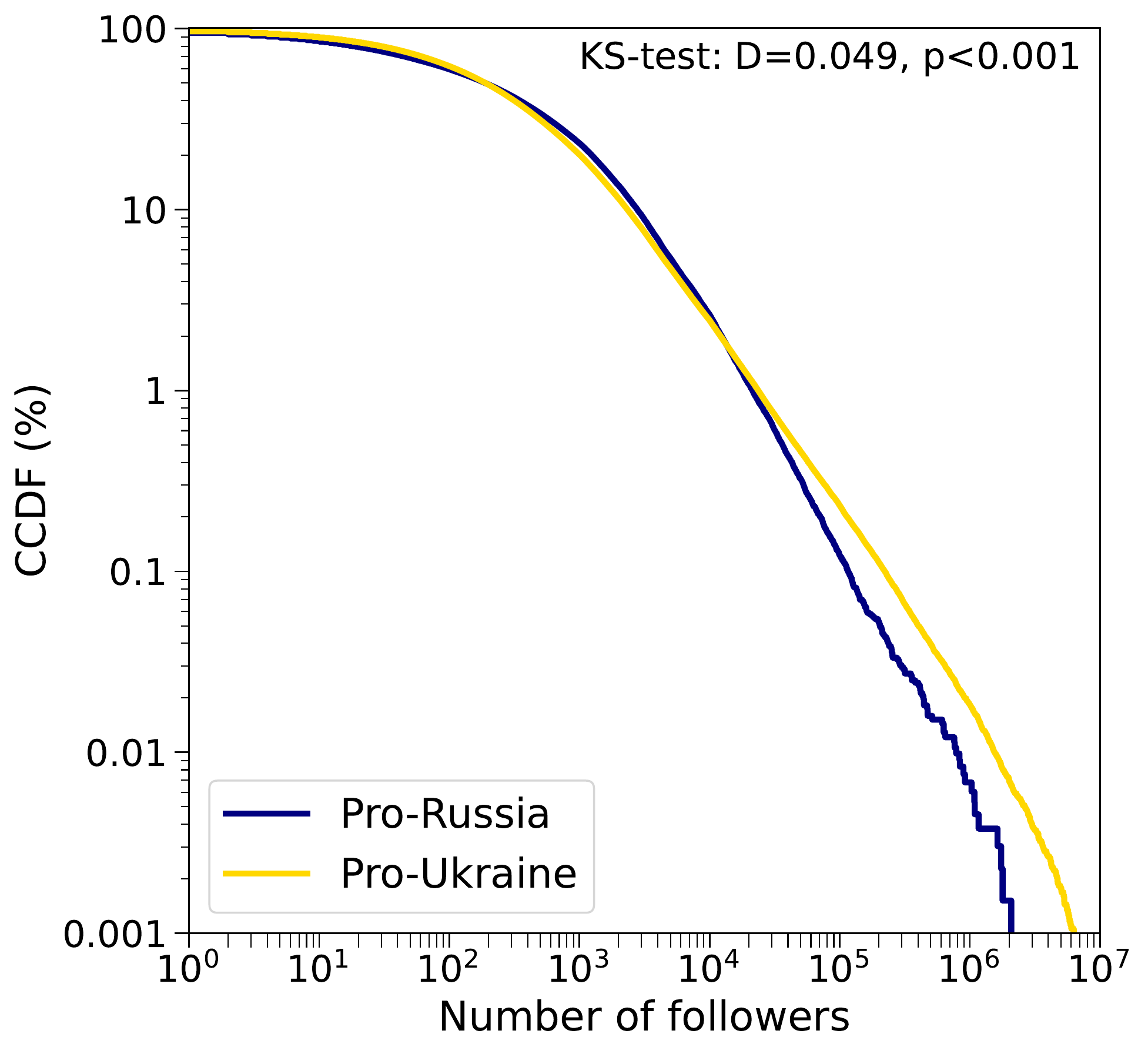}
        \end{figure}
    \end{minipage}
    \quad\quad
    \begin{minipage}{0.35\textwidth}
        \begin{figure}
            \centering
            \raggedright\figletter{b}\\
            \includegraphics[width=\textwidth]{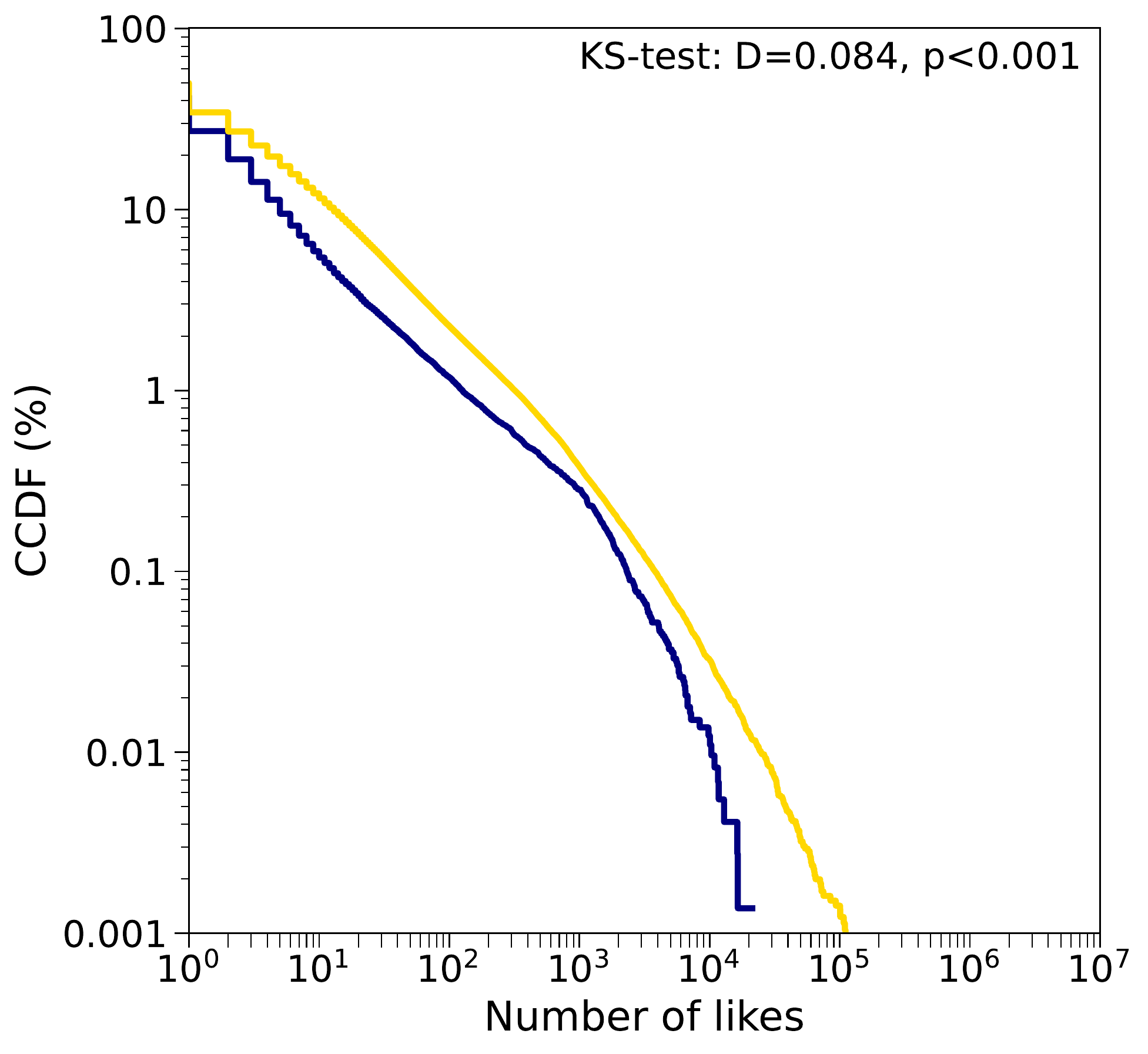}
        \end{figure}
    \end{minipage}
    
    \begin{minipage}{0.35\textwidth}
        \begin{figure}
            \centering
            \raggedright\figletter{c}\\
            \includegraphics[width=\textwidth]{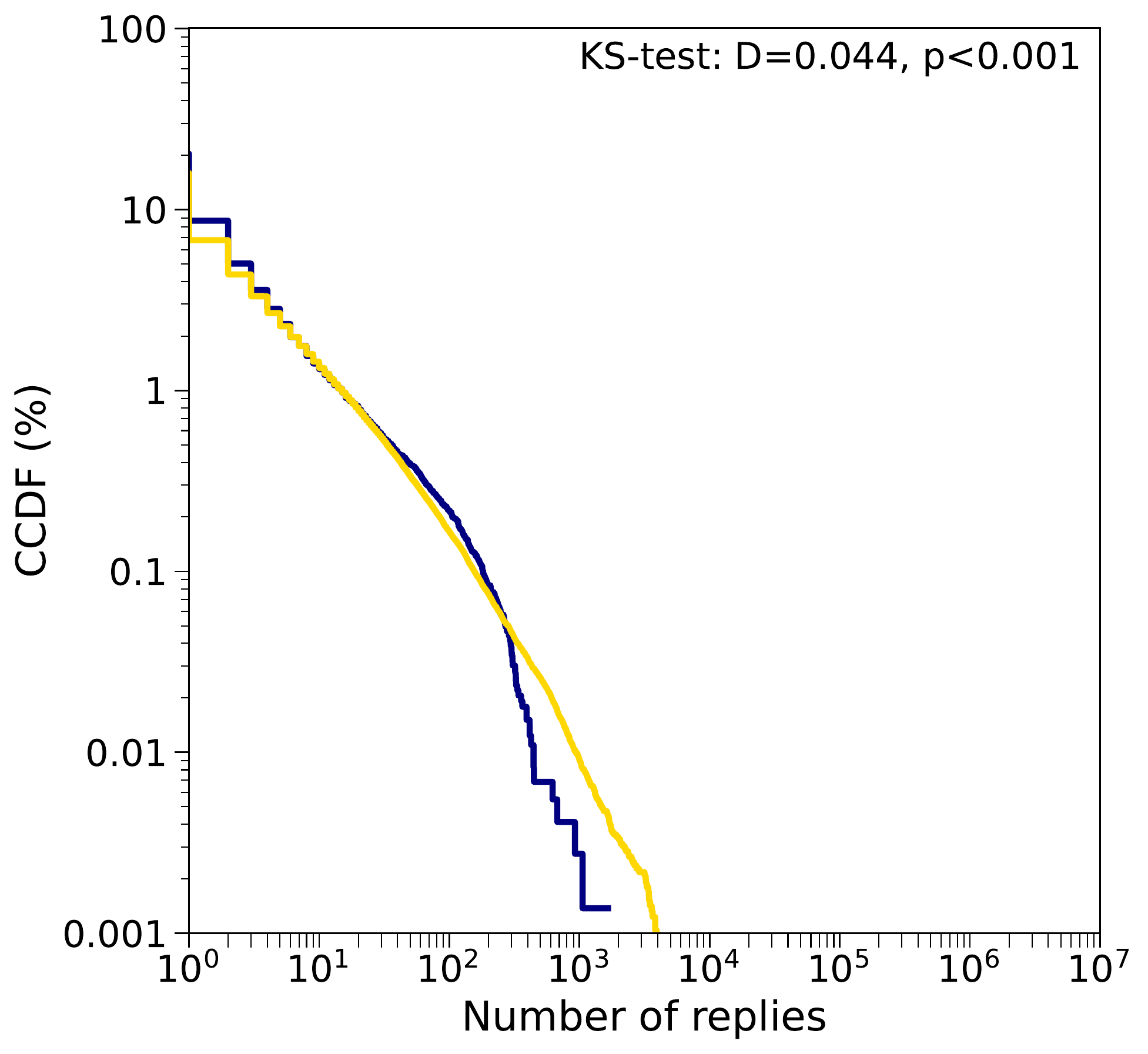}
        \end{figure}
    \end{minipage}
    \quad\quad
    \begin{minipage}{0.35\textwidth}
        \begin{figure}
            \centering
            \raggedright\figletter{d}\\
            \includegraphics[width=\textwidth]{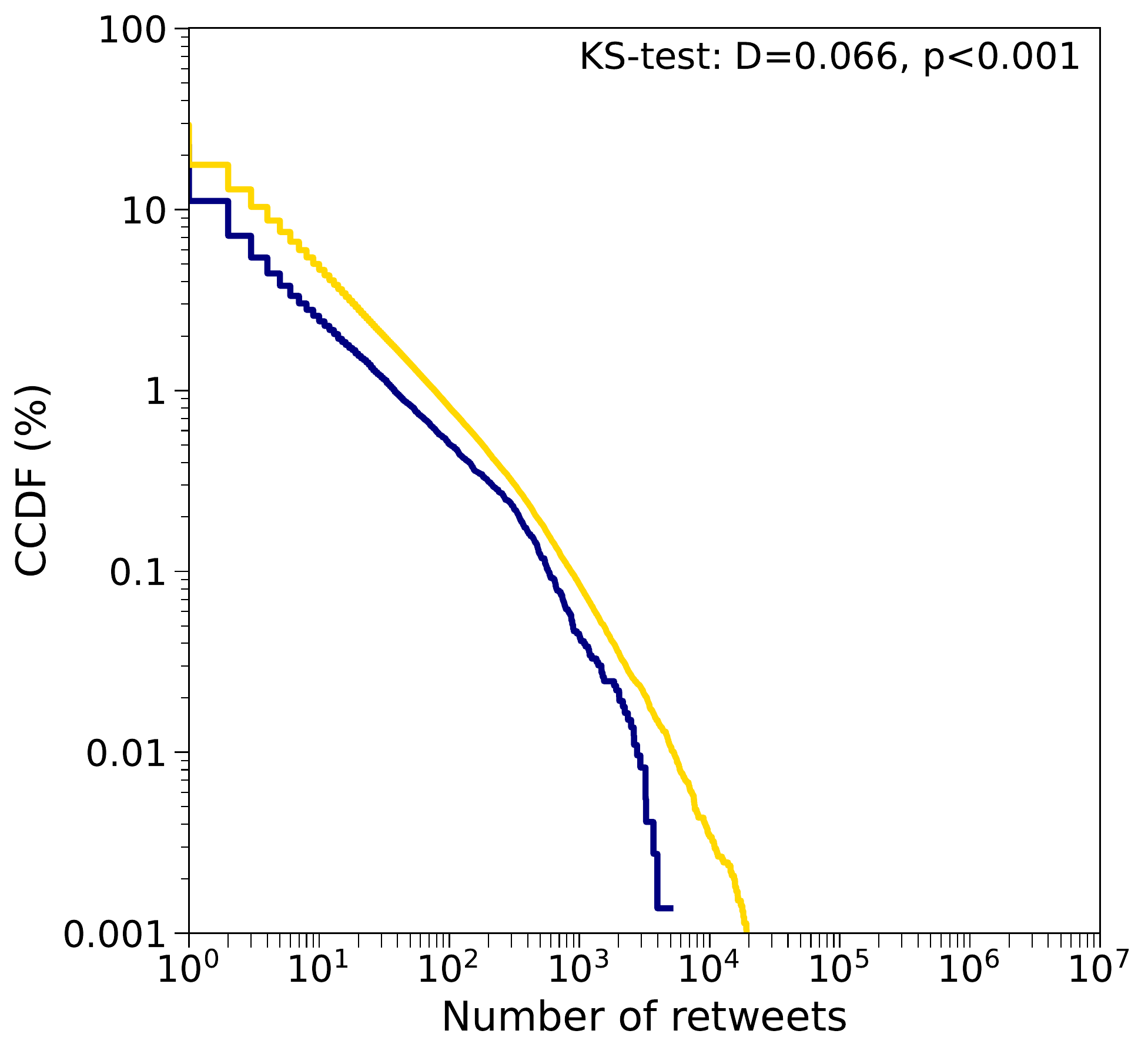}
        \end{figure}
    \end{minipage}
\caption{\textbf{Online virality of pro-Russian vs. pro-Ukrainian support.} Here, we compare complementary cumulative distribution functions (CCDFs) for: \figletter{a},~the number of followers; \figletter{b},~the number of likes; \figletter{c}~the number of replies; and \figletter{d},~the number of retweets. The former was computed at the user level, while the latter three were computed at the tweet level. Statistical comparisons are based on Kolmogorov-Smirnov (KS) tests \cite{Massey.1951}. All distributions span several orders of magnitude, implying that there was a substantial share of pro-Russian messages that went viral.}
\label{fig:ccdfs_ru_uk}
\end{figure}

To further quantitatively characterize the spreading dynamics of pro-Russian support, we collected an additional dataset with pro-Ukrainian messages that were posted on Twitter between February 2022 through July 2022 (see Supplementary~Table~\ref{tab:ukrainian_tweets}). We first compared the number of bots spreading pro-Russian messages (\num{20.28}\%) with the number of bots spreading pro-Ukrainian messages (\num{14.25}\%). Here, we find that pro-Ukrainian support was spread by significantly less bots than pro-Russian support (Kolmogorov-Smirnov (KS) test \cite{Massey.1951}: $D= 0.062$, $p < 0.001$). We then compared spreaders of pro-Russian vs. pro-Ukrainian support in terms of the number of followers (Figure~\ref{fig:ccdfs_ru_uk}a): Pro-Russian supporters had a substantially smaller number of followers with a mean of only \num{1690} followers, whereas the mean number of followers was \num{2248} for pro-Ukrainian supporters (KS test: $D= 0.049$, $p < 0.001$). The number of followers is typically interpreted as a proxy for the social influence of online users \cite{Cha.2010}, implying that spreaders of pro-Russian support had a comparatively smaller social influence than spreaders of pro-Ukrainian support.

We further find heterogeneity in the online virality of pro-Russian and pro-Ukrainian support. For this, we compared the number of likes, replies, and retweets that pro-Russian vs. pro-Ukrainian source tweets received (Figure~\ref{fig:ccdfs_ru_uk}b--d). On average, pro-Russian source tweets received \num{12.97} likes, \num{1.16} replies, and \num{3.38} retweets. The corresponding numbers were significantly smaller than for pro-Ukrainian source tweets, which, on average, received \num{28.35} likes, \num{1.22} replies, and \num{6.56} retweets (KS tests: $D= 0.084$, $p < 0.001$; $D= 0.044$, $p < 0.001$; and $D= 0.066$, $p < 0.001$, respectively). Thus, pro-Russian support tended to be less viral than pro-Ukrainian support. Note however that, for both pro-Russian and pro-Ukrainian support, we observe very broad distributions spanning several orders of magnitude. Hence, there was still a substantial proportion of pro-Russian messages that went viral.

\subsection*{Cross-country heterogeneity in the exposure to pro-Russian support}

\begin{figure}[H]
\centering
    \begin{minipage}{\textwidth}
    \centering
        \begin{minipage}{\textwidth}
            \begin{figure}
                \centering
                \raggedright\figletter{a}\\
                \includegraphics[width=\textwidth]{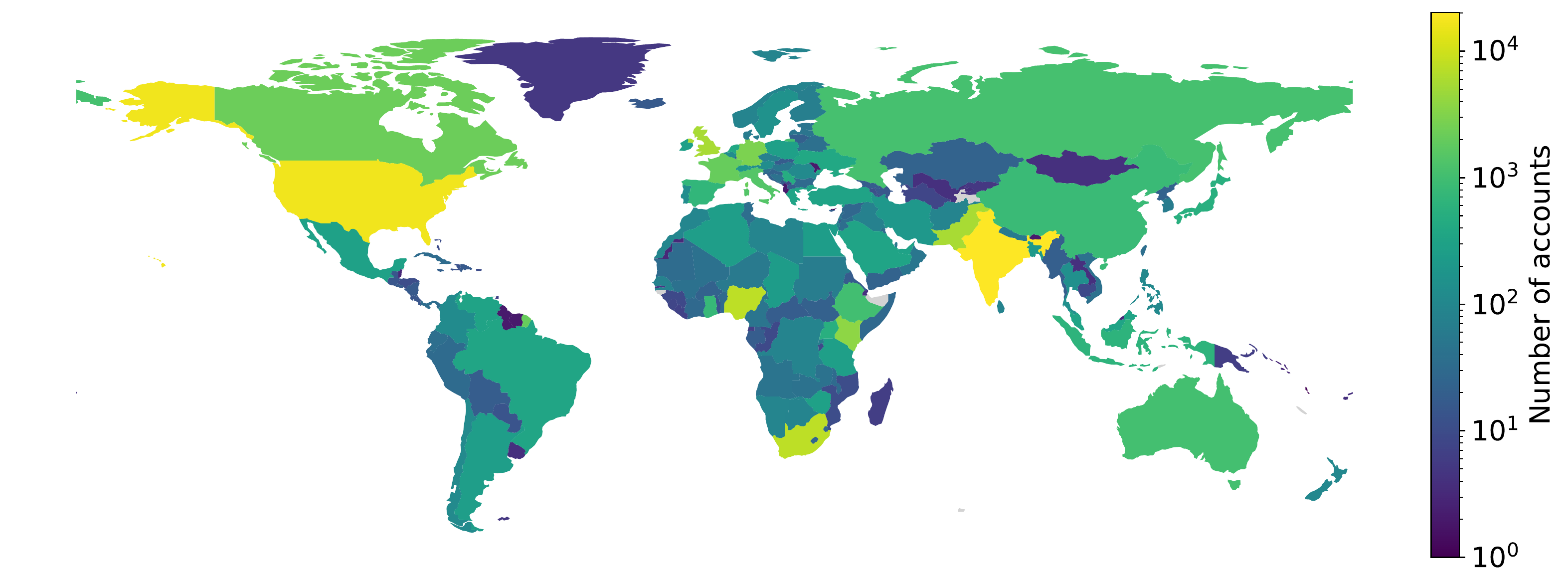}
            \end{figure}
        \end{minipage}
        
        \begin{minipage}{\textwidth}
            \begin{figure}
                \centering
                \raggedright\figletter{b}\\
                \includegraphics[width=\textwidth]{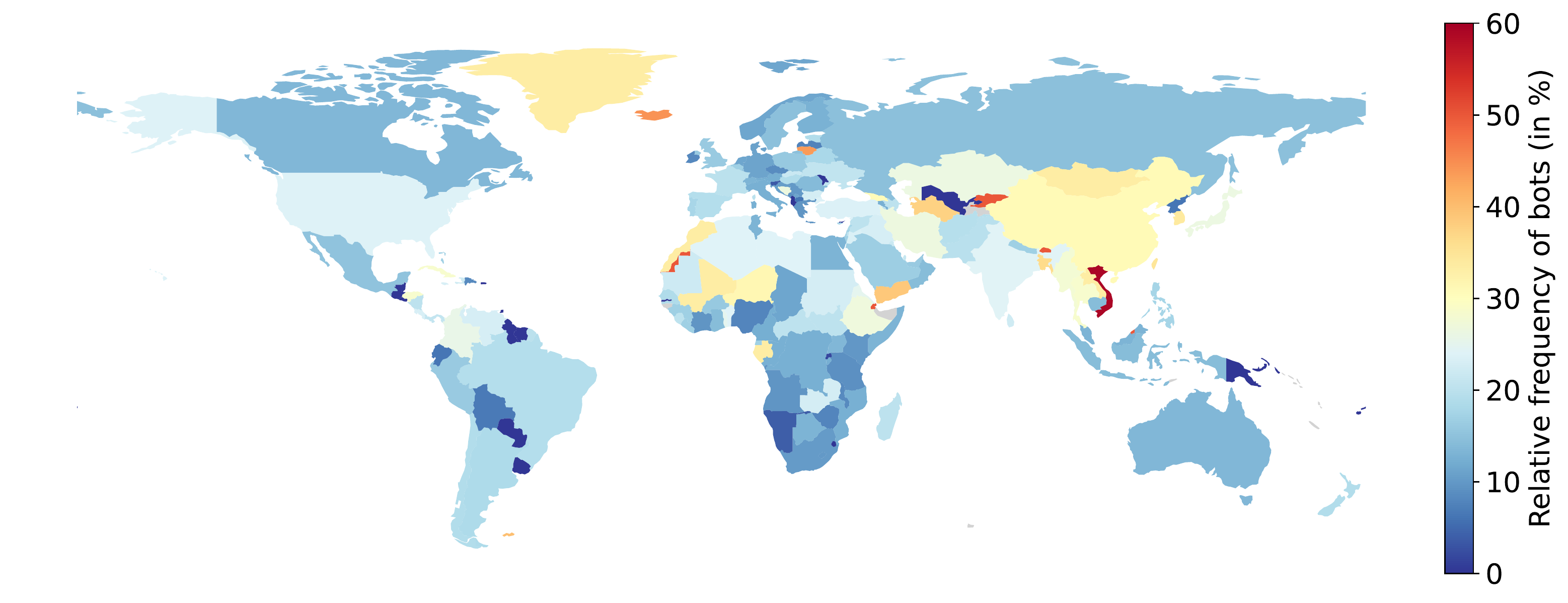}
            \end{figure}
        \end{minipage}
    \end{minipage}
\caption{\textbf{Cross-country differences in the spread of pro-Russian support.} Here, we inferred the geographic location of accounts (see \nameref{sec:methods}). \figletter{a},~Number of users per country (log scale). \figletter{b},~Relative frequency of bots per country (in \%).}
\label{fig:worldmap_main}
\end{figure}

To analyze the cross-country heterogeneity in pro-Russian support, we inferred the geographic location of the underlying user accounts (see \nameref{sec:methods}). Evidently, the countries with the most accounts spreading pro-Russian messages were India, the United States, South Africa, and Nigeria (Figure~\ref{fig:worldmap_main}a). The pronounced role of these English-speaking countries in spreading pro-Russian messages may be partially explained by the use of English hashtags as search queries. However, these countries also show a high percentage of pro-Russian supporters in comparison to the overall number of Twitter users in that country (see Table~\ref{tab:users_per_country}). Moreover, pro-Russian support was disproportionally high in countries that abstained from voting on the United Nations Resolution \mbox{ES-11/1} (such as India, South Africa, and Pakistan) relative to other English-speaking countries (such as the United States, the United Kingdom, and Australia). Subsequently, we computed the relative frequency of bots across countries (Figure~\ref{fig:worldmap_main}b). Several of the countries with many pro-Russian messages also showed a pronounced role of likely bot activity: \num{24.2}\% of the accounts in India were bots, \num{23.9}\% in the United States, \num{10.2}\% in South Africa, and \num{7.9}\% in Nigeria. 
The patterns remained robust across different methods for inferring geographic locations (Supplementary~Figure~\ref{fig:worldmap_nofollowers}). We also conducted a robustness check in which the locations of humans, bots, and accounts without bot scores were mapped separately and found robust patterns (see Supplementary~Figure~\ref{fig:worldmap_supp}).

\begin{table}[H]
	\centering
	\footnotesize
	\singlespacing
    \begin{tabular}{lrr}
        \toprule
        {Country} & Twitter users (in millions) & pro-Russian supporters (in \%) \\
        \midrule
            Nigeria        &          0.32 &  2.290 \\
            South Africa   &          2.85 &  0.263 \\
            Pakistan       &          3.40 &  0.161 \\
            India          &         23.60 &  0.085 \\
            United Kingdom &         18.40 &  0.030 \\
            Canada         &          7.90 &  0.028 \\
            United States  &         76.90 &  0.021 \\
            Indonesia      &         18.45 &  0.003 \\
            Saudi Arabia   &         14.10 &  0.002 \\
            Mexico         &         13.90 &  0.002 \\
            Turkey         &         16.10 &  0.002 \\
            Brazil         &         19.05 &  0.002 \\
            Japan          &         58.95 &  0.001 \\
        \bottomrule
    \end{tabular}
    \caption{Total number of Twitter users per country (in millions) and the relative frequency of pro-Russian supporters in our dataset. The total number of Twitter users is based on data from 2022 \cite{Statista.Twitter.User.2022, Datareportal.SouthAfrica.2022, Datareportal.Nigeria.2022, Datareportal.Pakistan.2022}. We selected the ten leading countries with the highest number of Twitter users. In addition, we included Nigeria, South Africa, and Pakistan, due to their relevance to our analysis.}
    \label{tab:users_per_country}
\end{table}

Overall, countries that abstained from the UN vote had the highest relative frequency of bots (\num{20.3}\%), in comparison to countries that voted against (\num{14.9}\%) or approved (\num{16.6}\%) the UN Resolution \mbox{ES-11/1} (one-way ANOVA test: $F=\num{84.73}$; $p<0.001$). Hence, countries abstaining from the UN vote (e.g., India, South Africa) have been prime targets of bots circulating pro-Russian support. Supplement~\ref{supp:content} provides a content analysis that further substantiates the connection between countries and the UN vote.

\begin{figure}[H]
\centering
    \begin{minipage}{\textwidth}
    \centering
        \begin{minipage}{\textwidth}
            \begin{figure}
                \centering
                \raggedright\figletter{a}\\
                \includegraphics[width=\textwidth]{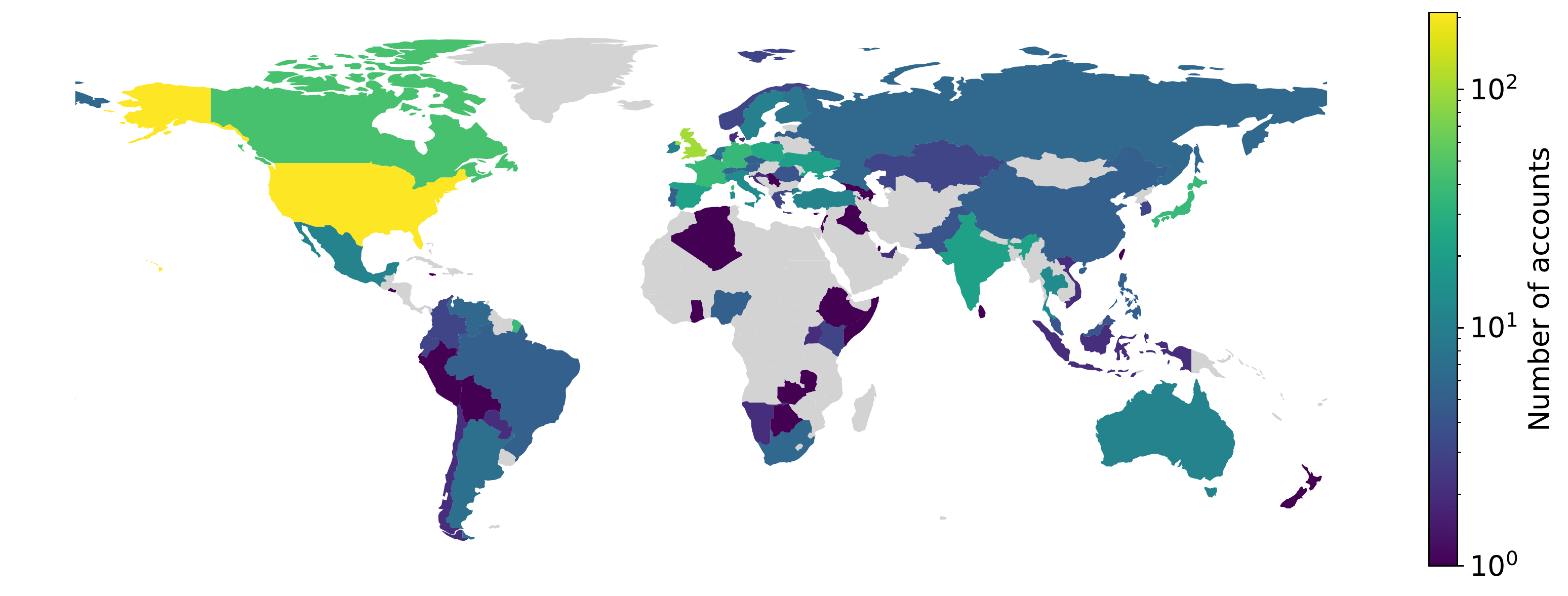}
            \end{figure}
        \end{minipage}
        
        \begin{minipage}{\textwidth}
            \begin{figure}
                \centering
                \raggedright\figletter{b}\\
                \includegraphics[width=\textwidth]{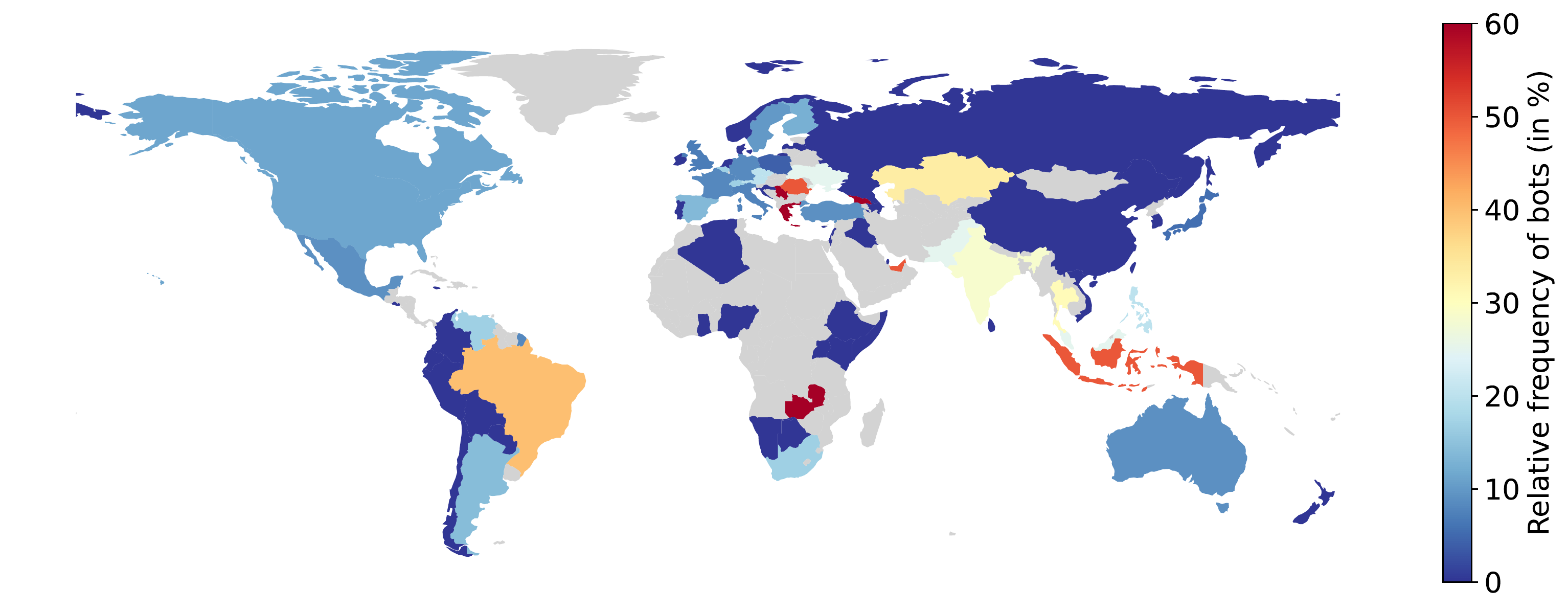}
            \end{figure}
        \end{minipage}
    \end{minipage}
\caption{\textbf{Cross-country differences in the spread of pro-Ukrainian support.} Here, we inferred the geographic location of a subsample of pro-Ukrainian accounts (see \nameref{sec:methods}). \figletter{a},~Number of users per country (log scale). \figletter{b},~Relative frequency of bots per country (in \%).}
\label{fig:worldmap_ukraine}
\end{figure}

We also compared the cross-country heterogeneity of pro-Russian support to pro-Ukrainian support (see Figure~\ref{fig:worldmap_ukraine}). We find a larger focus of pro-Ukrainian support in the U.S. and European countries. Countries that were highly active in spreading pro-Russian support such as South Africa, Pakistan, and Nigeria were not as active in spreading pro-Ukrainian support. Furthermore, we compared the relative frequency of bots of pro-Ukrainian supporters. Similarly to pro-Russian support, we found a pronounced bot activity in India (\num{28.57}\%) and South Africa (\num{16.67}\%). In contrast, the United States and Nigeria showed less to no bot activity (\num{11.42}\% and \num{0}\%, respectively).

\subsection*{Retweet network}

\begin{figure}[H]
\centering
    \begin{minipage}{\textwidth}
    \centering
        \begin{minipage}{0.45\textwidth}
            \begin{figure}
                \centering
                \raggedright\figletter{a}\\
                \hspace{1cm}\includegraphics[width=.8\textwidth]{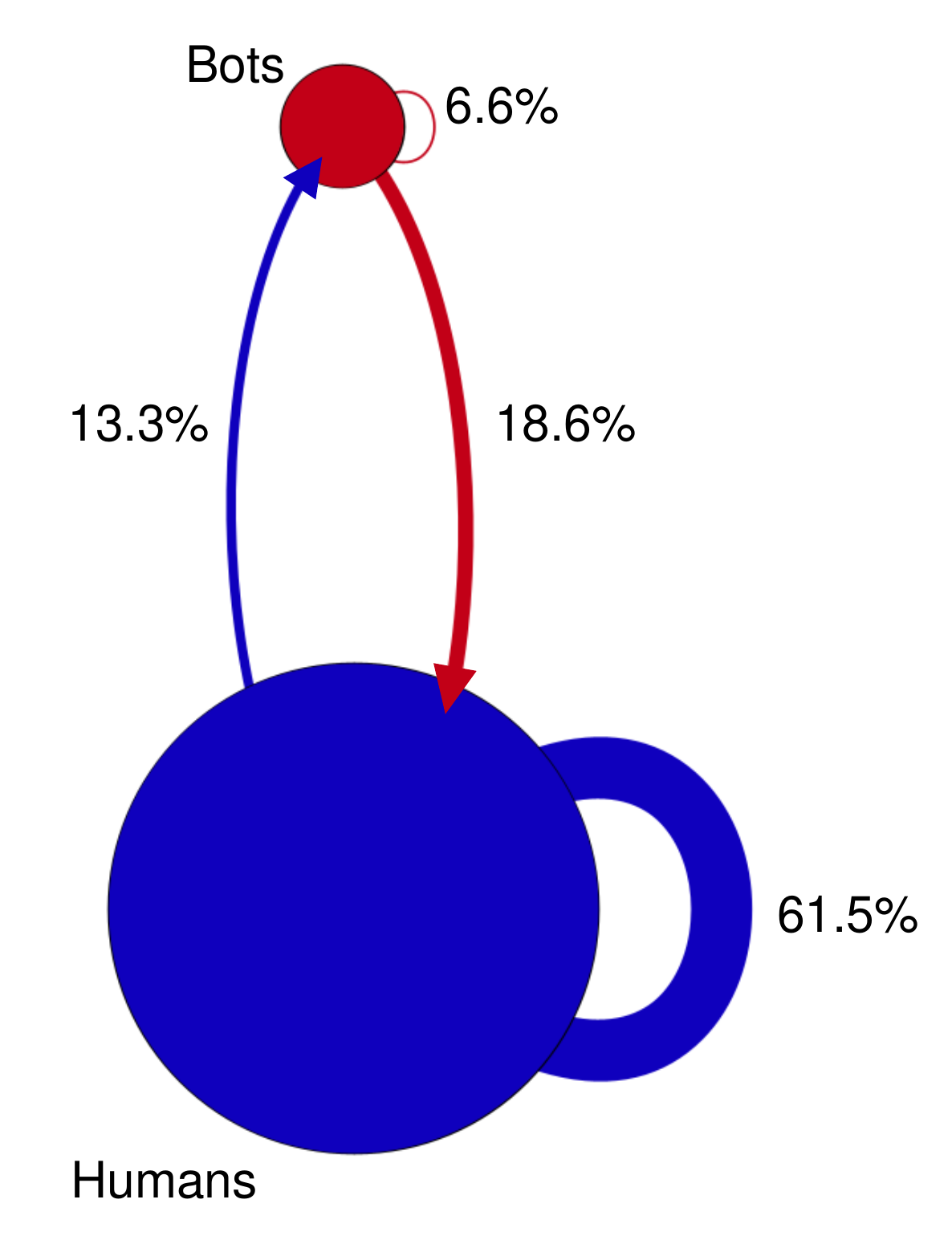}
            \end{figure}
        \end{minipage}
        \qquad
        \begin{minipage}{0.49\textwidth}
            \begin{figure}
                \centering
                \raggedright\figletter{b}\\
                \includegraphics[width=\textwidth]{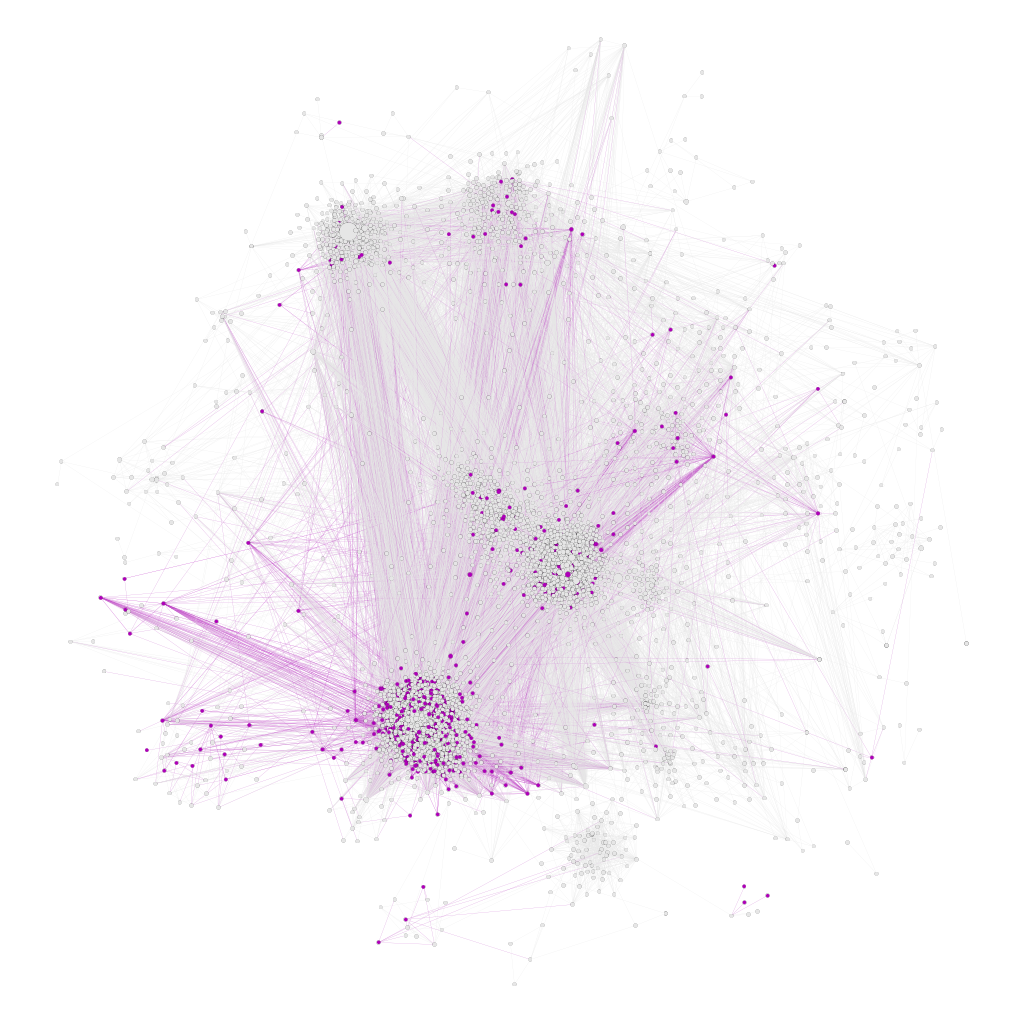}
            \end{figure}
        \end{minipage}
    \end{minipage}
    
    \begin{minipage}{\textwidth}
    \centering
        \begin{minipage}{0.49\textwidth}
            \begin{figure}
                \centering
                \raggedright\figletter{c}\\
                \includegraphics[width=\textwidth]{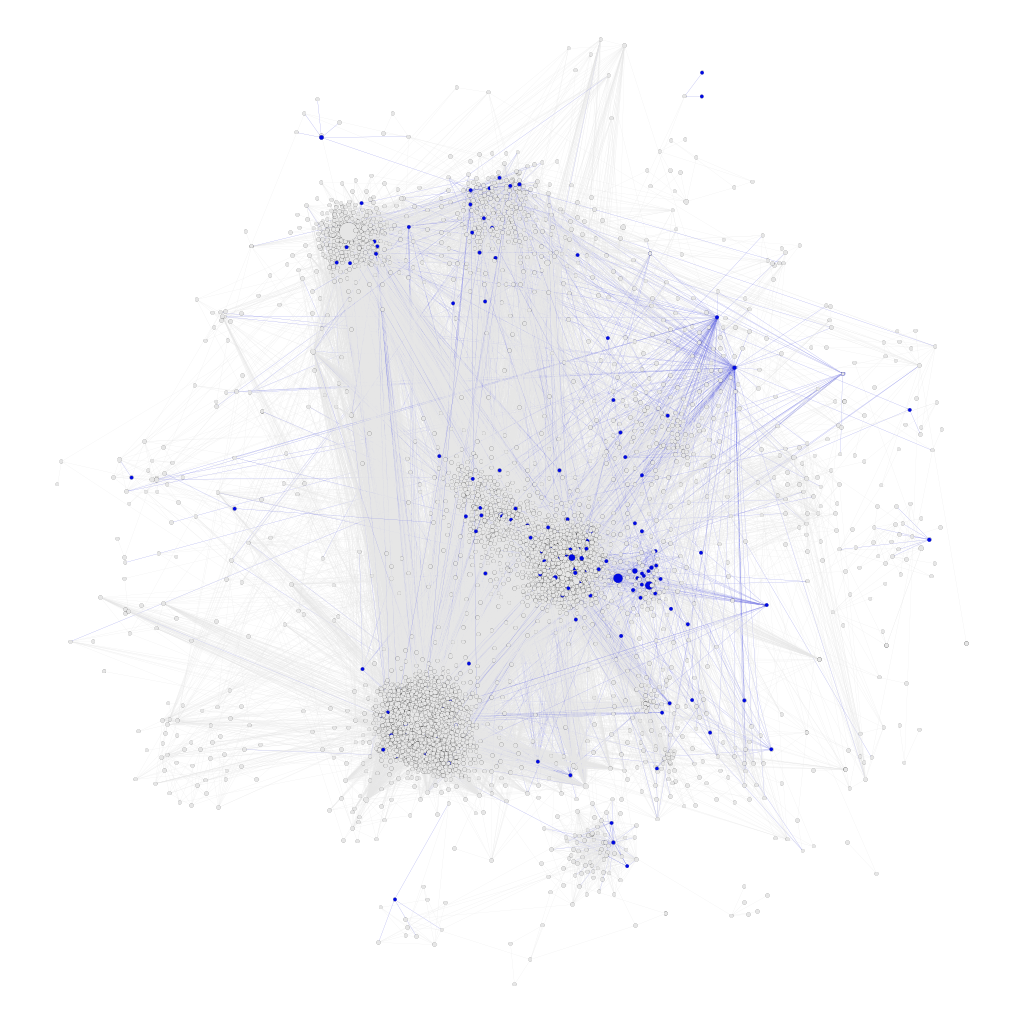}
            \end{figure}
        \end{minipage}
        \begin{minipage}{0.49\textwidth}
            \begin{figure}
                \centering
                \raggedright\figletter{d}\\
                \includegraphics[width=\textwidth]{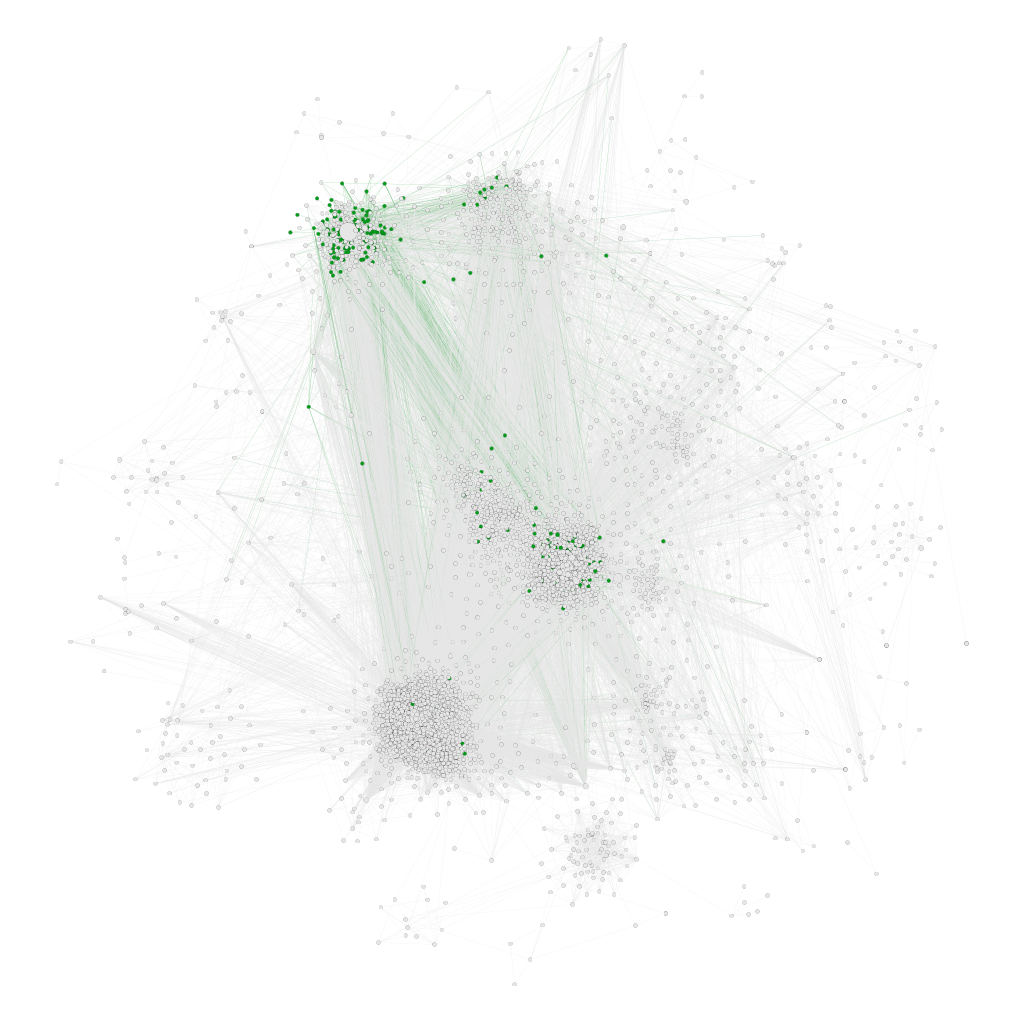}
            \end{figure}
        \end{minipage}
    \end{minipage}
\caption{\textbf{Retweeting network.} \figletter{a},~Spreading patterns between humans vs. bots (blue = humans, red = bots). The node size represents the number of humans vs. bots. The edges represent the direction and relative frequency of retweets. \figletter{b},~Retweet network with accounts from India colored in purple. \figletter{c},~Retweet network with accounts from the U.S. colored in blue. \figletter{d},~Retweet network with accounts from South Africa colored in green.  The retweet networks were visualized using Gephi \cite{Gephi.2009} (see \nameref{sec:methods}).
}
\label{fig:network}
\end{figure}

We analyzed the network diffusion patterns of pro-Russian support and especially how bots promoted its spread. First, we examined the retweet dynamics with which pro-Russian messages were disseminated across different account types (Figure~\ref{fig:network}a). humans tended to primarily retweet other humans rather than bots. bots, in return, tended to mainly retweet humans but retweeted other bots only rarely. This indicates that bots drove the spread of pro-Russian support primarily by exposing humans to human-generated, pro-Russian messages. 

The retweet network of individual accounts revealed several clusters in which pro-Russian messages primarily circulated (Figure~\ref{fig:network}b--d). By matching accounts to their geographic location, we find that some of the clusters were of large geographic homogeneity. In particular, we could map two of the clusters to users from India and South Africa, both of which were two major countries that abstained from the UN vote. These countries exhibited relatively isolated retweet networks in which pro-Russian messages were able to infiltrate the local online communities with little external influence. In comparison, accounts from the U.S. did not show the same geographic clustering but were more broadly scattered over the retweet network. This suggests that there may have been differences in the coordination behind the pro-Russian support across countries as India and South Africa were specifically targeted by pro-Russian supporters. Accounts from the U.S. retweeted accounts from all over the network, whereas accounts from South Africa and India discussed the invasion mostly with accounts from their country. The content analysis in Supplement~\ref{supp:content} further substantiates that discussions in India and South Africa were held at a local scale and focused on national issues.
We also performed a robustness check of the retweeting networks on the accounts that did not have bot information and corroborated our findings (see Supplementary~Figure~\ref{fig:network_undefined}). 

\newpage 

\thispagestyle{empty}


\subsection*{Amplification of pro-Russian support spreading through bots}

We further examined how bots contributed to the spreading of pro-Russian support (e.g., by automatically making pro-Russian hashtags go viral or retweeting other accounts) and, to this end, analyzed differences in the online behavior of humans vs. bots. Bots were responsible for only \num{20.82}\% of the source tweets, while \num{79.18}\% of the source tweets originated from humans (see Supplementary~Figure~\ref{fig:ccdfs_bot_human}). Hence, most of the content generation was done by humans. 
However, even though \num{20.28}\% of the accounts were categorized as bots in our sample, they were responsible for 25.72\% of the retweets. As a measure of popularity, we analyzed the number of likes that messages of humans and bots received. Messages from bots received \num{17.46}\% of the likes that pro-Russian messages received overall. Hence, messages from bots were slightly less popular than messages from humans (Mann-Whitney $U$ test: $U = 2 \cdot 10^9$; $p<0.001$ with $\mu_\textrm{bot} = 9.75$ and $\mu_\textrm{human} = 10.02$).

                

\begin{figure}[H]
\centering
    \begin{minipage}{\textwidth}
    \centering
        \begin{minipage}{0.4\textwidth}
            \begin{figure}
                \centering
                \raggedright\figletter{a}\\
                \includegraphics[width=\textwidth]{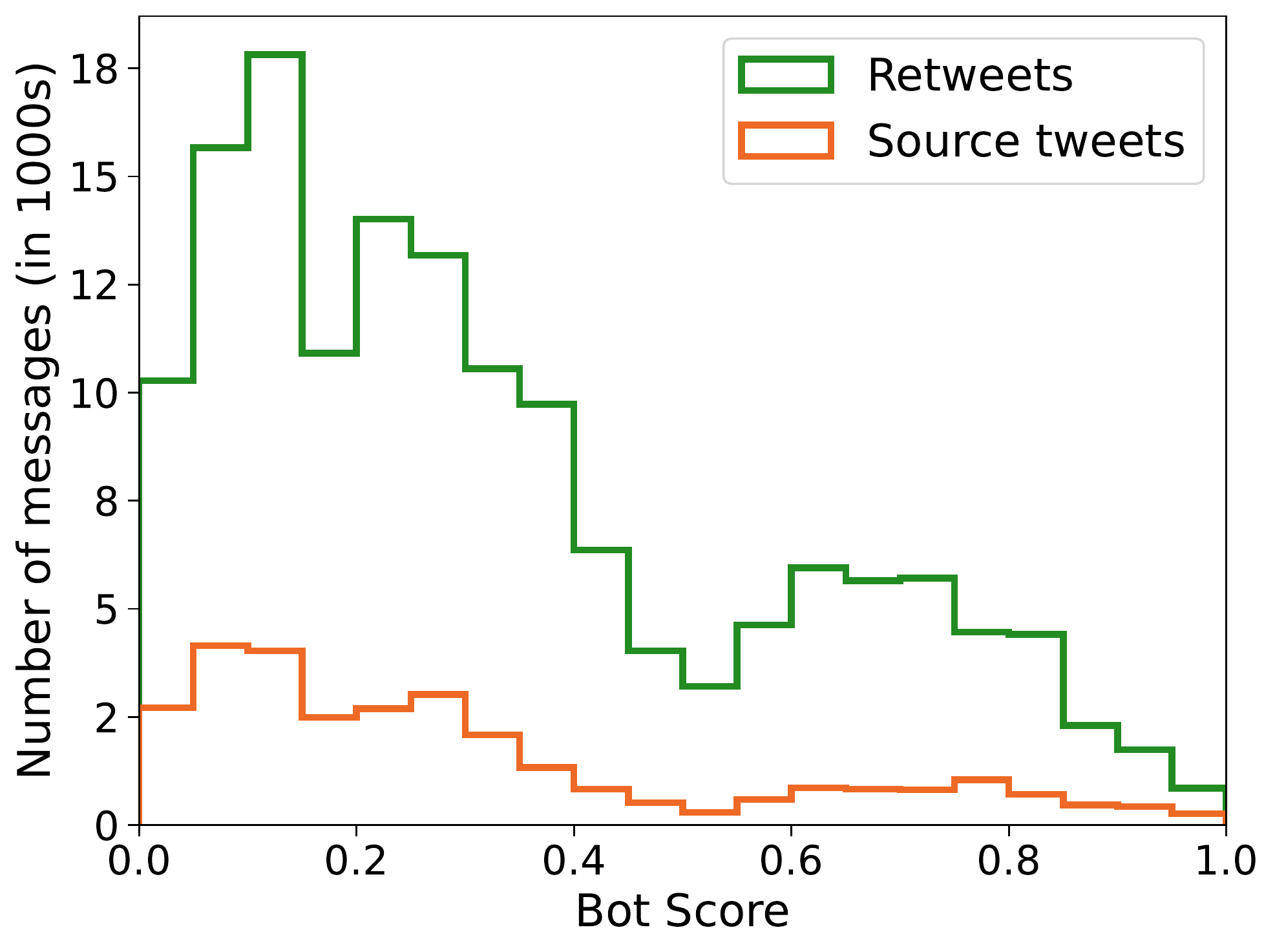}
            \end{figure}
        \end{minipage}
        \qquad
        \begin{minipage}{0.49\textwidth}
            \begin{figure}
                \centering
                \raggedright\figletter{b}\\
                \includegraphics[width=\textwidth]{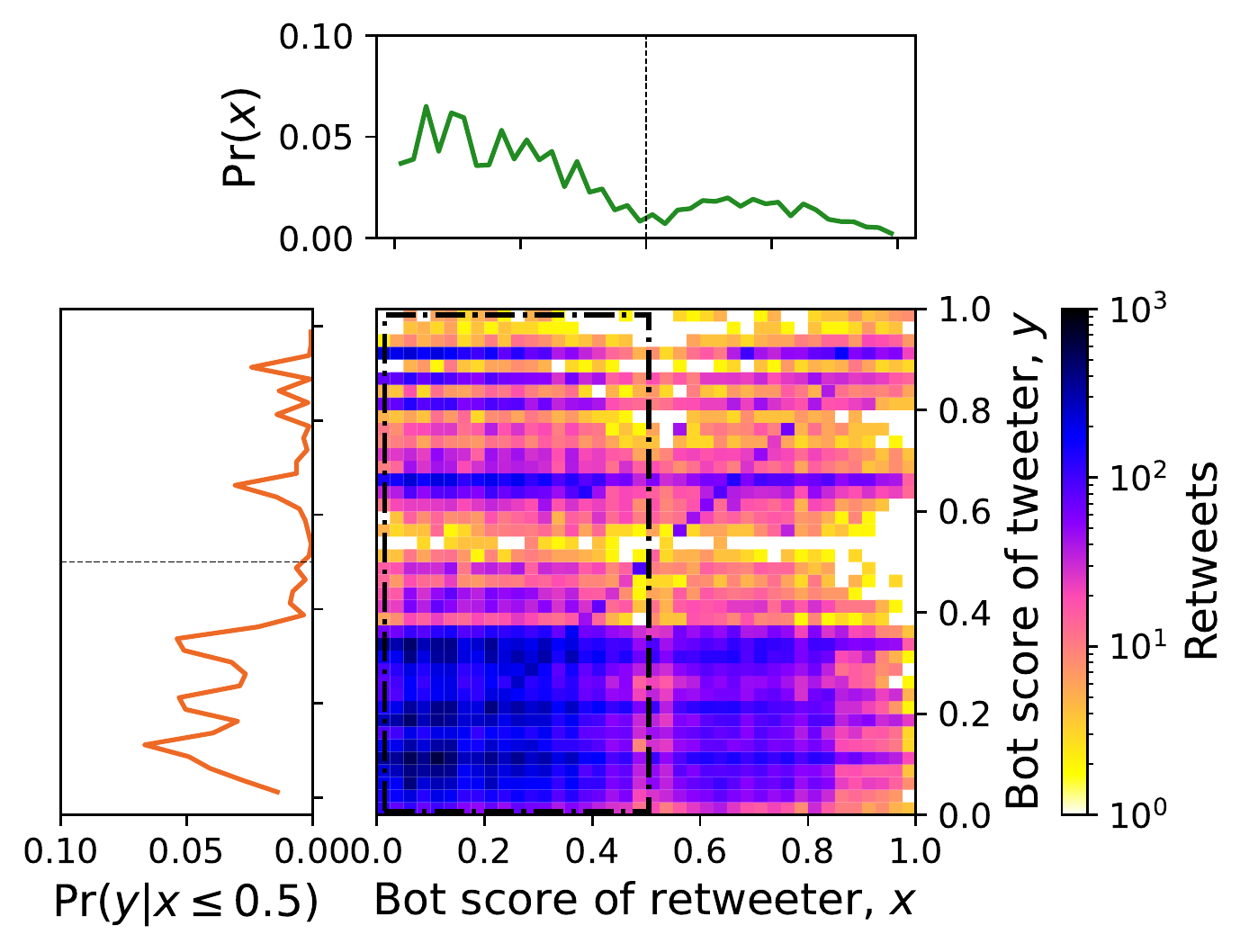}
            \end{figure}
        \end{minipage}
    \end{minipage}
\caption{\textbf{Impact of humans and bots.} \figletter{a},~Distribution of bot scores for source tweets and retweets. The two groups had significantly different bot scores (Mann-Whitney $U$ test: $U = 2 \cdot 10^9$; $p<0.001$, and Mood's median test: $\chi = 666.84 $; $p<0.001$ with median bot score of source tweeters $ = 0.22$ and median bot score of retweeters $ = 0.27$), implying that retweeters were more likely to be bots. \figletter{b},~Joint distribution of bot scores of authors of source tweet-retweet pairs (heatmap). The top subplot shows the distribution of bot scores for retweeters. The left subplot shows the distribution of bot scores for accounts that were retweeted by accounts classified as humans (using a threshold of 0.5). We find that most source tweets were posted by humans. They were also active retweeters, but so were bots. Different from the spread of low-credibility content \cite{Shao.2018}, we find that a significant proportion of retweeters were bots and that they tended to retweet humans rather than other bots. 
}
\label{fig:retweeter_bot_scores}
\end{figure}

We further explored the messaging activity of humans vs. bots. Specifically, we studied the distribution of bot scores across authors of source tweets and retweets (Figure~\ref{fig:retweeter_bot_scores}a). Here, we again find that humans took a leading role in content creation. We also explored how humans interacted with messages shared by bots. This provides insights into whether bots were able to elicit human interactions such as retweeting. For this, we computed the distribution of bot scores for each source tweet--retweet pair and thus analyzed who retweets whom (Figure~\ref{fig:retweeter_bot_scores}b). Generally, humans did most of the tweeting (Figure~\ref{fig:retweeter_bot_scores}b, top). humans were also active in retweeting but bots were relatively more active (see Supplementary~Figure~\ref{fig:ccdfs_bot_human}). Moreover, many accounts retweeted themselves to amplify their own messages, a tactic that was commonly used by bots (\num{23.5}\% of the \num{1653} accounts that retweeted themselves were bots). The results confirmed our findings from the retweeting network: humans tended to retweet other humans, while bots were more inclined to retweet humans. However, humans rarely retweeted bots. This is a crucial difference from earlier work on low-credibility content for which humans have been found to frequently retweet bots \cite{Shao.2018}, implying that it is difficult for bots to make pro-Russian messages go viral among humans.  

\begin{figure}[H]
\centering
    \begin{minipage}{\textwidth}
    \centering
        \begin{minipage}{0.45\textwidth}
            \begin{figure}
                \centering
                \raggedright\figletter{a}\\
                \includegraphics[width=\textwidth]{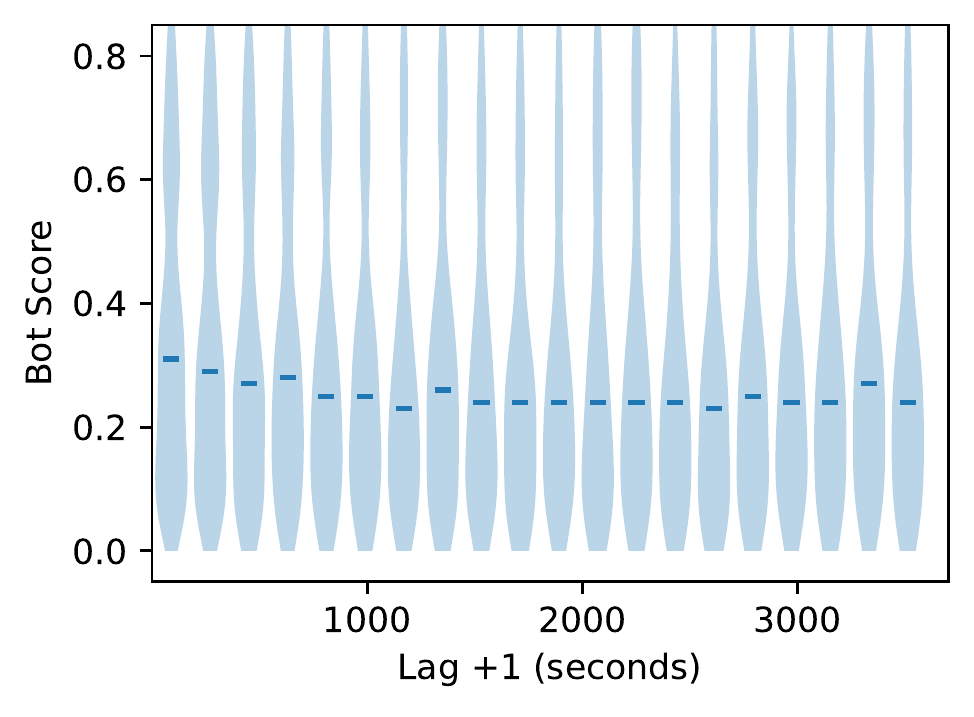}
            \end{figure}
        \end{minipage}
        \qquad
        \begin{minipage}{0.45\textwidth}
            \begin{figure}
                \centering
                \raggedright\figletter{b}\\
                \includegraphics[width=\textwidth]{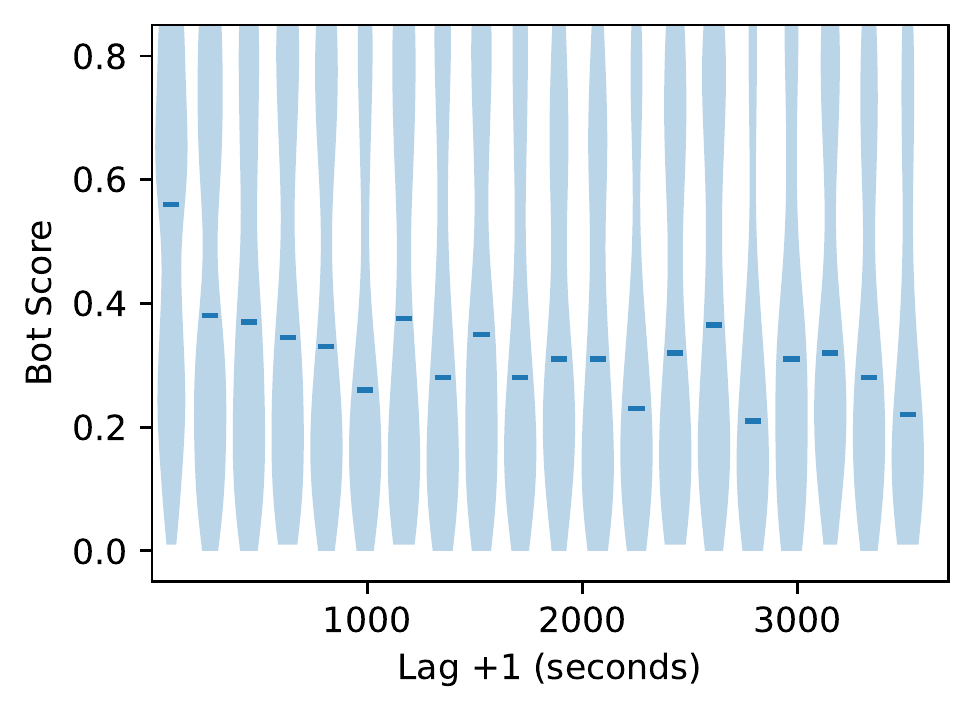}
            \end{figure}
        \end{minipage}
    \end{minipage}
    
    \begin{minipage}{0.5\textwidth}
    \centering
        \begin{figure}
            \centering
            \raggedright\figletter{c}\\
            \includegraphics[width=\textwidth]{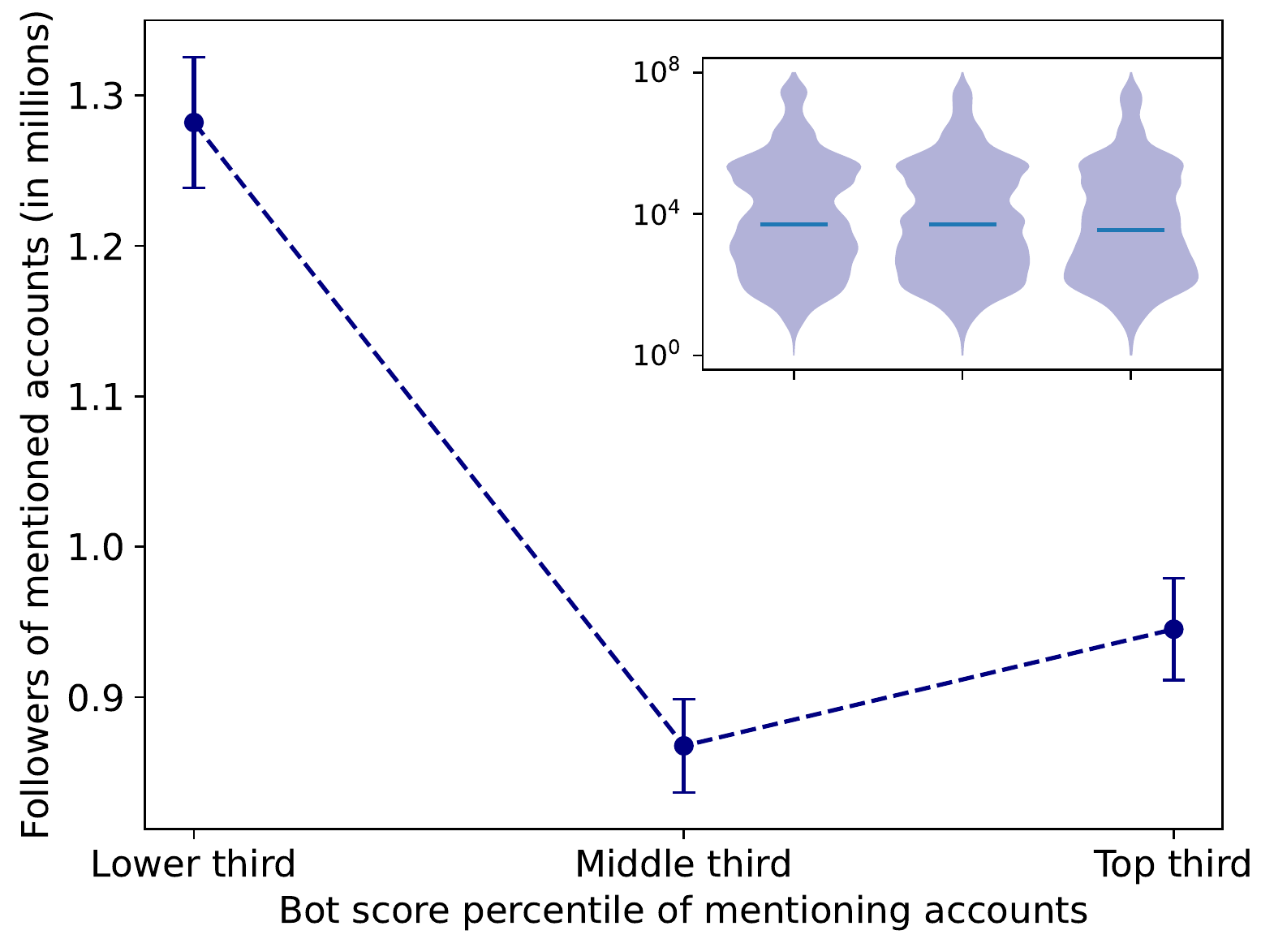}
        \end{figure}
    \end{minipage}
\caption{\textbf{Bot strategies.} \figletter{a},~Distribution of bot scores of accounts that retweeted human-made source tweets grouped by different time lags between source tweet and the corresponding retweet. \figletter{b},~Distribution of bot scores of accounts that retweeted likely bot-made source tweets grouped by different time lags between the source tweet and the corresponding retweet. Hence, bots (but not humans) tended to retweet bots to promote the early diffusion of pro-Russian messages. \figletter{c},~Here, we plot the average number of followers of mentioned accounts to analyze whether bots specifically targeted influential users. The mentioning accounts are grouped by their bot score percentile. Error bars indicate the standard errors. Inset: violin plot showing the distribution of follower counts for the mentioned user accounts in each bot score group. In violin plots, the width of a contour represents the probability of the corresponding value, and the median is marked by a colored line.
}
\label{fig:bot_strategy}
\end{figure}

Given this evidence, we further examined whether there were different temporal dynamics in the retweeting behavior of bots and humans. For this, we compared the bot score distribution of retweeters across different time lags for retweets (Figure~\ref{fig:bot_strategy}). We find that humans were retweeted equally fast by bots and humans (Figure~\ref{fig:bot_strategy}a), while bots were retweeted by other bots with a disproportionately small time lag (Figure~\ref{fig:bot_strategy}b). This suggests that bots systematically retweeted other bots early in the diffusion to promote the proliferation of pro-Russian support. 

Previous work found that a key strategy for bots is to spread content by mentioning influential accounts (\eg, \emph{\say{@UN}},\emph{\say{@cnnbrk}}, or \emph{\say{@RusEmbEthiopia}}), in the hope that they reshare and thus boost credibility \cite{Shao.2018}. To systematically analyze whether pro-Russian bots employ such a mentioning strategy, we computed the mean number of followers of the mentioned accounts (Figure~\ref{fig:bot_strategy}c). We find that humans tended to mention accounts with substantially more followers than bots. Recall that the number of followers is a common proxy for the social influence of online users \cite{Cha.2010}, which implies that bots tended to mention users with a smaller social influence in their messages. Notably, this finding differs from earlier research studying the spread of low-credibility content through bots, where bots -- and not humans -- target influential users to make messages strategically go viral \cite{Shao.2018}. 

An alternative proxy for the social influence of users is their centrality in a retweet network, computed as their PageRank \cite{Stella.2018}. Consistent with the above findings, we find that bots mentioned users with lower PageRank (mean PageRank of \num{0.002}) than humans (mean PageRank of 0.0022). This difference is statistically significant (Mann-Whitney $U$ test: $U = 2 \cdot 10^{10}$; $p<0.001$) and, again, differs from earlier findings, where bots have been found to target influential users at the center of retweeting networks when promoting the spread of inflammatory content \cite{Stella.2018}.

\section*{Discussion}
\label{sec:discussion}

The massive spread of online propaganda has been identified as a major threat to democracies \cite{Aral.2019}. While propaganda is a tool that has been used since ancient times, social media has made its spreading faster and more scalable, thereby presenting particularly fertile ground for sowing propaganda. Prior research provides evidence of systematic social media propaganda campaigns that aim to influence geopolitical events such as elections \cite{Eady.2023, Bail.2020, Shao.2018, Guess.2020}. Online propaganda has also become a concerning tool in modern warfare. Here, a particular threat is that social media amplifies the spread of misinformation and helps propaganda campaigns to shape false narratives around wars \cite{Scott.2022}. So far, however, there is little systematic, scientific research that analyzed the spread of pro-Russian support during the 2022 Ukraine invasion, which is our contribution. Unlike earlier research on historical tactics of the IRA \cite{Eady.2023, Shao.2018, Badawy.2018, Guess.2020, Luceri.2020, Bessi.2016, Dutta.2021, Arif.2018}, we focus on a recent foreign influence operation that employed state-of-the-art and novel tactics to proliferate propaganda (e.g., by making large-scale use of automation through bots). 

We find robust support for a Russian propaganda campaign, defined as systematic and coordinated efforts to manipulate beliefs and behaviors in the propagandists' interests \cite{Jowett.Chapter1.2012}. Pro-Russian messages have been spread on Twitter disproportionately through bots, which interacted in highly-connected retweet networks. The retweet networks showed distinctive clusters in countries that are of key interest for Russian politics (\eg, India and South Africa) and thus suggest a coordinated effort. The accumulation of messages on the day of the UN vote on Resolution \mbox{ES-11/1} gives rise to concerns that countries that abstained from the UN voting were targeted by Russian propaganda efforts. Strikingly, many bots that spread pro-Russian messages were created shortly before the UN vote, which indicates an intentional and planned manipulation of public opinion on Twitter as part of a Russian propaganda campaign. 

Our findings demonstrate that bots are an important driver in the early diffusion of bot-created propaganda on social media. Bots were more active retweeters than humans and acted together in a coordinated manner. Unlike spreaders of low-credibility content \cite{Shao.2018} and inflammatory content \cite{Stella.2018}, bots mentioned users with less social influence than humans when spreading pro-Russian messages. A possible explanation for this strategy behind Russian propaganda is that, because bots were rarely retweeted by humans (cf. Figure~\ref{fig:retweeter_bot_scores}b), they did not target individuals. Instead, bots primarily aimed to expose users to organic, pro-Russian messages from humans. By creating traffic around Russian propaganda, certain hashtags appeared as so-called ``trending topics'' on the front page of Twitter and were thus visible to all users \cite{TheEconomist.2022, Atlantic.2022, BBC.2022}. This is especially alarming, since repeated exposure can lead people to perceive misinformation as accurate \cite{Pennycook.2018}.

Crucial differences between the spread of propaganda and the spread of low-credibility content \cite{Shao.2018, Stella.2018, Vosoughi.2018} by bots become evident. On the one hand, we identified bots as amplifiers of propaganda rather than content creators. bots in propaganda were more inclined to retweet than to produce ``original'' content (\eg, source tweets). On the other hand, bots did not specifically target influential users. Instead, they aimed at broad exposure to maximize the number of people that see their message. Previously, such an amplification strategy has been conjectured to be a mature tactic of the IRA \cite{MisinformationReview.2022}. The likely goal is to augment the prominence and activity level of organic accounts that naturally act in ways that are aligned with the objectives of the propaganda campaign. 

As with other research, ours is not free of limitations, which presents opportunities for future research. First, our results are based on a single social media platform. However, Twitter is a platform with a particularly large and international audience, which makes it a fertile ground for planting propaganda and, hence, presents a common focus in earlier research \cite{Caldarelli.2020, Alieva.2022, Bail.2020, Badawy.2018, Dutta.2021}. Second, our data covers mostly messages in English since we searched messages based on English hashtags. However, these hashtags went reportedly viral in March 2022 \cite{TheEconomist.2022, Atlantic.2022, BBC.2022} and, subsequently, were widely used as search terms as well as to strategically flag corresponding messages. Third, the pro-Ukrainian support on Twitter is much larger than the pro-Russian support in absolute terms. This is likely the case since the main user base of Twitter is located in the West, which mostly supports Ukraine in the conflict. By primarily analyzing pro-Russian support, we focus on a minority of all tweets around the Russo-Ukrainian war. However, there is anecdotal evidence that there is a coordinated propaganda campaign behind the pro-Russian support on Twitter, and not behind the pro-Ukrainian support. Fourth, another limitation of our study is the possibility that Twitter may have removed some particularly egregious pro-Russian messages through content moderation efforts. However, messages that were removed by Twitter are also those that were hindered to go viral and that humans were thus not exposed to. Fifth, the accuracy of our analysis depends on the accuracy of other tools such as Botometer \cite{Varol.2017}. However, these tools have been shown to achieve a high accuracy \cite{Varol.2017} and are widely used in research \cite{Shao.2018, Bessi.2016, Ferrara.2017, SuarezSerrato.2016}. Sixth, while the scale of the pro-Russian support on Twitter is impressive in absolute terms (e.g., reached $\sim$\num{14.4}~million users), it may not have infiltrated online communities to an extent that swayed public opinion. Research largely still lacks an understanding of the real-world effects of social media propaganda \cite{Bail.2020}, which future work should explore. In particular, additional research with complementary research methods (e.g., survey approaches \cite{Eady.2023}) is needed to better understand the impact of exposure to propaganda on opinion formation and public discourse.

Our results have direct implications for society and democracies. First, our results are alarming as social media platforms present substantial vulnerabilities that propaganda campaigns can exploit strategically. Without significant effort by social media platforms to curb the spread of disinformation, toxic content can spread widely and virally \cite{Bar.2023b, Prollochs.2021, Prollochs.2021b, Prollochs.2023, Robertson.2023, Naumzik.2022}. Here, more research is needed to understand the mechanism behind the pro-Russian propaganda campaign \cite{Geissler.2023}, as well as machine learning for detection \cite{Maarouf.2023}. Second, our results suggest that an effective countermeasure to curb the spread of propaganda is to reduce the influence of bots. Here, it may be likely that counter-measures from fake news mitigation can be adapted \cite{Pennycook.2021, Gallotti.2020, Ducci.2020}; yet this requires further research to establish the effectiveness of such interventions. Third, propaganda on social media may influence public opinion and increase political division. It is thus important that policy-makers are aware of the potential threats that social media propaganda poses to modern societies. As such, it will be critical to continuously monitor and actively counter the proliferation of online propaganda in the future.



\newpage
\section*{List of abbreviations}

\noindent
\textbf{AUROC} \quad area under the receiver operating curve

\noindent
\textbf{IRA}\quad Internet Research Agency

\noindent
\textbf{KS test} \quad Kolmogorov-Smirnov test

\noindent
\textbf{U.K.} \quad United Kingdom

\noindent
\textbf{UN} \quad United Nations

\noindent
\textbf{U.S.} \quad United States of America


\newpage
\section*{Declarations}

\vspace{0.4cm}
\noindent
\textbf{Availability of data and material.} The data and code that support the findings of our study are available on GitHub (\url{https://github.com/DominiqueGeissler/Russian_Propaganda_on_social_media}). 

\vspace{0.4cm}
\noindent
\textbf{Competing interests.} The authors declare no competing interests.

\vspace{0.4cm}
\noindent
\textbf{Funding.} Not applicable.

\vspace{0.4cm}
\noindent
\textbf{Author contributions.} All authors contributed to conceptualization, results interpretation, and manuscript writing. DG contributed to data analysis. All authors approved the manuscript.

\vspace{0.4cm}
\noindent
\textbf{Acknowledgements.} Codes for plotting are based on Shao et~al.~\cite{Shao.2018}, which we gratefully acknowledge.


\newpage
\bibliography{literature}


\newpage 
\appendix

\renewcommand{\thetable}{S\arabic{table}}
\renewcommand{\thefigure}{S\arabic{figure}}
\setcounter{figure}{0}
\setcounter{table}{0}

\begin{minipage}[t]{\textwidth}
\nolinenumbers
\begin{center}
\huge\bfseries Supplementary Information
\end{center}
\vspace{1cm}
\end{minipage}


\newpage
\section*{Supplementary tables}

\subsection*{Supplementary Table 1: Examples of pro-Russian messages}
\label{supp:pro-russian_tweets}

\begingroup
\scriptsize\singlespacing
\begin{longtable}{p{.5cm} p{15cm}}
\toprule
{\#} & Message \\
\midrule
1  &  @RWApodcast I literally love Putin. The most honest leader in the world.  \texttt{\#istandwithrussia} \\
2  & America's position is known. Putin removes Pant Biden's \texttt{\#StandWithRussia} \\
3  & The notion used by the West (Nato) to attack Libya which you support: "their leader killed his own people" is the same one Russia uses in \texttt{\#Ukraine} - so why don't you support Russia's actions since a leader who kills his people must be killed? \texttt{\#IStandWithPutin} \texttt{\#istandwithrussia} \\
4  & \texttt{\#IStandWithPutin} \texttt{\#IStandWithPutin} \texttt{\#istandwithrussia} \texttt{\#istandwithrussia} \texttt{\#IStandWithPutin} \texttt{\#istandwithrussia} \texttt{\#IStandWithPutin}  \\
5  &  Let's United against Western countries  \texttt{\#IStandWithPutin} \\
6  &  @PalmerReport \texttt{\#IStandWithPutin}  \texttt{\#StandWithRussia}  is the only sensible trend I've seen so far... The propaganda that ukraine is using just showing how brave they are its bullshit. Crying but racists as fuck. May one day historians add a topic on \texttt{\#thefallofukraine} id be glad \\
7  &  Daily mail, it deals better with sewage, you are more professional there. You're funny, retarded reporters \texttt{\#Russia} \texttt{\#Ukraine} \texttt{\#Ukrania} \texttt{\#UkraineWar} \texttt{\#EU} \texttt{\#NATO} \texttt{\#USA} @UN @cnnbrk \texttt{\#IStandWithRussia} @BBC\_ua  \texttt{\#UkraineRussiaWar} @MailOnline \texttt{\#StandWithUkriane} \texttt{\#Kyiv} \texttt{\#Ukraine} \\
8  &  The threat to world peace is the USA and not Russia. Mandela said it and it still holds true. \texttt{\#iSupportRussia} \texttt{\#iStandWithRussia} \texttt{\#RacistEU} \\
9  &  Nazis lost access to the sea! \texttt{\#Russia} \texttt{\#StandWithRussia} \texttt{\#IStandWithPutin}  \\
10 &  The oppressors must be irritated \texttt{\#IStandWithPutin} \\
11 & @SheripetersonS @KylaInTheBurgh Lol.. Its better that Trump is not the President otherwise it would have not been so easy for putin to get Ukraine you fools.. U trust biden?? Biden keeps supporting china, during biden putin invaded Ukraine.. \texttt{\#istandwithrussia} \texttt{\#IStandWithPutin} \\
12 &  @CroatiaTruth @TheDailyShow So that makes okay to discriminate others based on race? It's really hard to sympathize with Ukraine at this point and moving towards \texttt{\#IStandWithPutin} \\
13 &  \texttt{\#IStandWithPutin} Putin is a brave and clever man. He is fighting this war for safe future of Russia \texttt{\#IStandWithPutin} \\
14 &  @heinz\_hartz Your leader is a a cocaine addict \texttt{\#istandwithrussia} \\
15 &  @KyivIndependent America \& the whole of NATO together Europe took advantage of Ukraine for their own benefit, got Ukraine and Russia to fight and USA was selling weapons/natural gas to Europe. Europe got hostile from Russia, and no one understood this \& that \texttt{\#Zelenskiy} is stupid \texttt{\#IStandWithPutin} \\
16 &  \texttt{\#istandwithrussia} \texttt{\#IStandWithPutin}  \texttt{\#iSupportRussia}   They always wanted to destroy Russia and they are doing it now.. \\
17 &  @davis\_valence @jazzamerica1 @BrotherWarfare You tell me, master. You're the brains here, we're all dummies. \texttt{\#Putin} \texttt{\#IStandWithPutin} \texttt{\#istandwithrussia} \texttt{\#NaziUkraine} \texttt{\#Russian} \texttt{\#RussianUkrainianWar} \texttt{\#NATO} \texttt{\#Israel} \texttt{\#Israeli} World War \\
18 &  @realGonzaloLira Ukrainian Nazis WILL BURN IN HELL, ALONG WITH NATO LEADERS WHO IS PUMPING WEAPONS INTO THE HANDS OF NAZIS \texttt{\#istandwithrussia} \\
19 &  @AlphaGlobalInc Money grabbers have been grooming Ukraine to rape \texttt{\#Russia} that's a fact. The \texttt{\#megaclubs} who supply \texttt{\#Ukraine} weapons and money are pimps stealing all your childrens lives, eventually. \texttt{\#StandWithRussia} \\
20  &  @CGMeifangZhang USA IS A KILLER! NATO IS A KILLER!  \texttt{\#IStandWithPutin}  \texttt{\#istandwithserbia}  \texttt{\#istandwithlivia}  \texttt{\#istandwithsiria} \texttt{\#istandwithchina} \texttt{\#stopnato}  \\
\bottomrule
    \caption{Examples of pro-Russian messages.}
    \label{tab:russian_tweets}
\end{longtable}
\endgroup

\newpage

\subsection*{Supplementary Table 2: Examples of pro-Ukrainian messages}
\label{supp:pro-ukrainian_tweets}

\begingroup
\scriptsize\singlespacing
\begin{longtable}{p{.5cm} p{15cm}}
\toprule
{\#} & Message \\
\midrule
1  & @ReutersWorld \texttt{\#DefeatPutin} \texttt{\#IStandWithPutin} \texttt{\#Donbass} \texttt{\#PutinHitler} \texttt{\#PutinIsaWarCriminal}  \\
2  &  \texttt{\#IStandWithPutin} \texttt{\#StopPutinNOW} \texttt{\#stopputin} \texttt{\#StopRussia} \texttt{\#Belarus}  \\
3  &  @carlosp202 @KyivIndependent My God only cowards shoot civilians especially children and women  that indicates that Putin is losing . \texttt{\#PutinWarCriminal} \texttt{\#StopRussianAggression} \texttt{\#StandWithRussia} \\
4  &  RT @davy\_mkisii: Better stop this war before kenya joins in . \texttt{\#RussianUkrainianWar}  \texttt{\#IStandWithPutin}  \texttt{\#StopPutinNOW}  \\
5  &   @TallDesmond If you are \texttt{\#IStandWithPutin} because the “West” has done some questionable stuff then you really need to examine your logic (and morals) \texttt{\#IStandWithUkraine} \texttt{\#PutinWarCriminal} \texttt{\#PutinHitler} \\
6  &  @POTUS @BorisJohnson @AndrzejDuda @EmmanuelMacron @OlafScholz \texttt{\#StopRussia} \texttt{\#StandWithUkraine} \texttt{\#StandWithPutin} \texttt{\#SaveUkraineNow}  \\
7  & @mccaffreyr3 Its all about money. So which country will Putin (Russia - \$10K GDP) attack next? \texttt{\#IStandWithUkriane}  I dont \texttt{\#IStandWithPutin} I dont \texttt{\#istandwithrussia} \texttt{\#Ukraine} \texttt{\#UkraineUnderAttack} \texttt{\#UkraineInvasion} \texttt{\#UkraineRussiaWar}\\
8  &  Isondo liyajika They're the ones seeking refuge now \texttt{\#UkraineUnderAttack} \texttt{\#UkraineRussianWar} \texttt{\#istandwithrussia} \\
9  & @blackintheempir Many people on Twitter are too smart to fall for Western propaganda But too stupid to not fall for Russian propaganda \texttt{\#fake} \texttt{\#Zelenskyy} Kremlin \texttt{\#war} \texttt{\#PutinWarCrimes} \texttt{\#IStandWithPutin} \texttt{\#UkraineInvasion}\\
10 & @ZMiasojedow Lie! it's \texttt{\#RussianPropaganda} \& hate for UA \texttt{\#MariupolMassacre} \texttt{\#MariupolGenocide} \texttt{\#NaziRussia} \texttt{\#DemilitarizeRussia} \texttt{\#DenazifyRussia} \texttt{\#StopRussianAggression} \texttt{\#StopRussianGenocideInUkraine} \texttt{\#StandWithUkraine} \texttt{\#StandUpForUkraine} \texttt{\#NaziRussianArmy} Don't \texttt{\#standwithrussia}\\
11 & @KyivIndependent Moments of the \texttt{\#Putin} regime  terrorists airstrike near the Kharkiv city  \texttt{\#PutinHitler} \texttt{\#StopPutinNOW} \texttt{\#IStandWithPutin} \texttt{\#StopPutinNOW} \texttt{\#Ukraine} \texttt{\#Lviv} \texttt{\#KyivNow}\\
12 &  @FoxNews \texttt{\#PutinHitler} \texttt{\#PutinIsaWarCriminal} \texttt{\#RussiaInvadedUkraine} \texttt{\#RussianArmy} \texttt{\#IStandWithPutin} \\
13 & Hit by an air strike today. 1,200 civilians, including children, were sheltering in it. \texttt{\#RussianUkrainianWar} \texttt{\#RussiaInvadedUkraine} \texttt{\#UkraineUnderAttack} \texttt{\#StandWithUkriane} \texttt{\#StopRussia} Only devil can \texttt{\#StandWithPutin} \\
14 &  By \texttt{\#PutinLies}, \texttt{\#putin} invent new enemies among \texttt{\#Russians} so \texttt{\#Russian} hate each other and missed \texttt{\#PutinIsaWarCriminal} while make everyone poor, shameful, guilty. Be angry, but not to your thinking brave citizens \texttt{\#Ukrainian} \texttt{\#UkraineRussianWar} \texttt{\#IStandWithPutin} \texttt{\#IStandWithUkriane}  \\
15 &  \texttt{\#istandwithrussia} \texttt{\#IStandWithPutin} trending, whilst the rest of the world is accusing him of world crime \texttt{\#RussianUkrainianWar} \texttt{\#StopPutinNOW}  \\
16 &  @IAPonomarenko @olgatokariuk To every russian out there. If you do not condemn the atrocities made by your boys and men in Ukraine, you are complicit.  \texttt{\#fckrussia} \texttt{\#IStandWithPutin} \texttt{\#WarCrimes} \texttt{\#Russians} \\
17 &  \texttt{\#IStandWithPutin} \texttt{\#StopPutinNOW} \texttt{\#stopputin} \texttt{\#StopRussia} \texttt{\#Belarus}  \\
18 &  \texttt{\#PutinWarCriminal} \texttt{\#PutinLies} \texttt{\#PutinsGOP} \texttt{\#Ukraine} \texttt{\#Russia} \texttt{\#IStandWithRussia} \texttt{\#UkraineRussia} \texttt{\#UkraineInvasion} \texttt{\#UkraineUnderAttack} \texttt{\#IStandWithUkraine}  \texttt{\#UkraineWar} \texttt{\#UkraineRussiaConflict} \texttt{\#Putin} shame on you @mgimo \\
19 & @Indddy77 Hey Russian bot troll - \texttt{\#IStandWithPutin} going to \texttt{\#TheHague} and imprisoned for war crimes.  \texttt{\#PutinWarCriminal} \texttt{\#Putin} \texttt{\#RussianBot} \texttt{\#RussianTroll}  \\
20  &  @MtwanaXabiso @EmbassyofRussia What's more powerful than a missile, fuckface? Slava Ukraina! \texttt{\#StandWithUkraine} \texttt{\#IStandWithUkraine} \texttt{\#FuckRussia} \texttt{\#RussianArmy} \texttt{\#RussianAirForce }\texttt{\#UkraineRussianWar} \texttt{\#UkraineUnderAttack} \texttt{\#Ukraine} \texttt{\#Russia} \texttt{\#UkraineRussiaCrisis} \texttt{\#StandWithRussia} \texttt{\#IStandWithRussia}   \\
\bottomrule
\caption{Examples of pro-Ukrainian messages.}
    \label{tab:ukrainian_tweets}
\end{longtable}
\endgroup

\newpage

\subsection*{Supplementary Table 3: Top hashtags in pro-Russian messages}
\label{supp:pro-russian_hashtags}

\begin{table}[H]
	\centering
	\footnotesize
	\singlespacing
    \begin{tabular}{llrr}
        \toprule
        {} &               Hashtag & \#Accounts &  Freq. \\
        \midrule
        1 &      \texttt{\#istandwithputin} &      73091 &     180937 \\
        2 &     \texttt{\#istandwithrussia} &      44747 &     126209 \\
        3 &  \texttt{\#russianukrainianwar} &      15757 &      46534 \\
        4 &        \texttt{\#russiaukraine} &      14050 &      18896 \\
        5 &               \texttt{\#russia} &      12234 &      20598 \\
        6 &      \texttt{\#standwithrussia} &       9972 &      23402 \\
        7 &              \texttt{\#ukraine} &       7223 &      13472 \\
        8 &                \texttt{\#putin} &       7118 &      11743 \\
        9 &     \texttt{\#ukrainerussiawar} &       6551 &      32382 \\
        10&       \texttt{\#standwithputin} &       5027 &      10231 \\
        \bottomrule
    \end{tabular}
    \caption{Most frequent hashtags in pro-Russian messages. Also shown is the number of accounts who tweeted them and the overall number of occurrences in the dataset.}
    \label{tab:russian_hashtags}
\end{table}

\newpage

\subsection*{Supplementary Table 4: Anti-Russian hashtags}
\label{supp:anti-russian_hashtags}

\begin{table}[H]
	\centering
	\footnotesize
	\singlespacing
    \begin{tabular}{ll}
        \toprule
        {} &               Hashtag \\
        \midrule
        1 & \texttt{\#stopputinnow} \\
        2 & \texttt{\#stoprussia} \\
        3 & \texttt{\#stopputin} \\
        4 & \texttt{\#fuckputin} \\
        5 & \texttt{\#putinwarcriminal} \\
        6 & \texttt{\#stopwar} \\
        7 & \texttt{\#ukraineunderattack} \\
        8 & \texttt{\#putinwarcrimes} \\
        9 & \texttt{\#putinisawarcriminal}  \\
        10 & \texttt{\#warcrimes} \\
        11 & \texttt{\#fckptn} \\
        12 & \texttt{\#noflyzone} \\
        13 & \texttt{\#fuckrussia} \\
        14 & \texttt{\#standwithrussiansagainstputin} \\
        15 & \texttt{\#attackputin} \\
        16 & \texttt{\#stopthewar} \\
        17 & \texttt{\#russianpropaganda}  \\
        18 & \texttt{\#defeatputin}  \\
        19 & \texttt{\#pissonputin}\\
        \bottomrule
    \end{tabular}
    \caption{List of anti-Russian hashtags used to filter messages with an anti-Russian stance.}
    \label{tab:anti-russian_hashtags}
\end{table}

\newpage

\subsection*{Supplementary Table 5: Exclusion of news media outlets}
\label{supp:western_news_media}

\begin{table}[H]
	\centering
	\footnotesize
	\singlespacing
    \begin{tabular}{llll}
        \toprule
        {\#} & Account & \# & Account  \\
        \midrule
        1 & @NBCNews        & 23 & @qWKRG \\
        2 & @thehill        & 24 & @NewsNation \\
        3 & @thetimes       & 25 & @KGETnews\\
        4 & @BITech         & 26 & @WSPA7 \\
        5 & @YahooNews      & 27 & @WJBF \\
        6 & @nytimesbusiness& 28 & @FOX23News \\
        7 & @Techmeme       & 29 & @WSAV \\
        8 & @derStandardat  & 30 & @WEHTWTVWlocal \\
        9 & @NBCNewsWorld   & 31 & @DavidClinchNews \\
        10 & @WTNH          & 32 & @WFRVLocal5 \\
        11 & @8NewsNow      & 33 & @wnct9 \\
        12 & @kron4news     & 34 & @KAMRLocal4News \\
        13 & @TheWrap       & 35 & @NBC6News \\
        14 & @WKRN          & 36 & @CW39Houston \\
        15 & @crikey\_news  & 37 & @WETM18News \\
        16 & @WFLA          & 38 & @WVNS59News \\
        17 & @whnt          & 39 & @TexomasHomepage \\
        18 & @webstandardat & 40 & @KTABTV \\
        19 & @abc27News     & 41 & @KMSSTV \\
        20 & @8NEWS         & 42 & @KRBCnews \\
        21 & @WJTV          & 43 & @KREX5\_Fox4 \\
        22 & @WTEN          & 44 & @CabbageTV \\
        \bottomrule
    \end{tabular}
    \caption{List of Western news media outlets with verified accounts on Twitter that were excluded due to their journalistic nature (e.g., they reported on that some pro-Russian hashtags went viral or that Twitter took action against Russian propaganda but without disseminating Russian propaganda themselves).}
    \label{tab:western_news_media}
\end{table}

\newpage

\subsection*{Supplementary Table 6: Anti-Ukrainian hashtags}
\label{supp:anti-ukrainian_hashtags}

\begin{table}[H]
	\centering
	\footnotesize
	\singlespacing
    \begin{tabular}{ll}
        \toprule
        {} &               Hashtag \\
        \midrule
        1 & \texttt{\#stopukrainianaggression} \\
        2 & \texttt{\#russianlivesmatter} \\
        3 & \texttt{\#zelenskywarcriminal} \\
        4 & \texttt{\#nazisinukraine} \\
        5 & \texttt{\#denazifyukraine} \\
        \bottomrule
    \end{tabular}
    \caption{List of anti-Ukrainian hashtags used to filter messages with an anti-Ukrainian stance.}
    \label{tab:anti-ukrainian_hashtags}
\end{table}

\newpage

\subsection*{Supplementary Table 7: Influential users}
\label{supp:influential_users}

\begingroup
\footnotesize\singlespacing
\begin{longtable}{p{2.5cm} p{10cm} >{\centering}p{2cm} >{\centering\arraybackslash}p{1.5cm}}
\toprule
 Username & Profile description & \#Followers & Verified  \\
\midrule
 Misha Collins & Actor, baker, candlestick maker 
 Cell: (323)405-9939 
 he/him  & \num{2836321} & \includegraphics[width=0.5cm]{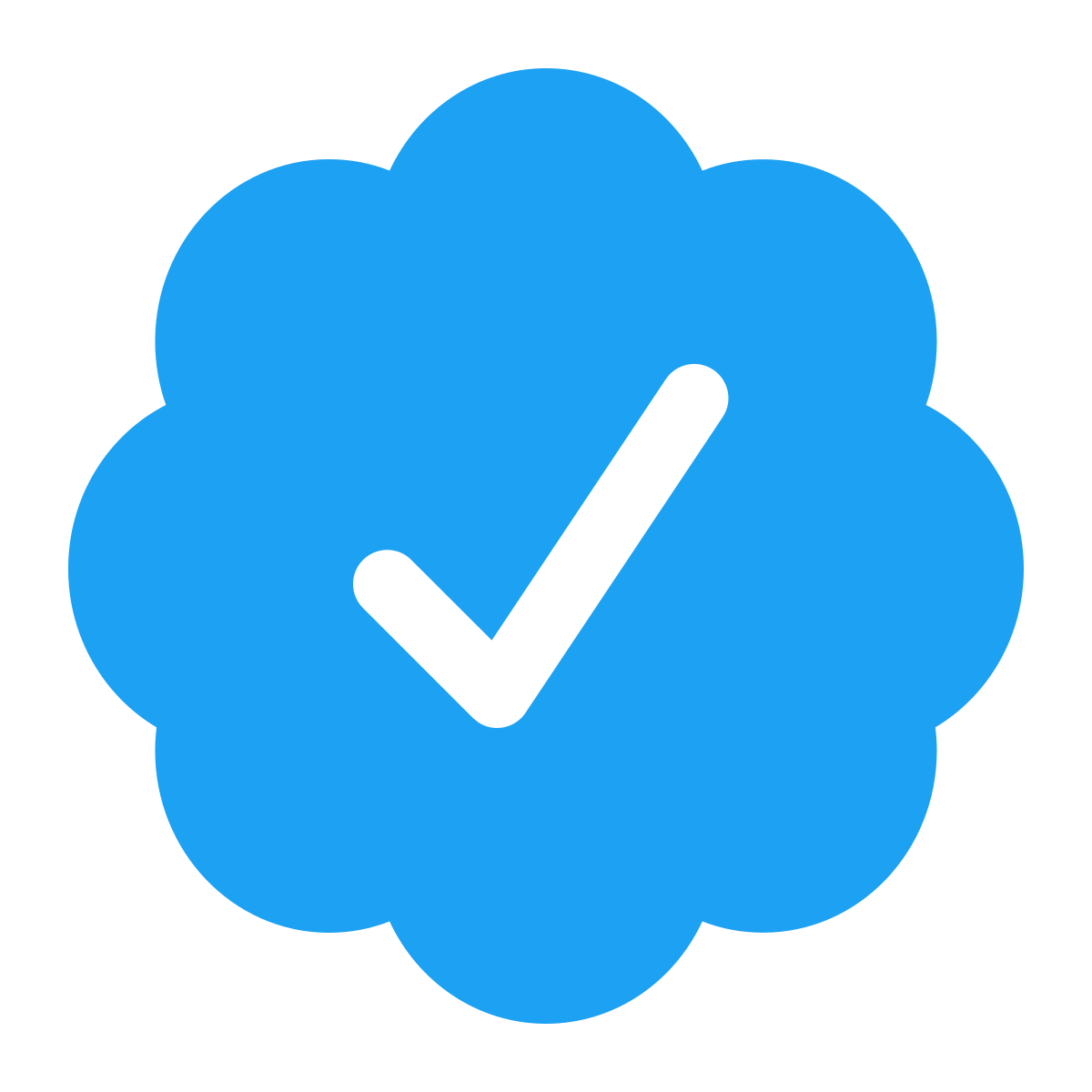}       \\
 Firstpost & Incisive opinions, in-depth analysis and views that matter. & \num{2077445} & \includegraphics[width=0.5cm]{figures/Twitter_Verified_Badge.svg.png}       \\
 Robert ALAI & I am for humanity in whatever we do. We must rethink Nairobi and make it humane. 
 Email: me@robertalai.com & \num{1783408} & \includegraphics[width=0.5cm]{figures/Twitter_Verified_Badge.svg.png}       \\
 John Cusack & Apocalyptic shit disturber and elephant trainer & \num{1736601} & \includegraphics[width=0.5cm]{figures/Twitter_Verified_Badge.svg.png}       \\
 Gharidah Farooqi & I AM \textasciitilde{} Steel Magnolia. Millennial. Feminist. Activist. Journalist since 2003. Program 'G For Gharidah' on News One - Monday to Thursday. Tweets only personal. & \num{1629142}  & \includegraphics[width=0.5cm]{figures/Twitter_Verified_Badge.svg.png}       \\
 Donald B Kipkorir  & KTK Advocates: Corporate Law \& Commercial Practice , IFLR1000 Recognized Law Firm, Member Of https://t.co/OYIejTcFud  .. Roman Catholic, Monarchist, Strong Rule & \num{1165978} & \includegraphics[width=0.5cm]{figures/Twitter_Verified_Badge.svg.png}       \\
 ChrisExcel & I'm a Savage.. I'm an AssHole I'm A King!!! The only legal Catfish BLACK TWITTER President & \num{1088859} & --- \\
 extra3 & Der Irrsinn der Woche. Not established since 1976 / Impressum: https://t.co/DGrtF5g13t, https://t.co/2rwf6gQZpJ & \num{1080864} & \includegraphics[width=0.5cm]{figures/Twitter_Verified_Badge.svg.png} \\
 Jon Cooper & Former National Finance Chair of Draft Biden 2016, Long Island Campaign Chair for @BarackObama \& Majority Leader of Suffolk County Legislature, NY. @DukeU alum & \num{1030357} & \includegraphics[width=0.5cm]{figures/Twitter_Verified_Badge.svg.png} \\
 SA Breaking News & All the latest breaking news from across South Africa in one stream. info@sabreakingnews.co.za & \num{915583} & --- \\
\bottomrule
    \caption{\textbf{Influential users.} The number of followers is used as a proxy for social influence of Twitter users \cite{Cha.2010}. Listed are the top-10 users with the largest number of followers (in decreasing order). To further characterize influential accounts, we manually inspected messages spread by these accounts. The top-3 accounts, for example, pursue different communication strategies. The first critically engages in the discussion on Twitter; the second primarily reports the hashtag \texttt{\#istandwithputin} to trend in India; and the third actively disseminated pro-Russian messages. For example, the account posted \emph{``The threat to world peace is the USA and not Russia. Mandela said it and it still holds true. \texttt{\#iSupportRussia} \texttt{\#iStandWithRussia} \texttt{\#RacistEU}''}. The account also retweeted Russian propaganda: \emph{``RT \@DonaldBKipkorir: US \& NATO want to destroy Russia the way they did Afghanistan, Iraq, Yemen, Lebanon, Somalia \& Libya ... US \& Europe media is curating false \& alternative narrative on the war in Ukraine .. Russia \& her people have legitimate \& strategic interest in Ukraine .. \texttt{\#IStandWithPutin}}.''}
    \label{tab:influential_indegree_users}
\end{longtable}
\endgroup

\newpage
\section*{Supplementary Figures}

\subsection*{Supplementary Figure 1: Message volume of bots and humans over time}
\label{supp:message_volume_bothuman}

\begin{figure}[H]
\centering
    \begin{minipage}{\textwidth}
    \centering
        \begin{figure}
            \centering
            \includegraphics[width=.75\textwidth]{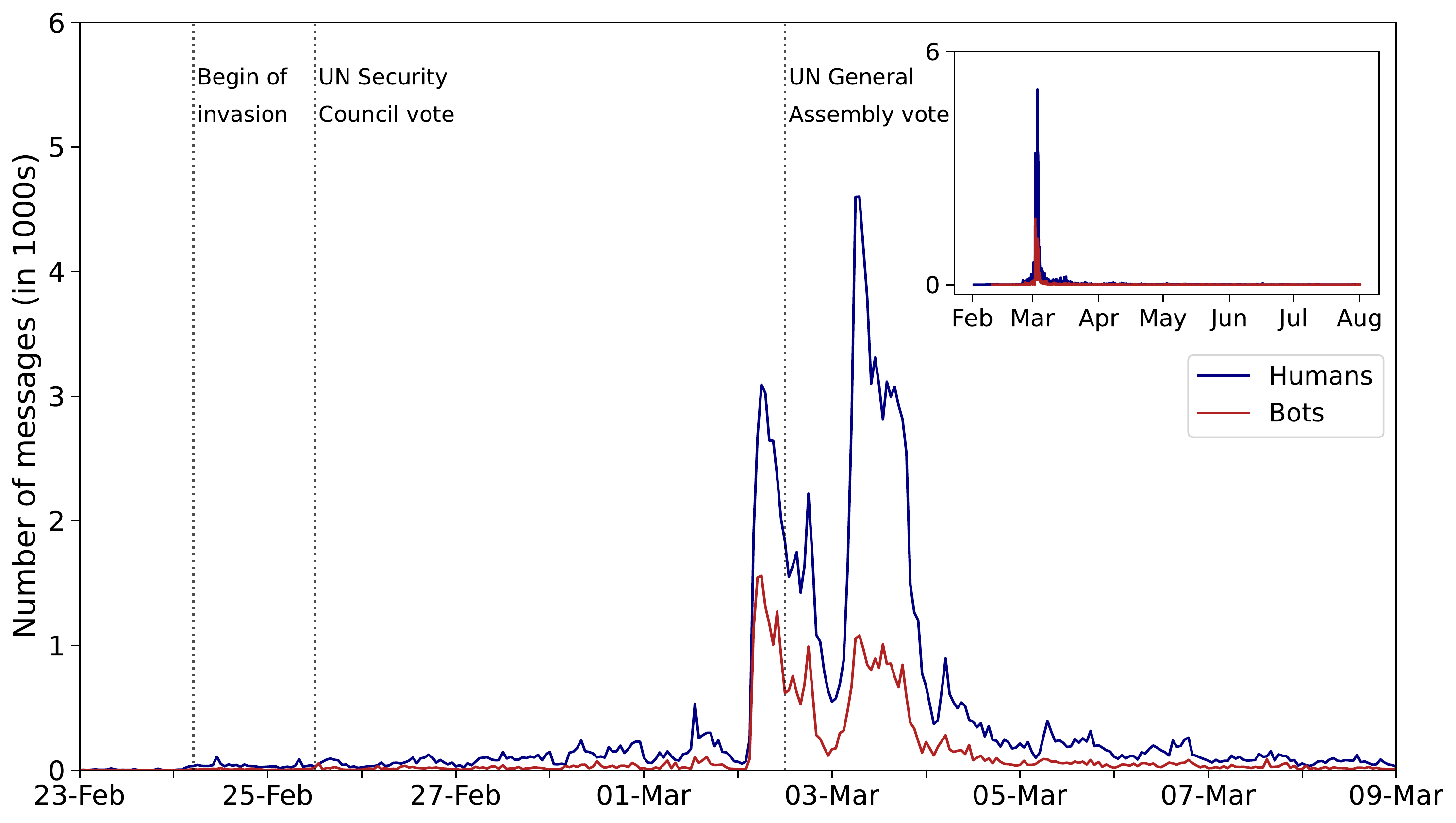}
        \end{figure}
    \end{minipage}
\caption{\textbf{Temporal dynamics of pro-Russian bots and humans.} The plot shows the number of pro-Russian messages from bots and humans during the first two weeks of the invasion. The peaks on March~2 and 3 coincide with the overall peaks of messages and corroborate our findings. Inset: volume of pro-Russian messages for the entire time period of the dataset.}
\label{fig:message_volume_bothuman}
\end{figure}

\subsection*{Supplementary Figure 2: Creation dates of pro-Russian supporter without bot information}
\label{supp:spreaders_pro_russia_undefined}

\begin{figure}[H]
\centering
    \begin{minipage}{0.45\textwidth}
        \begin{figure}
            \centering
            \raggedright\figletter{a}\\
            \includegraphics[width=\textwidth]{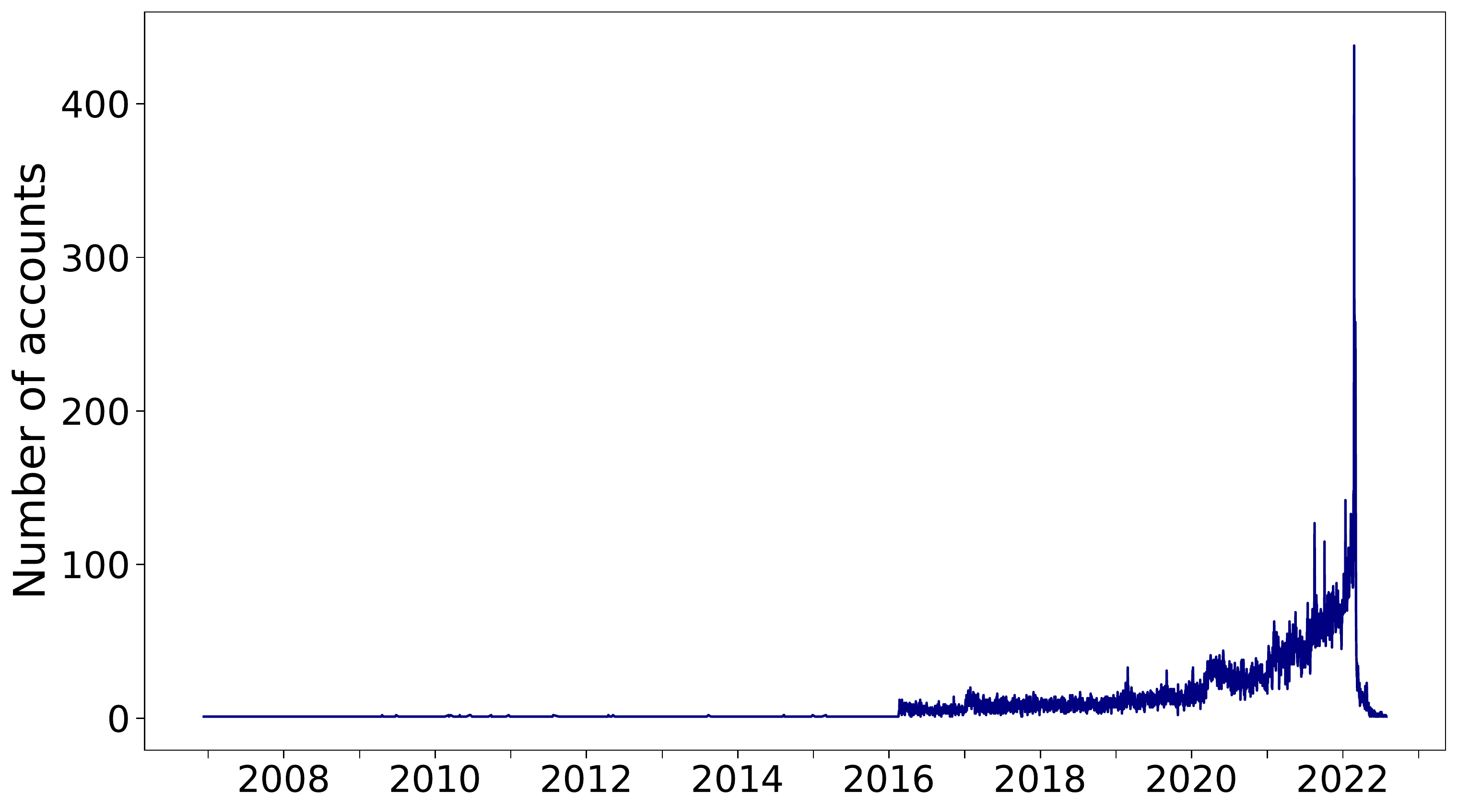}
        \end{figure}
    \end{minipage}
    \quad\quad    
        \begin{minipage}{0.45\textwidth}
        \begin{figure}
            \centering
            \raggedright\figletter{b}\\
            \includegraphics[width=\textwidth]{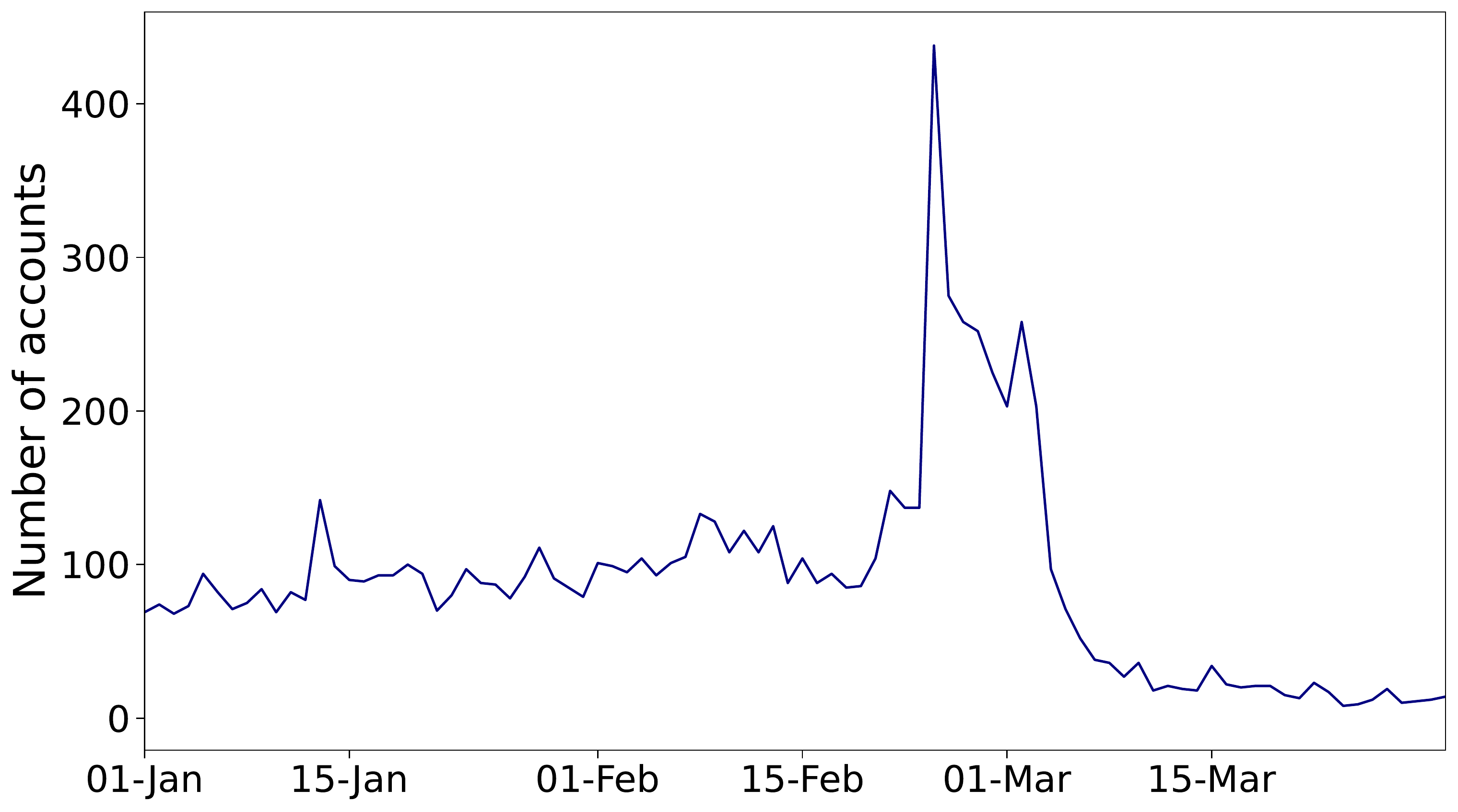}
        \end{figure}
    \end{minipage}
\caption{\textbf{Spreaders of pro-Russian messages without bot information.} \figletter{a},~Dates on which accounts were created. Here, the time axis starts with the inception of Twitter in 2006. In contrast to above, accounts without bot score information were only created after 2016, while the vast majority was created shortly before the invasion. \figletter{b},~Dates on which accounts were created. Here, the time axis starts shortly before the beginning of the 2022 Russian invasion.}
\label{fig:spreaders_pro_russia_undefined}
\end{figure}

\subsection*{Supplementary Figure 3: Robustness check for location analysis }
\label{supp:worldmaps_nofollowers}

\begin{figure}[H]
\centering
    \begin{minipage}{\textwidth}
    \centering
        \begin{minipage}{\textwidth}
            \begin{figure}
                \centering
                \raggedright\figletter{a}\\
                \includegraphics[width=\textwidth]{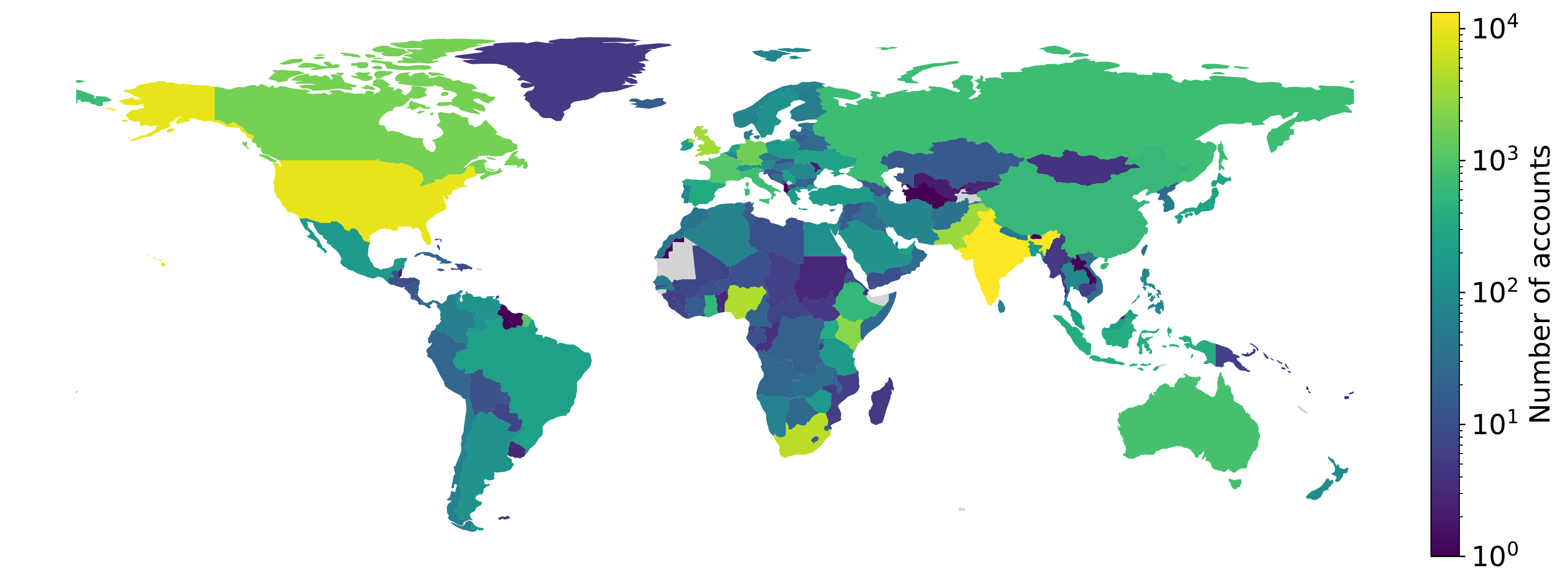}
            \end{figure}
        \end{minipage}
        
        \begin{minipage}{\textwidth}
            \begin{figure}
                \centering
                \raggedright\figletter{b}\\
                \includegraphics[width=\textwidth]{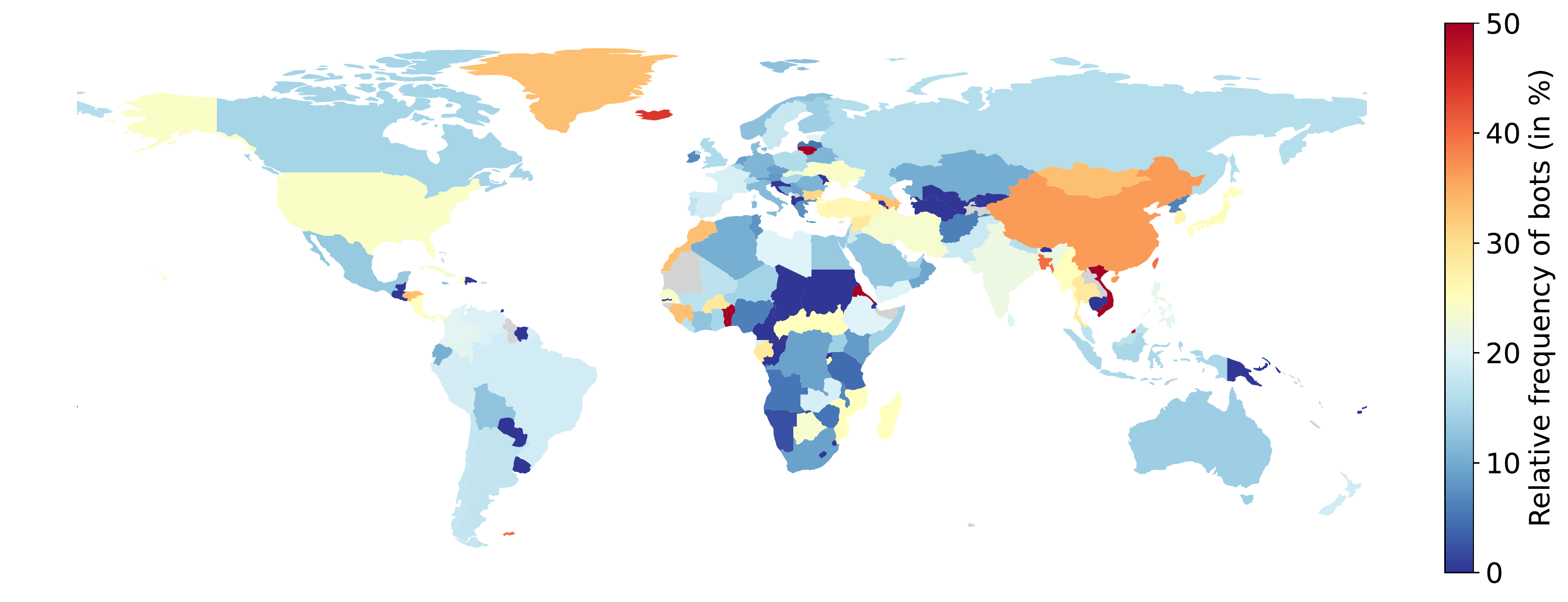}
            \end{figure}
        \end{minipage}
    \end{minipage}
\caption{\textbf{Robustness check for location analysis.} Here, we perform a robustness check using a different approach where we infer the geographic location of accounts via the self-reported location in a user's profile and via the geolocations in messages (that is, without using the heuristics based on the geographic location of followers). Shown are the differences in Russian propaganda across countries based on: \figletter{a},~Number of users per country (log scale). \figletter{b},~Relative frequency of bots per country (in \%). Here, all bots were excluded where the geographic location could not be inferred. Overall, we find patterns similar to the location analysis in the main paper. }
\label{fig:worldmap_nofollowers}
\end{figure}

\subsection*{Supplementary Figure 4: Cross-country difference for humans, bots and accounts without bot information}
\label{supp:worldmaps}

\begin{figure}[H]
\centering
    \begin{minipage}{\textwidth}
    \centering
        \begin{minipage}{\textwidth}
            \begin{figure}
                \centering
                \raggedright\figletter{a}\\
                \includegraphics[width=0.8\textwidth]{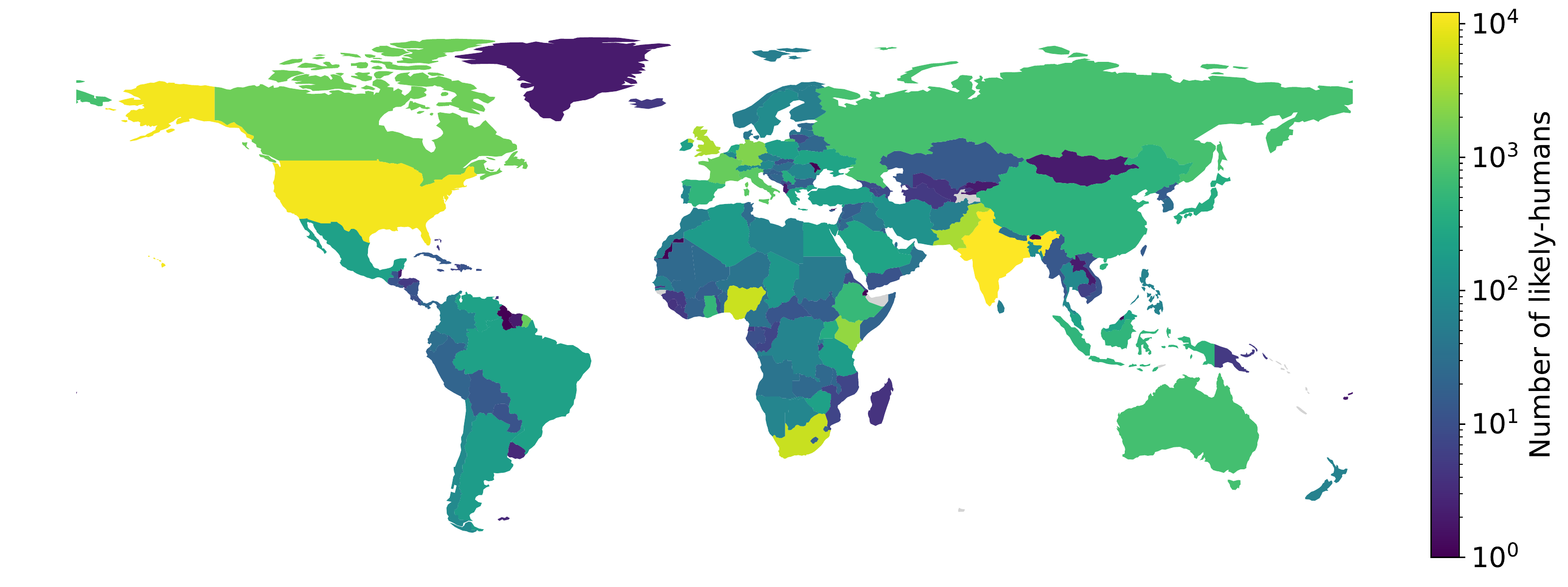}
            \end{figure}
        \end{minipage}
        
        \begin{minipage}{\textwidth}
            \begin{figure}
                \centering
                \raggedright\figletter{b}\\
                \includegraphics[width=0.8\textwidth]{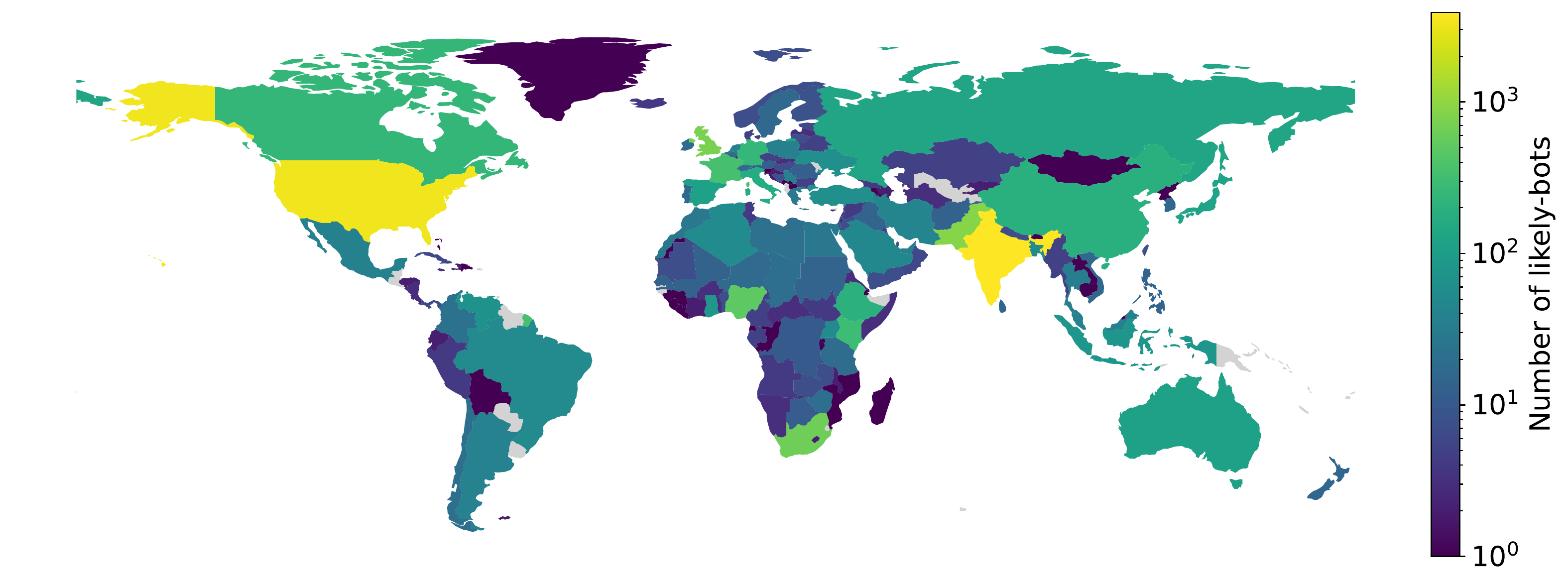}
            \end{figure}
        \end{minipage}

        \begin{minipage}{\textwidth}
            \begin{figure}
                \centering
                \raggedright\figletter{c}\\
                \includegraphics[width=0.8\textwidth]{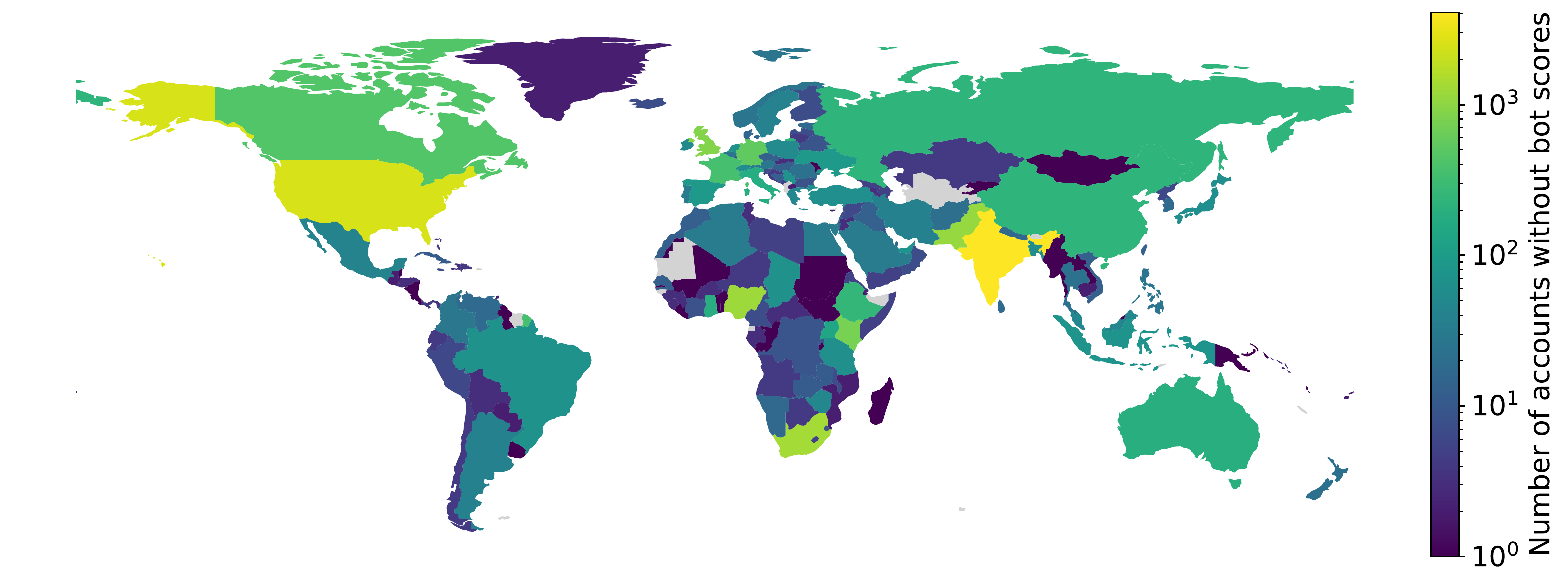}
            \end{figure}
        \end{minipage}
    \end{minipage}
\caption{\textbf{Cross-country differences in the spread of pro-Russian support by humans, bots, and accounts without bot information.} \figletter{a},~Number of humans per country (log scale). \figletter{b},~Number of bots per country (log scale). \figletter{c},~Number of accounts without bot information (log scale). None of the groups show deviating geospatial patterns.}
\label{fig:worldmap_supp}
\end{figure}

\subsection*{Supplementary Figure 5: Retweet networks of users without bot informations}
\label{supp:retweet_network_undefined}

\thispagestyle{empty}
\begin{figure}[H]
\centering
    \begin{minipage}{\textwidth}
    \centering
        \begin{minipage}{0.45\textwidth}
            \begin{figure}
                \centering
                \raggedright\figletter{a}\\
                \hspace{1cm}\includegraphics[width=.8\textwidth]{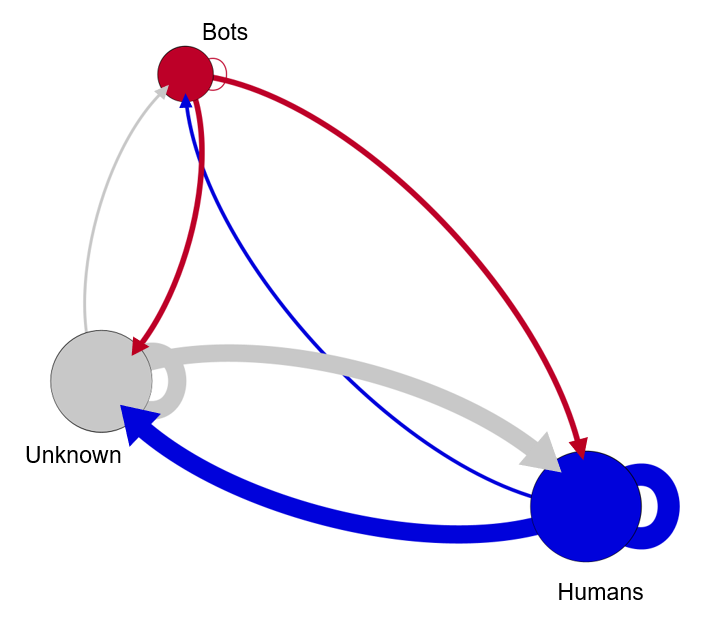}
            \end{figure}
        \end{minipage}
        \qquad
        \begin{minipage}{0.45\textwidth}
            \begin{figure}
                \centering
                \raggedright\figletter{b}\\
                \includegraphics[width=\textwidth]{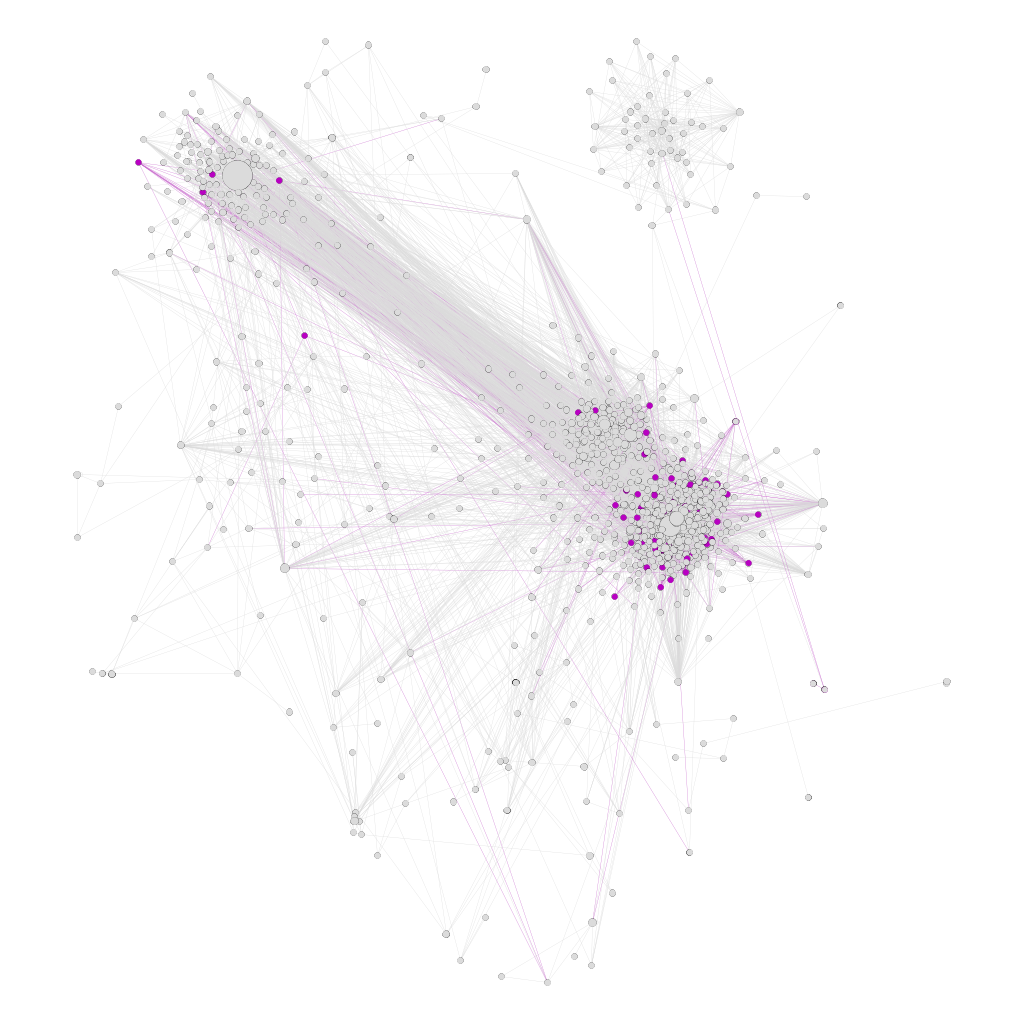}
            \end{figure}
        \end{minipage}
    \end{minipage}
    
    \begin{minipage}{\textwidth}
    \centering
        \begin{minipage}{0.45\textwidth}
            \begin{figure}
                \centering
                \raggedright\figletter{c}\\
                \includegraphics[width=\textwidth]{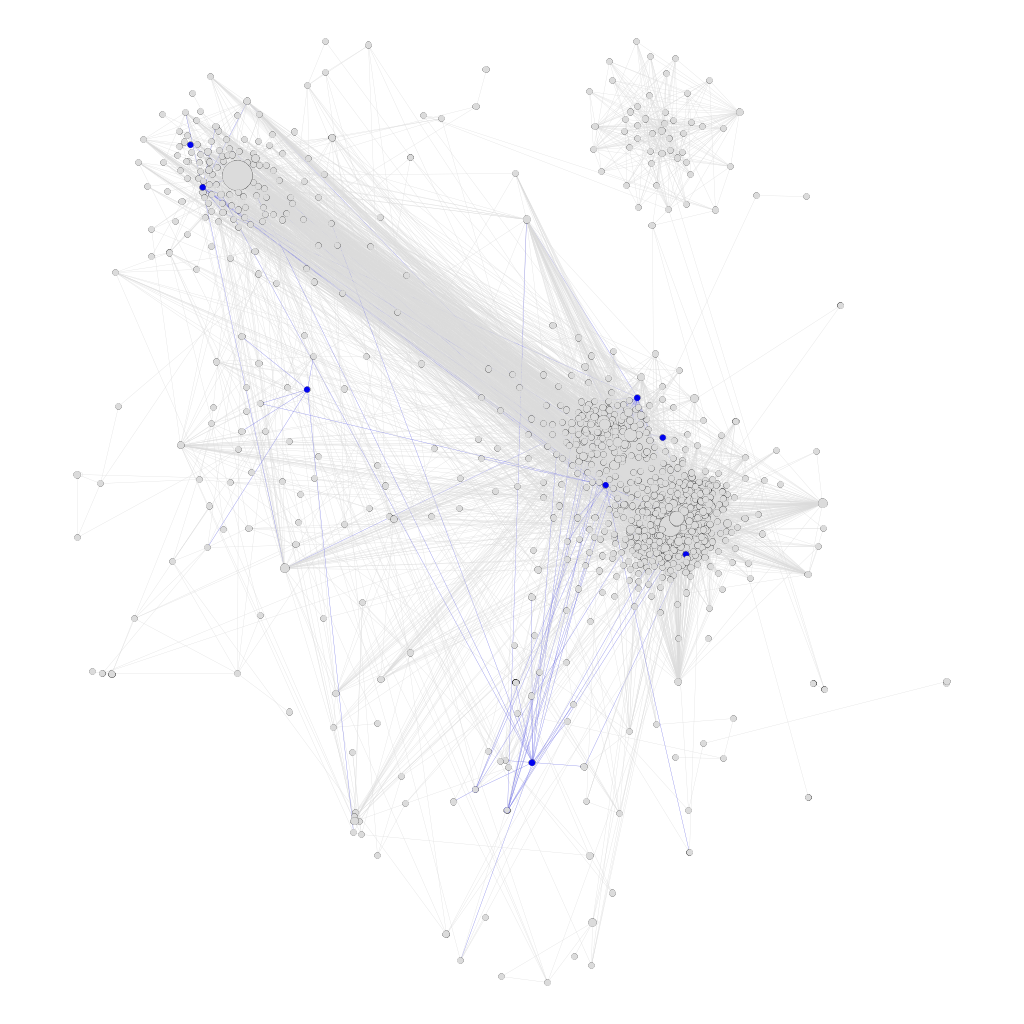}
            \end{figure}
        \end{minipage}
        \begin{minipage}{0.45\textwidth}
            \begin{figure}
                \centering
                \raggedright\figletter{d}\\
                \includegraphics[width=\textwidth]{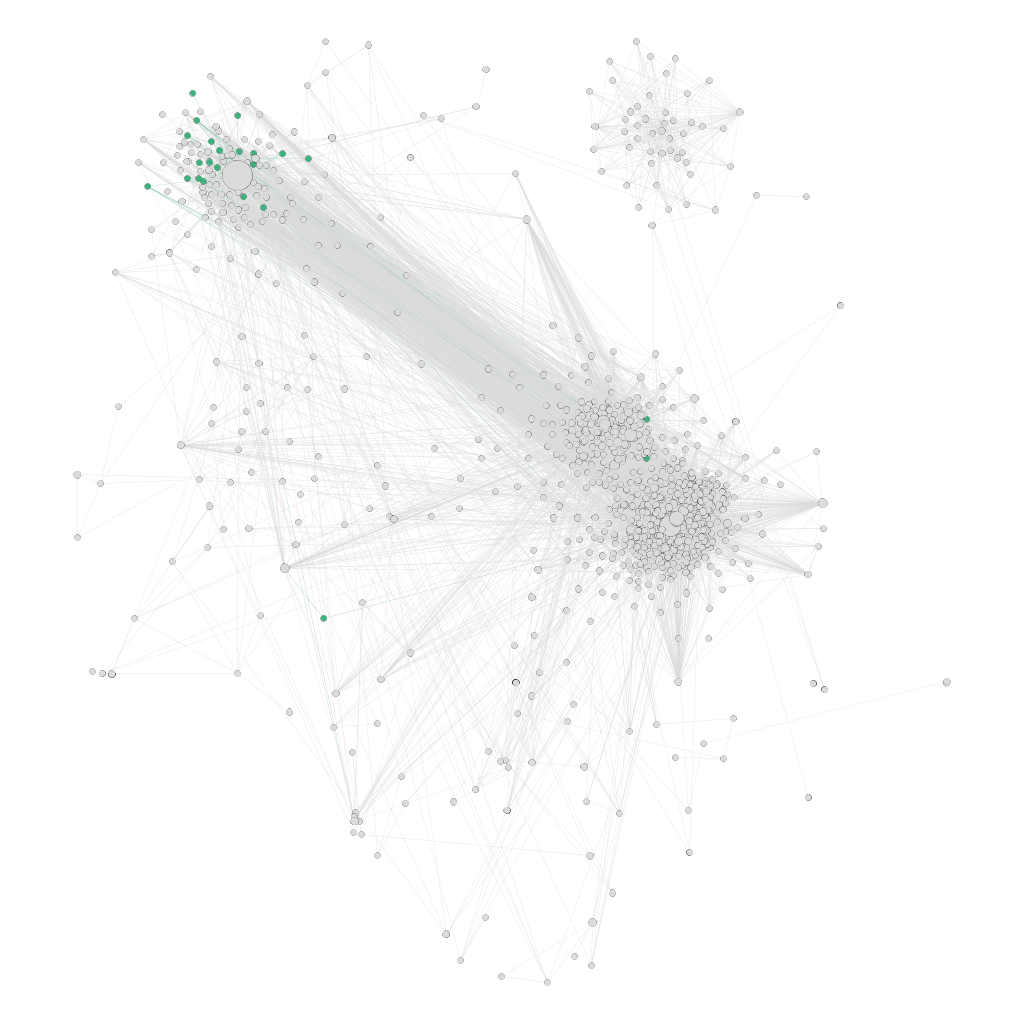}
            \end{figure}
        \end{minipage}
    \end{minipage}
\caption{\textbf{Retweeting network of users without bot information.} \figletter{a},~Spreading patterns between humans, bots and users without bot information (blue = humans, red = bots, grey = users without bot information). The node size represents the number of users per account type. The edges represent the direction and relative frequency of retweets. \figletter{b},~Retweet network with accounts from India colored in purple. \figletter{c},~Retweet network with accounts from the U.S. colored in blue. \figletter{d},~Retweet network with accounts from South Africa colored in green. The retweet networks were visualized using Gephi \cite{Gephi.2009} (see \nameref{sec:methods}). The networks corroborate the findings of the main analysis: Accounts from India and South Africa formed relatively isolated retweet clusters while accounts from the U.S. were more broadly scattered over the network.
}
\label{fig:network_undefined}
\end{figure}

\subsection*{Supplementary Figure 6: Online virality of bots and humans}
\label{supp:ccdfs_bot_human}

\begin{figure}[H]
\centering
    \begin{minipage}{0.29\textwidth}
    \centering
        \begin{figure}
            \centering
            \raggedright\figletter{a}\\
            \includegraphics[width=\textwidth]{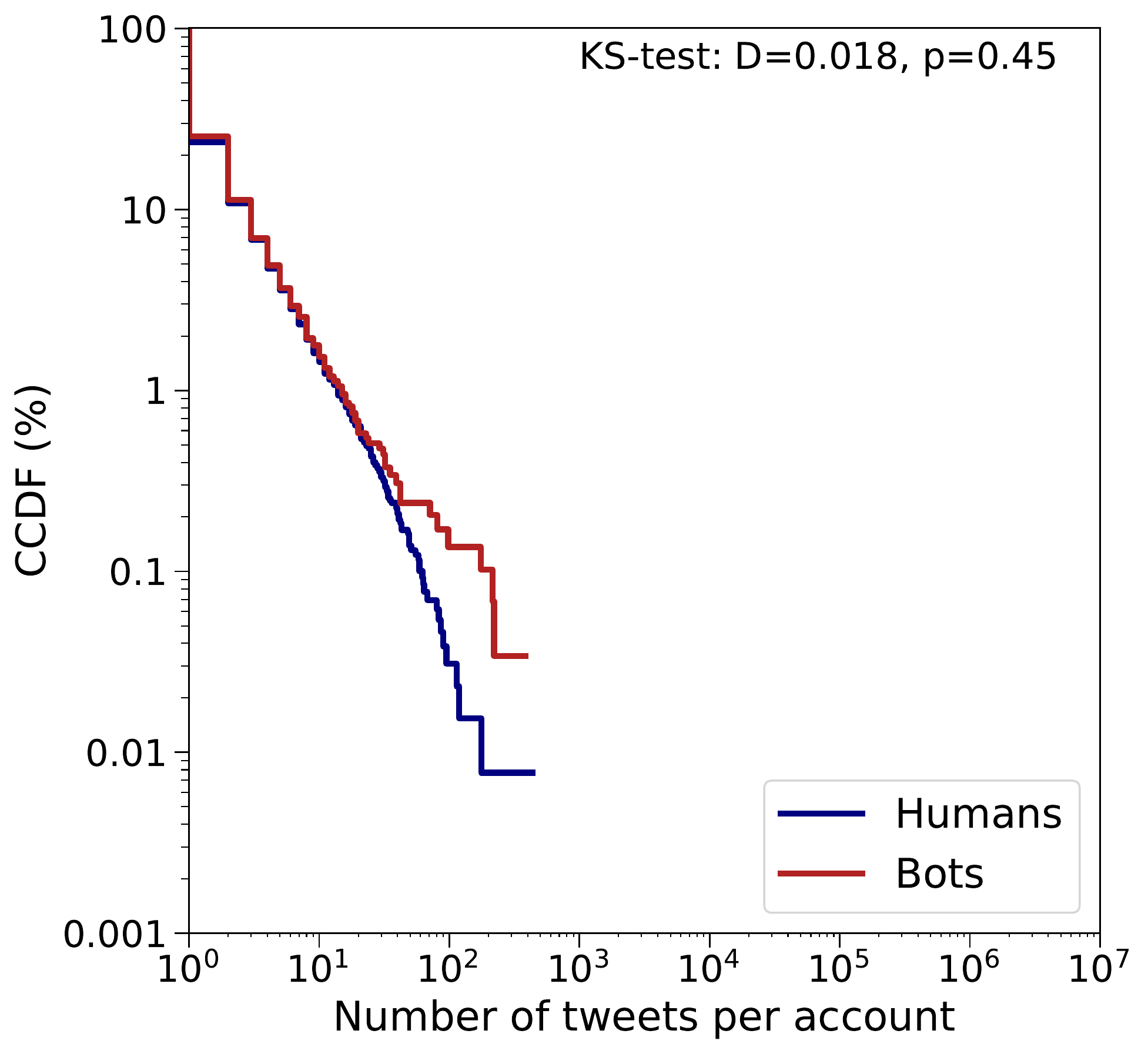}
        \end{figure}
    \end{minipage}
    \quad\quad
    \begin{minipage}{0.29\textwidth}
        \begin{figure}
            \centering
            \raggedright\figletter{b}\\
            \includegraphics[width=\textwidth]{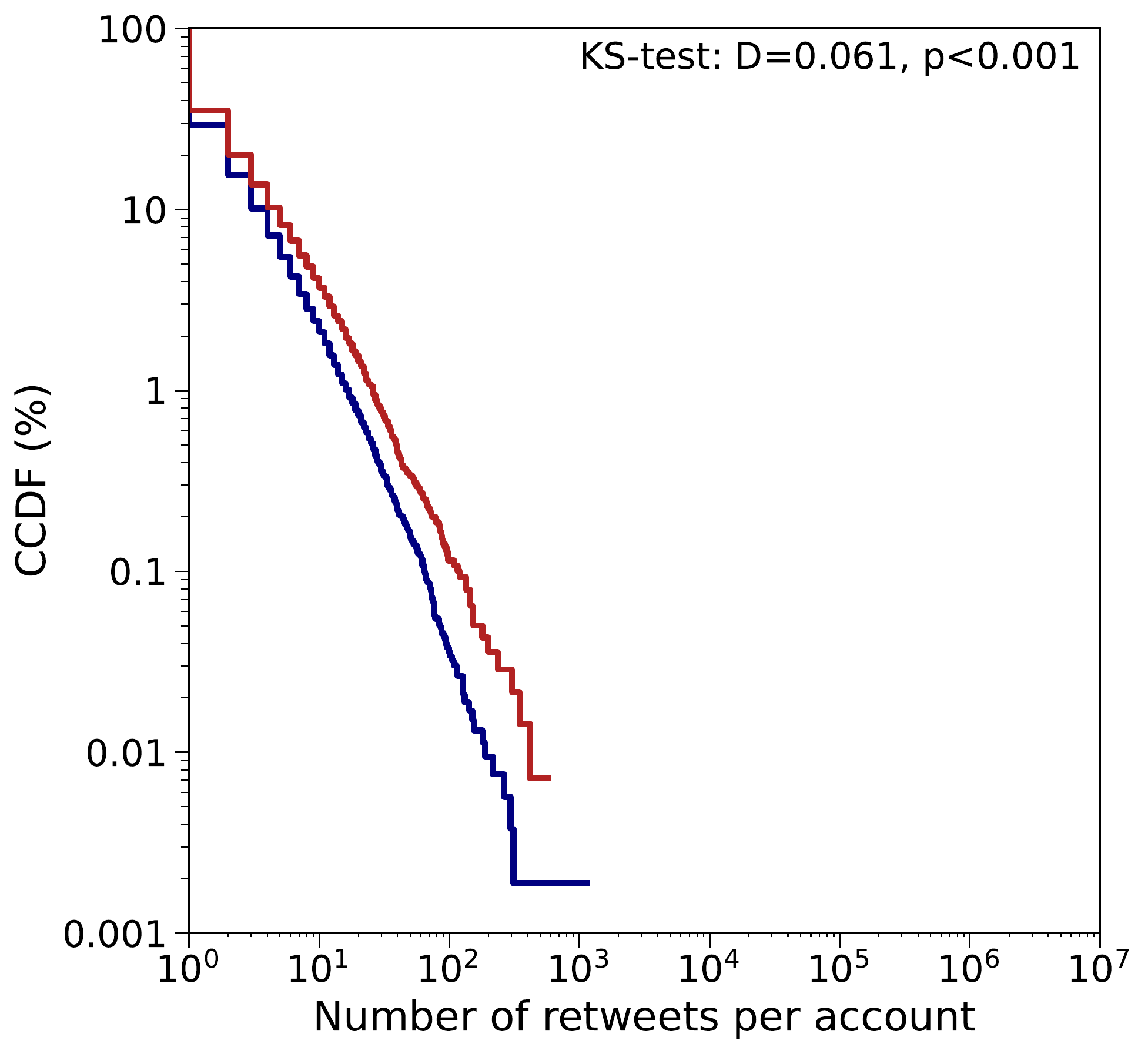}
        \end{figure}
    \end{minipage}
    \quad\quad
    \begin{minipage}{0.29\textwidth}
        \begin{figure}
            \centering
            \raggedright\figletter{c}\\
            \includegraphics[width=\textwidth]{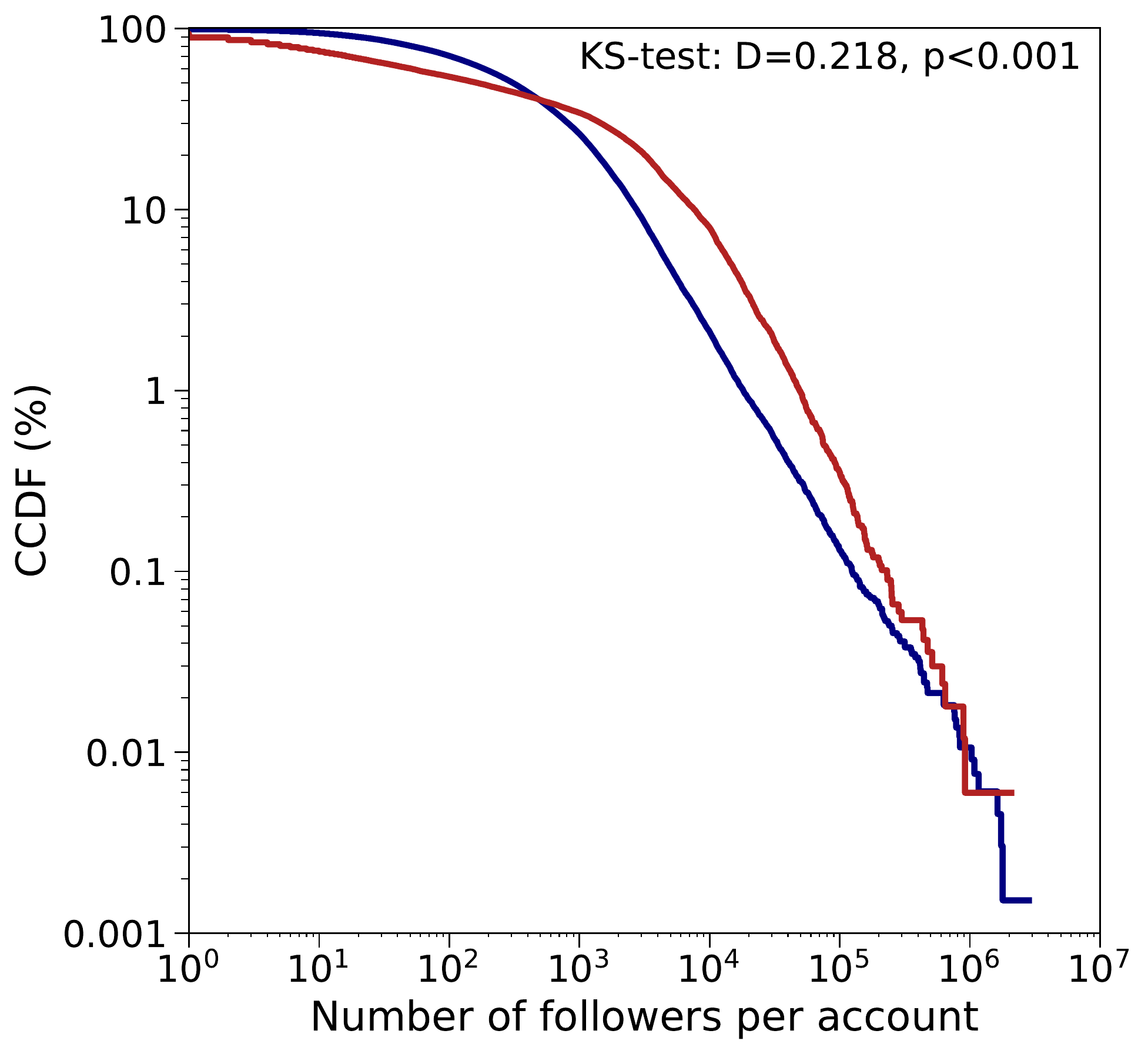}
        \end{figure}
    \end{minipage}
    
    \begin{minipage}{0.29\textwidth}
    \centering
        \begin{figure}
            \centering
            \raggedright\figletter{d}\\
            \includegraphics[width=\textwidth]{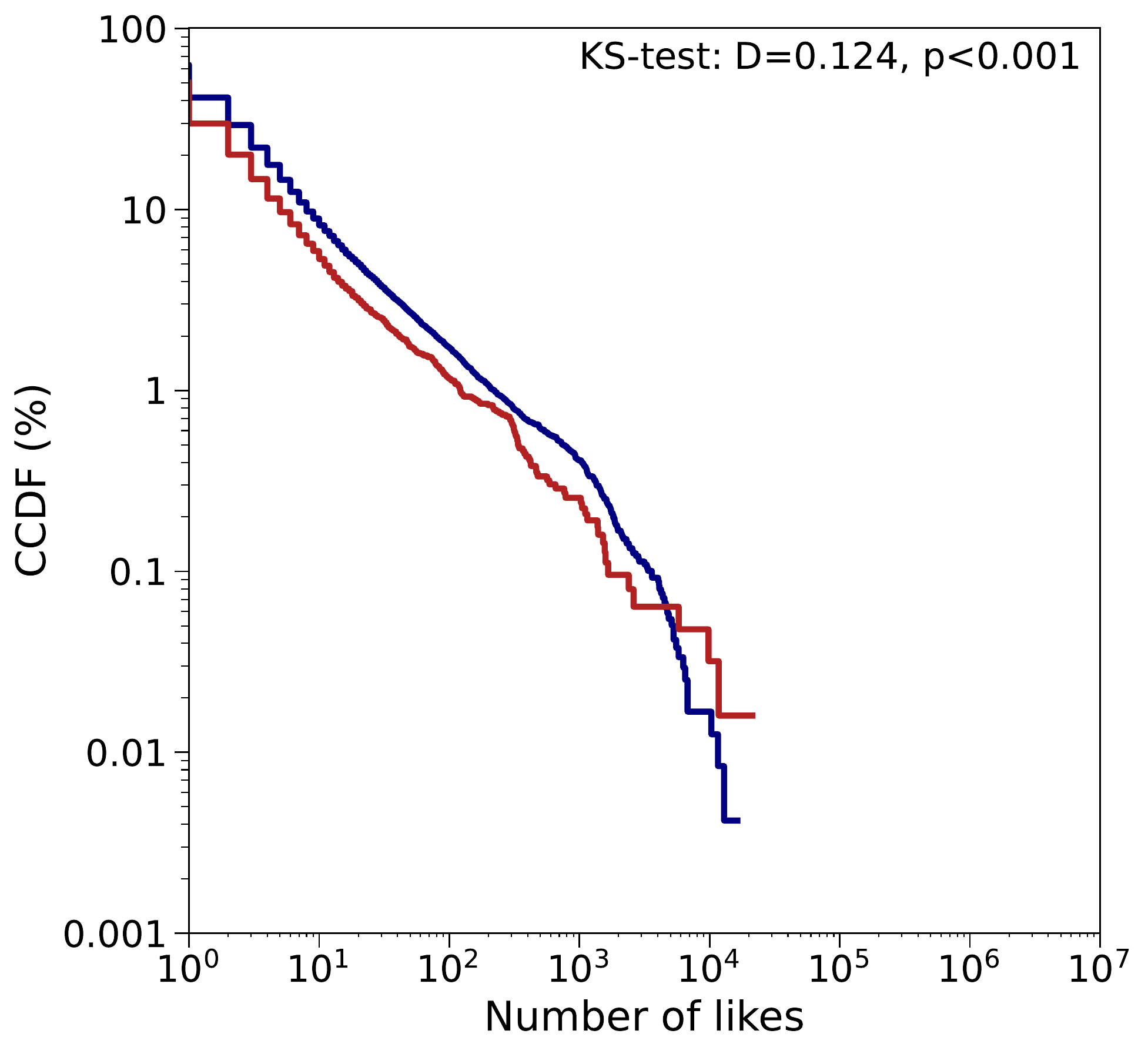}
        \end{figure}
    \end{minipage}
    \quad\quad
    \begin{minipage}{0.29\textwidth}
        \begin{figure}
            \centering
            \raggedright\figletter{e}\\
            \includegraphics[width=\textwidth]{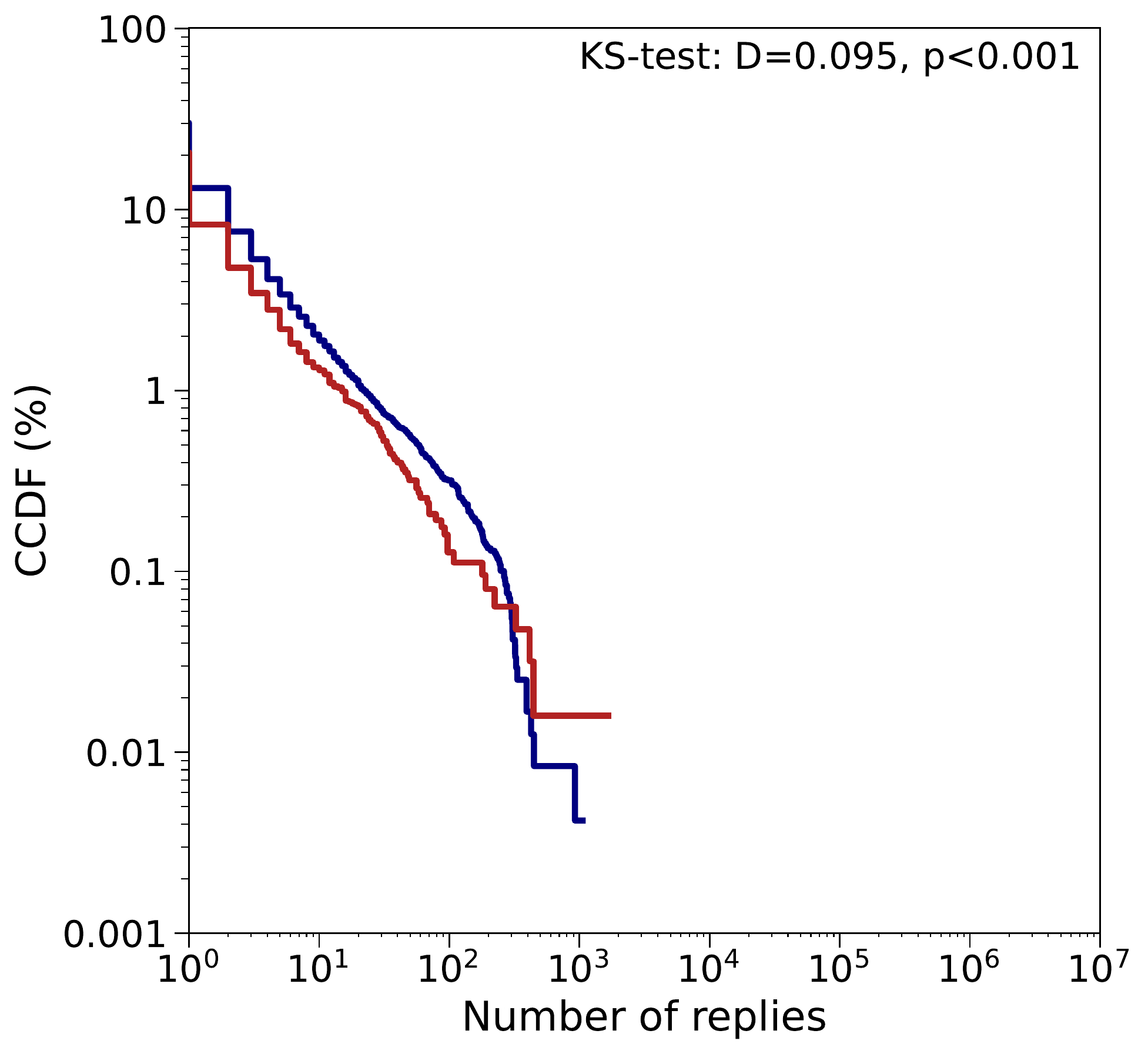}
        \end{figure}
    \end{minipage}
    \quad\quad
    \begin{minipage}{0.29\textwidth}
        \begin{figure}
            \centering
            \raggedright\figletter{f}\\
            \includegraphics[width=\textwidth]{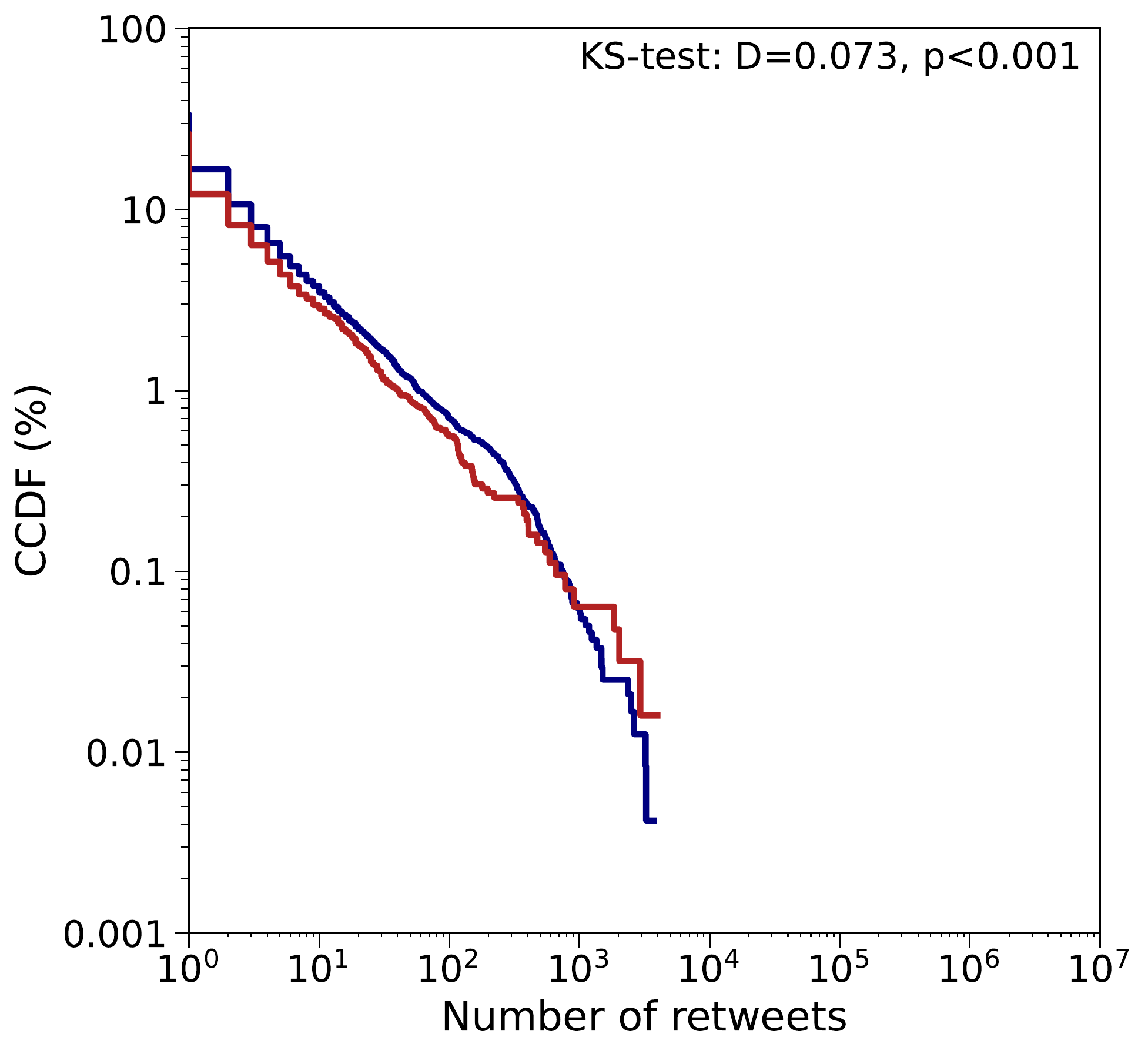}
        \end{figure}
    \end{minipage}
\caption{\textbf{Online virality of bots and humans.} Here, we compare complementary cumulative distribution functions (CCDFs) for humans vs. bots across the following dimensions: \figletter{a},~the number of source tweets per account; \figletter{b},~the number of retweets per account; \figletter{c},~the number of followers per account; \figletter{d},~the number of likes; \figletter{e},~the number of replies; and \figletter{f},~the number of retweets. The statistics in (a--c) are computed at the user level, while the statistics in (d--f) are computed at the interaction level. Statistical comparisons are based on a Kolmogorov-Smirnov (KS) test \cite{Massey.1951}. 
}
\label{fig:ccdfs_bot_human}
\end{figure}

\newpage
\section{Supplementary Materials: Content analysis of the Russian propaganda campaign}
\label{supp:content}


To analyze the narrative in the Russian propaganda campaign, we performed a quantitative and qualitative study of the underlying content. Thereby, we substantiate two of our previous observations related to (1)~the subject and (2)~the intended recipient (``target''). In terms of (1), we see that Russian propaganda made several claims that discredit Western countries and organizations. In terms of (2), we see that Russian propaganda was especially active around the UN vote and in specific countries. This suggests that the propaganda campaign was intended to steer the public opinion in those countries.  

\underline{Subject.} We analyzed the frequency of keyword terms related to different geographic regions and international bodies. For this, we first tokenized the messages, then performed pattern matching, and eventually counted the number of matches. Here, we focused on the following entities: the EU (via the search terms ``eu'' and ``european union''), NATO (via ``nato''), the UN (via ``un'' and ``united nations''), the USA (via ``usa''), the West in general (via ``west''). In our dataset, we observe that the following entities are common (reported as the relative share of messages that include the search terms): NATO (\num{7.1}\%), the West (\num{5.4}\%), and USA (\num{4.1}\%).

We then manually analyzed a random sample of \num{100} messages to better understand the underlying narrative. Here we find that messages frequently discredit Western countries and organizations. For example, one account posted: \emph{\say{Russian People were with South Africans during difficult times of Apartheid, they never deserted us till today, they are indeed Africans in heart in Russian County, they assisted us when the rest were supporting our oppressors. \texttt{\#IStandWithRussia}}} Another account then added: \emph{\say{@WailetVersteeg @KremlinRussia\_E \texttt{\#IStandWithPutin} Because enemy of my enemy is my friend! 95\% of mass murder(war) injustice oppression etc in Africa \& world since \num{1000} years ago, is caused by the West (NATO) in their quest to enslave others \& milk their natural resources! \texttt{\#CREATORCRACY} \texttt{\#EmbraceTruth} \texttt{\#BIAFRA}}}. This thus gives an example that connects to Western countries and institutions. 

The qualitative analysis further revealed that the Russian propaganda campaign frequently picked up existing as well as historical narratives (e.g., Apartheid, slavery, oppression, imperial times, mass murders). A similar observation was made earlier for propaganda originating from Russian traditional media \cite{Sloane.2022, Yablokov.2022}. 

\underline{Intended recipient.} To analyze the connection between the Russian propaganda campaign and the UN General Assembly vote on Resolution \mbox{ES-11/1}, we took a closer look at the countries that voted against or abstained from the vote \cite{UN.GA.2022}. Overall, there were 5 countries voting against (Belarus, Eritrea, North Korea, Russia, Syria) and 35 countries abstaining (e.g., South Africa, India, Pakistan, China, and Iraq). Here, we again analyzed the frequency of specific search terms that related to the previous countries of interest. Specifically, we tokenized the messages and then counted the relative frequency of messages that included a country name (in English). Common country names were: Iraq (\num{3.3}\%), Syria (\num{2.7}\%), and India (\num{2.5}\%). Evidently, several of the commonly mentioned countries were also those that were prone to abstain from the UN vote.

We further manually analyzed the above random sample of \num{100} messages to better understand the underlying narrative. There was, for example, a stream of messages that mentioned the support for Putin by South Africans due to Russia's help during the time of Apartheid. As another example, we find that the messages frequently make the connection between India and Russia by linking to historical support of Russia during imperial times and in the UN. Here, one user posted for example: \emph{\say{Yes! We are friends with the US and the West, but Russia is our brother!!! Russia have always supported India in UN no matter what the issue is. Today, India remaining abstain alongside China \& the UAE says a lot. We remember everything you’ve done for us. \texttt{\#istandwithrussia}}}. This provides further quantitative and qualitative evidence that connects the Russian propaganda campaign with the UN vote. 

In sum, the findings of the above analysis are two-fold: (1)~We find that, as the underlying mechanism, the campaign is -- to a large extent -- focused on eliciting negative opinions towards Western countries and institutions. (2)~We further find that it targets specifically countries that were prone to abstain from the UN vote. 

\newpage

\newpage
\section{Supplementary Materials: Sentiment analysis}
\label{supp:sentiment}

We further analyzed the content of pro-Russian source tweets with regard to sentiment and emotions to examine whether humans and bots use a different tone when disseminating propaganda. We first preprocessed the source tweets by removing numbers, mentions, hashtags, and links from the content. Subsequently, we used the NRC lexicon \cite{Mohammad.2013} to assign scores to the source tweets related to both sentiment and different emotions. Specifically, the scores capture the following dimensions: positive, negative, and neutral sentiment as well as the eight emotions of the NRC Lexicon (fear, anger, trust, surprise, sadness, disgust, joy, and anticipation). To do so, we classified source tweets as positive or negative depending on the predominant sentiment score, and as neutral when the scores are equal. 

We show the percentage of positive, negative, and neutral source tweets for humans and bots in Supplementary~Figure~\ref{fig:sentiment}a. Overall, the differences between humans and  bots were comparatively small. In particular, the hypothesis that bots may strategically opt for negative language was not supported. A similar observation has been made in earlier research \cite{Smart.2022}. Importantly, this is different from bots spreading, for example, inflammatory content, where negative language is more common \cite{Stella.2018}. As an additional analysis, we also computed the mean percentage of emotions in the source tweets of  humans and  bots (Supplementary~Figure~\ref{fig:sentiment}b). Again, the scores were fairly similar for both  humans and  bots. 

\begin{figure}[H]
\centering
    \begin{minipage}{\textwidth}
    \centering
        \begin{minipage}{0.49\textwidth}
            \begin{figure}
                \centering
                \raggedright\figletter{a}\\
                \includegraphics[width=\textwidth]{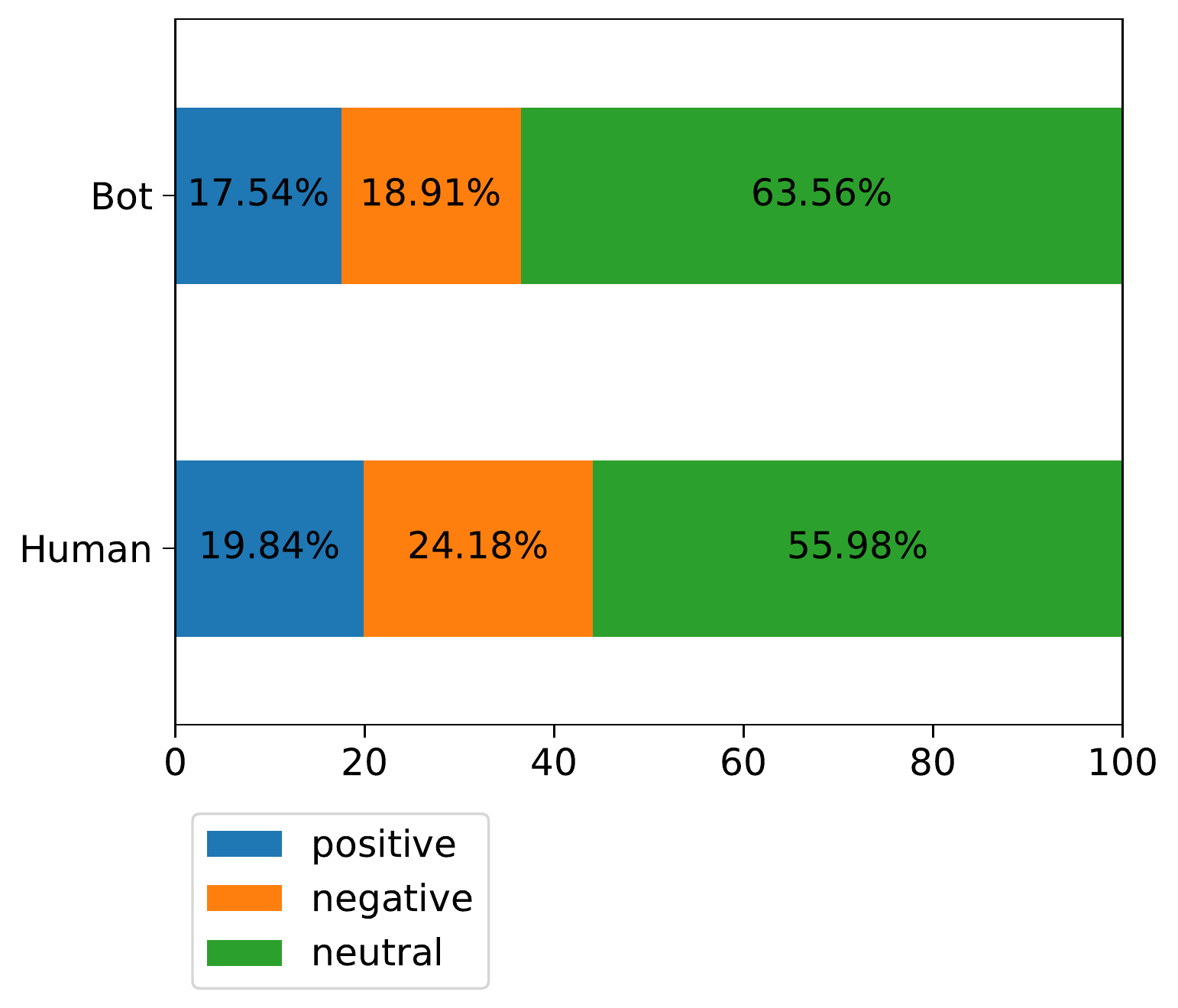}
            \end{figure}
        \end{minipage}
        \begin{minipage}{0.49\textwidth}
            \begin{figure}
                \centering
                \raggedright\figletter{b}\\
                \includegraphics[width=\textwidth]{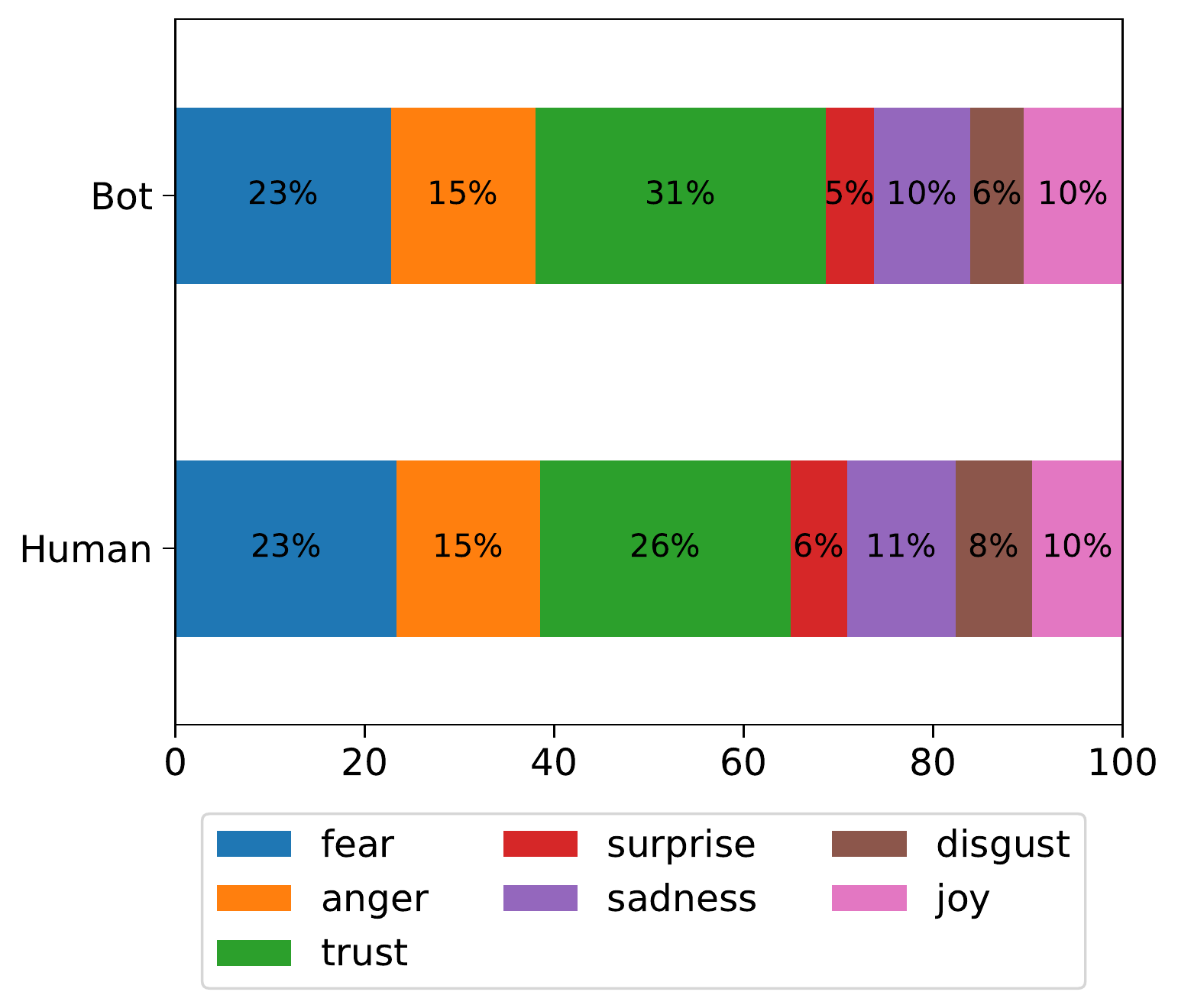}
            \end{figure}
        \end{minipage}
    \end{minipage}
\caption{\textbf{Sentiment and emotions of source tweets.} Here, we classified source tweets of  humans and  bots into different sentiment and emotion categories using the NRC lexicon \cite{Mohammad.2013}. Shown are:
\figletter{a},~relative frequency of source tweets with predominantly positive, negative, or neutral sentiment for  humans and  bots; and \figletter{b},~relative frequency of the different emotions categories in source tweets from  bots and  humans.}
\label{fig:sentiment}
\end{figure}

\end{document}